\documentclass{emulateapj}
\usepackage{apjfonts}
\usepackage{rotating}
\newcommand{\beginsidefig}{\begin{sidewaysfigure*}}
\newcommand{\xendsidefig}{\end{sidewaysfigure*}}
\newcommand{\figexpand}{\epsscale{1.15}}
\newcommand{\plotter}{\plotone}

\newcommand{\etal}{et al.}
\newcommand{\mbh}{M_{\rm BH}}
\newcommand{\mstar}{M_{\ast}}
\newcommand{\lstar}{L_{\ast}}

\newcommand{\fgas}{f_{\rm gas}}

\newcommand{\msun}{M_{\sun}}
\newcommand{\tH}{t_{\rm H}}
\newcommand{\tmerger}{t_{\rm merger}}

\newcommand{\mhalo}{M_{\rm halo}}
\newcommand{\mgal}{M_{\rm gal}}

\newcommand{\mmerger}{M_{\rm merger}}
\newcommand{\paperone}{Paper \textrm{I}}
\newcommand{\papertwo}{Paper \textrm{II}}

\shorttitle{Co-Evolution of Quasars, Black Holes, and Galaxies \textrm{I}}
\shortauthors{Hopkins \etal}
\slugcomment{Submitted to ApJ, June 8, 2007}
\begin{document}

\title{A Cosmological Framework for the Co-Evolution of Quasars, Supermassive Black
Holes, and Elliptical Galaxies: \textrm{I}. Galaxy Mergers \& Quasar Activity}
\author{Philip F. Hopkins\altaffilmark{1}, 
Lars Hernquist\altaffilmark{1}, 
Thomas J. Cox\altaffilmark{1}, 
\&\ Du{\v s}an Kere{\v s}\altaffilmark{1}
}
\altaffiltext{1}{Harvard-Smithsonian Center for Astrophysics, 
60 Garden Street, Cambridge, MA 02138}

\begin{abstract}
We develop a model for the cosmological role of mergers in the evolution of 
starbursts, quasars, and spheroidal galaxies. By combining theoretically well-constrained 
halo and subhalo mass functions as a function of redshift and 
environment with empirical halo occupation models, we can estimate where 
galaxies of given properties live at a particular epoch. This allows us to 
calculate, in an {\em a priori} cosmological manner, where major galaxy-galaxy 
mergers occur and what kinds of galaxies merge, at all redshifts. 
We compare this with the observed mass functions, clustering, fractions as a function 
of halo and galaxy mass, and small-scale environments of mergers, and show 
that this approach yields robust estimates in good agreement with
observations, and can be extended to predict detailed properties 
of mergers. Making the simple ansatz that 
major, gas-rich
mergers cause quasar activity (but not strictly assuming they are the only 
triggering mechanism), we demonstrate that this model naturally reproduces 
the observed rise and fall of the quasar luminosity density from $z=0-6$, as well as
quasar luminosity functions, fractions, host galaxy colors, and clustering 
as a function of redshift and luminosity.
The recent observed excess of quasar clustering on small scales at $z\sim0.2-2.5$ 
is a natural prediction of our model, as mergers will preferentially occur in regions 
with excess small-scale galaxy overdensities.
In fact, we demonstrate that quasar environments at all observed redshifts 
correspond closely to the empirically determined small group scale, where
major mergers of $\sim L_{\ast}$ gas-rich galaxies will be most efficient. 
We contrast this with a secular model in which quasar activity is driven by 
bars or other disk instabilities, and show that while these modes of fueling 
probably dominate the high-Eddington ratio population at Seyfert luminosities 
(significant at $z=0$), 
the constraints from quasar clustering, 
observed pseudobulge populations, and disk mass functions 
suggest that they are a small contributor to the $z\gtrsim1$ quasar luminosity density, 
which is dominated by massive BHs in predominantly classical 
spheroids formed in mergers. 
Similarly, low-luminosity Seyferts do not show a clustering excess on small scales, 
in agreement with the natural prediction of secular models, but bright quasars at all redshifts do so. 
We also compare recent observations of the colors of quasar host galaxies, and 
show that these correspond to the colors of recent 
merger remnants, in the transition region between the blue cloud and 
the red sequence, and are distinct from the colors of systems with observed 
bars or strong disk instabilities. Even the most extreme secular models, 
in which all bulge (and therefore BH) formation proceeds via disk instability, 
are forced to assume that this instability acts before the (dynamically inevitable) mergers, and 
therefore predict a history for the quasar luminosity density which is 
shifted to earlier 
times, in disagreement with observations. 
Our model provides a powerful means to predict the abundance and 
nature of mergers, and to contrast cosmologically motivated predictions of 
merger products such as starbursts and AGN. 
\end{abstract}


\section{Introduction}
\label{sec:intro}

\subsection{Motivation}
\label{sec:intro:motives}

Over the past decade, observations have established that supermassive
black holes likely reside in the centers of all galaxies with
spheroids \citep[e.g.,][]{KormendyRichstone95,Richstone98,KormendyGebhardt01}, 
and that the properties of these black holes
and their hosts are correlated.  These correlations take various
forms, relating the black hole mass to e.g.\ the mass \citep{magorrian,
mclure.dunlop:magorrian,marconihunt,haringrix}, 
velocity dispersion \citep{FM00,Gebhardt00,tremaine:msigma}, 
and concentration or Sersic index \citep{graham:concentration,graham:sersic} 
of the spheroid.  Most recently,
\citet{hopkins:bhfp} have demonstrated that these relationships are
not independent and can be understood as various projections of a
black hole fundamental plane analogous to the fundamental plane
for elliptical galaxies \citep{dressler87:fp,dd87:fp}. 
The striking similarity between these two fundamental planes
indicates that galaxy spheroids and supermassive black holes are not
formed independently, but originate via a common physical process.

Furthermore, although there may be some relatively weak evolution in the 
correlation between BH mass and host mass or velocity dispersion 
owing to changes in spheroid structural properties and 
internal correlations with redshift 
\citep[e.g.,][]{peng:magorrian.evolution,
shields03:msigma.evolution,shields06:msigma.evolution,walter04:z6.msigma.evolution,
salviander:msigma.evolution,woo06:lowz.msigma.evolution,hopkins:msigma.limit}, 
the fundamental plane appears to be 
preserved \citep{hopkins:bhfp}, and in any case {\em some} correlation 
exists at all redshifts. There are not, at any redshifts, bulgeless 
systems with large black holes or bulges without correspondingly 
large black holes. This empirically demonstrates that whatever 
process builds up black hole mass {\em must} trace the formation of 
spheroids (albeit with potentially redshift-dependent efficiency). 

These connections extend to other phenomena associated with galaxies
that have sometimes been interpreted as being independent.  For
example, by estimating the total energy radiated by quasars, \citet{soltan82} 
showed that nearly all the mass in supermassive black holes
must have been accumulated during periods of bright quasar activity.
This analysis has since been revisited on a number of occasions
\citep{salucci:bhmf,yutremaine:bhmf,marconi:bhmf,shankar:bhmf,yulu:bhmf}, 
with various assumptions for
quasar obscuration and bolometric corrections. 
\citet{hopkins:bol.qlf} 
have reformulated the Soltan argument from the evolution of the {\it
bolometric} quasar luminosity function (LF).  In their analysis,
Hopkins et al. combined observations of the quasar LF in a variety of
wavebands with purely empirical determinations of the luminosity
dependence of quasar obscuration and spectral emission to
infer the bolometric quasar LF.  By integrating this over luminosity
and redshift, it is then possible to obtain a {\it model-independent}
estimate of the total energy density of radiation from quasars.  The
cosmic black hole mass density then follows if black holes in quasars
accrete with constant radiative efficiency $\epsilon_{r}$ \citep{shakurasunyaev73}, 
by integrating $L_{\rm bol} = \epsilon_{r}\,
\dot{M}_{\rm BH}\,c^{2}$.  This yields a $z=0$ black hole mass density of 
\begin{equation}
\rho_{\rm BH}(z=0) = {4.81}^{+1.24}_{-0.99}\,{\Bigl(}\frac{0.1}
{\epsilon_{r}}{\Bigr)}\,h_{70}^{2}\times 10^{5}\,M_{\sun}\,{\rm M
pc^{-3}},
\end{equation}
consistent with estimates of $\rho_{\rm BH}(z=0)$ obtained from
local bulge mass, luminosity, and velocity dispersion functions
\citep[e.g.,][]{marconi:bhmf,shankar:bhmf}.

Taken together, the black hole fundamental plane
and the Soltan argument imply that the common
physical process which produces galaxy spheroids and supermassive
black holes also must be responsible for triggering {\em most} bright
quasars.  Moreover, there is compelling evidence that quasar activity
is preceded by a period of intense star formation in galaxy centers so
that, for example, ultraluminous infrared galaxies (ULIRGs) and
distant submillimeter galaxies (SMGs) would eventually evolve into
quasars \citep{sanders88:quasars,sanders88:warm.ulirgs,sanders96:ulirgs.mergers,
dasyra:pg.qso.dynamics}. Essentially all sufficiently deep studies of 
the spectral energy distributions (SEDs) of
quasar host galaxies
reveal the presence of young stellar populations 
indicative of a recent starburst 
\citep{brotherton99:postsb.qso,canalizostockton01:postsb.qso.mergers,
kauffmann:qso.hosts,yip:qso.eigenspectra,
jahnke:qso.host.sf,jahnke:qso.host.uv,sanchez:qso.host.colors,
vandenberk:qso.spectral.decomposition,barthel:qso.host.sf,zakamska:qso.hosts}. 
There further appears to be a 
correlation in the sense that the most luminous quasars have the youngest 
host stellar populations \citep{jahnke:qso.host.sf,vandenberk:qso.spectral.decomposition} and 
the greatest prominence of post-merger tidal features and disturbances 
\citep{canalizostockton01:postsb.qso.mergers,kauffmann:qso.hosts,
hutchings:redqso.lowz,hutchings:highz,hutchings:redqso.midz,
zakamska:qso.hosts,letawe:qso.merger.ionization}.  
These observations indicate that intense starbursts
must result from the same process as 
most quasars and supermassive black holes.

In the simplest interpretation, we seek an explanation for the various
phenomena summarized above such that they result from the {\it same
event}.  There are general, theoretical requirements that any such
event must satisfy.  In particular, it must be fast and violent, blend
together gas and stellar dynamics appropriately, and involve a supply
of mass comparable to that in large galaxies.  Why should this be the
case?

The accepted picture for the growth of supermassive black holes
is that the mass is primarily assembled by gas accretion \citep{Lynden-Bell69}. 
From the Soltan argument, we know that this mass
must be gathered in a time comparable to the lifetimes of bright
quasars, which is similar to the \citet{salpeter64} time $\sim 10^{7.5}$
years, for black holes accreting at the Eddington rate. Independent 
limits \citep[][and references therein]{martini04} from quasar 
clustering, variability, luminosity function evolution, and other methods 
demand a {\em total} quasar lifetime (i.e.\ duration of major growth for 
a given BH) of $\lesssim\,10^{8.5}\,{\rm yr}$. 
In order to explain the existence of black holes with masses
$\sim 10^{9} M_\odot$, the amount of
gas required is likely comparable to that contained in entire large
galaxies. Thus, the process we seek must be able to deliver
a galaxy's worth of gas to the inner regions of a galaxy on a
relatively short timescale, $\ll10^{9}$ years.

If this event is to simultaneously build galaxy spheroids, it must
involve stellar dynamics acting on a supply of stars similar to that
in large galaxies because the stellar mass is $\sim 1000$ times larger 
than that of the
black hole and it is believed that spheroids are assembled mainly
(albeit not entirely) 
through dissipationless physics (i.e.\ the movement of stars from 
a circular disk to random spheroid orbits).  A plausible candidate process is
violent relaxation \citep[e.g.][]{Lynden-Bell67} which has been
demonstrated to yield phase space distributions akin to those of
elliptical galaxies through large, rapid fluctuations in the
gravitational potential.  Violent relaxation operates on a timescale
similar to the free-fall time for self-gravitating systems, again
$\ll 10^{9}$ years for the bulk of the mass.

Motivated by these considerations, \citet{hopkins:qso.all} developed a
model where starbursts, quasars, supermassive black hole growth, and
the formation of red, elliptical galaxies are connected through an
evolutionary sequence, caused by {\it mergers} between {\it gas-rich}
galaxies.  There is, in fact, considerable observational evidence
indicating that mergers are responsible for triggering ULIRGs, SMGs,
and quasars \citep[see references in Hopkins et al.\ 2006a; for
reviews see][]{barneshernquist92, schweizer98,jogee:review}. 
Furthermore, the long-standing ``merger
hypothesis,'' which proposes that most elliptical galaxies formed in
mergers \citep{toomre72,toomre77}, is supported by the
structure of known ongoing mergers \citep[e.g.,][]{schweizer92,
rothberg.joseph:kinematics,rothberg.joseph:rotation} and the
ubiquitous presence of fine structures such as shells, ripples,
tidal plumes, nuclear light excesses, and
kinematic subsystems in ellipticals \citep[e.g.][]{schweizerseitzer92,
schweizer96}, 
which are signatures of mergers 
\citep[e.g.][]{quinn.84,hernquist.quinn.87,hernquist.spergel.92,
hernquist:kinematic.subsystems,mihos:cusps}.

Numerical simulations performed during the past twenty years verify
that {\it major} mergers of {\it gas-rich} disk galaxies can plausibly
account for these phenomena and have elucidated the underlying physics.
Tidal torques excited during a merger lead to rapid inflows of gas
into the centers of galaxies \citep{hernquist.89,barnes.hernquist.91,
barneshernquist96}. 
The amount of gas involved can be a large fraction of
that in the progenitor galaxies and is accumulated on roughly a
dynamical time in the inner regions, $\ll 10^9$ years \citep{hernquist.89}.
The resulting high gas densities trigger starbursts \citep{mihos:starbursts.94,
mihos:starbursts.96}, and feed rapid black hole growth \citep{dimatteo:msigma}.
Gas consumption by the starburst and dispersal of residual
gas by supernova-driven winds and feedback from black hole growth 
\citep{springel:red.galaxies} terminate star formation so that the remnant
quickly evolves from a blue to a red galaxy.  The stellar component of
the progenitors provides the bulk of the material for producing the
remnant spheroid \citep{barnes:disk.halo.mergers,barnes:disk.disk.mergers,
hernquist:bulgeless.mergers,hernquist:bulge.mergers}
through violent relaxation.

The simulations also place significant constraints on the types of
mergers that can initiate this sequence of events.  First, a major
merger is generally required in order for the tidal forces to excite a
sufficiently strong response to set up nuclear inflows of gas.
Although simulations involving minor mergers with mass ratios $\sim
10:1$ show that gas inflows can be excited under some circumstances
\citep[e.g.][]{hernquist.89,hernquist.mihos:minor.mergers,bournaud:minor.mergers}, a systematic study
indicates that such an outcome is limited to specific orbital
geometries \citep{younger:minor.mergers} and 
that the overall efficiency of triggering inflows declines rapidly 
with increasing mass ratio.  Thus, while the precise
definition of a major merger in this context is blurred by the
degeneracy between the mass ratio of the progenitors and the orbit of
the interaction, it appears that a mass ratio $\sim 3:1$ or smaller is
needed. 
This is further supported by
observational studies \citep{dasyra:mass.ratio.conditions,woods:tidal.triggering}, 
which find 
that strong gas inflows and nuclear starbursts are typically seen
only below these mass ratios, despite the much greater frequency of 
higher mass-ratio mergers. 

Second, the merging galaxies must contain a supply of {\it cold} gas,
which in this context refers to gas that is rotationally supported, in
order that the resonant response leading to nuclear inflows of gas in
a merger be excited.  Elliptical galaxies contain large quantities of
hot, thermally supported gas, but even major mergers between two such
objects will not drive the nuclear inflows of gas that fuel rapid
black hole growth.

It also must be emphasized that essentially all numerical studies 
of spheroid kinematics find that {\em only} mergers 
can reproduce the observed kinematic properties of elliptical 
galaxies and ``classical'' bulges \citep{hernquist.89,hernquist:bulgeless.mergers,
hernquist:bulge.mergers,barnes:disk.halo.mergers,barnes:disk.disk.mergers,
schweizer92,naab:boxy.disky.massratio,
naab:minor.mergers,naab:gas,naab:dry.mergers,naab:profiles,
bournaud:minor.mergers,jesseit:kinematics,cox:kinematics}. 
Disk instabilities and
secular evolution (e.g.\ bar instabilities, harassment, and other 
isolated modes) can indeed produce bulges, but these are invariably 
``pseudobulges'' \citep{schwarz:disk-bar,athanassoula:bar.orbits,
pfenniger:bar.dynamics,combes:pseudobulges,
raha:bar.instabilities,kuijken:pseudobulges.obs,oniell:bar.obs,athanassoula:peanuts}, 
with clearly distinct shapes (e.g.\ flattened or 
``peanut''-shaped isophotes), rotation properties (large $v/\sigma$), 
internal correlations (obeying different Kormendy and Faber-Jackson relations), 
light profiles (nearly exponential Sersic profiles), and colors and/or 
substructure from classical bulges 
\citep[for a review, see][]{kormendy.kennicutt:pseudobulge.review}. 
Observations indicate that 
pseudobulges constitute only a small fraction of the total mass density 
in spheroids \citep[$\lesssim10\%$; see][]{allen:bulge-disk,ball:bivariate.lfs,
driver:bulge.mfs}, becoming a large fraction of the bulge 
population only for small bulges in late-type hosts 
\citep[e.g.\ Sb/c, corresponding to typical $\mbh\lesssim10^{7}\,\msun$; see][and 
references therein]{carollo98, kormendy.kennicutt:pseudobulge.review}. 
Therefore, it is clear that although such processes may be important 
for the buildup of the smallest black hole and spheroid 
populations, secular evolution {\em cannot} be the agent 
responsible for the formation of most 
elliptical galaxies, or for the buildup of 
most black hole mass, or the triggering of bright quasar activity. 

We are thus led to suggest a generalization of the merger hypothesis
proposed by \citet{toomre77} whereby major mergers of {\it gas-rich} disk
galaxies represent the dominant process for producing the supermassive
black hole and spheroid populations in the Universe.  Then, by the
Soltan argument and the association of starbursts with quasars, it
follows that this must also be the primary mechanism for triggering
the most intense infrared luminous galaxies and the brightest quasars
and active galactic nuclei (AGN).  It is important to keep in mind
that this does not rule out other processes occurring at lower levels
and under other circumstances.  For example, we are not claiming that
all AGN result from mergers.  In fact, low levels of such activity, as in
Seyfert galaxies, often appear in undisturbed galaxies.  For these
objects, other modes of fueling are likely more significant, as in the
stochastic accretion scenario of \citet{hopkins:seyferts}.  The primary
requirement on our model is that the bulk of the supermassive black
hole mass density should have accumulated through gas-rich mergers,
consistent with the redshift evolution of the quasar population
\citep{hopkins:bol.qlf}.
Similarly,
spheroid evolution by gas-free (``dry'') mergers will go on, but does
not explain how stellar mass is initially moved onto the red sequence 
or how black hole mass is initially accreted.

\subsection{Outline}
\label{sec:intro:outline}

To test our hypothesis, we have developed methods for following the
growth of black holes in numerical simulations of galaxy mergers,
using a multiphase model for the star-forming gas that enables us to
consider progenitor disks with large gas fractions. Generically, we
find that major mergers of gas-rich galaxies evolve through distinct
phases that can plausibly be identified with the various observed
phenomena summarized above. 

Figure~\ref{fig:outline} presents a 
schematic outline of these phases. 
In this picture, galactic disks grow mainly
in quiescence, with 
the possibility of secular-driven bar or pseudobulge
formation, until the onset of a major merger. A significant, perhaps
even dominant fraction of Seyferts and 
low-luminosity quasars will almost certainly arise from this secular 
evolution, but the prevalence of pseudobulges only in the 
hosts of $\lesssim10^{7}\,\msun$
black holes suggests this is limited to luminosities $M_{B}\gtrsim-23$ 
(see the discussion in \S~\ref{sec:quasars:secular}). 

\beginsidefig
    \centering
    \figexpand
    \plotone{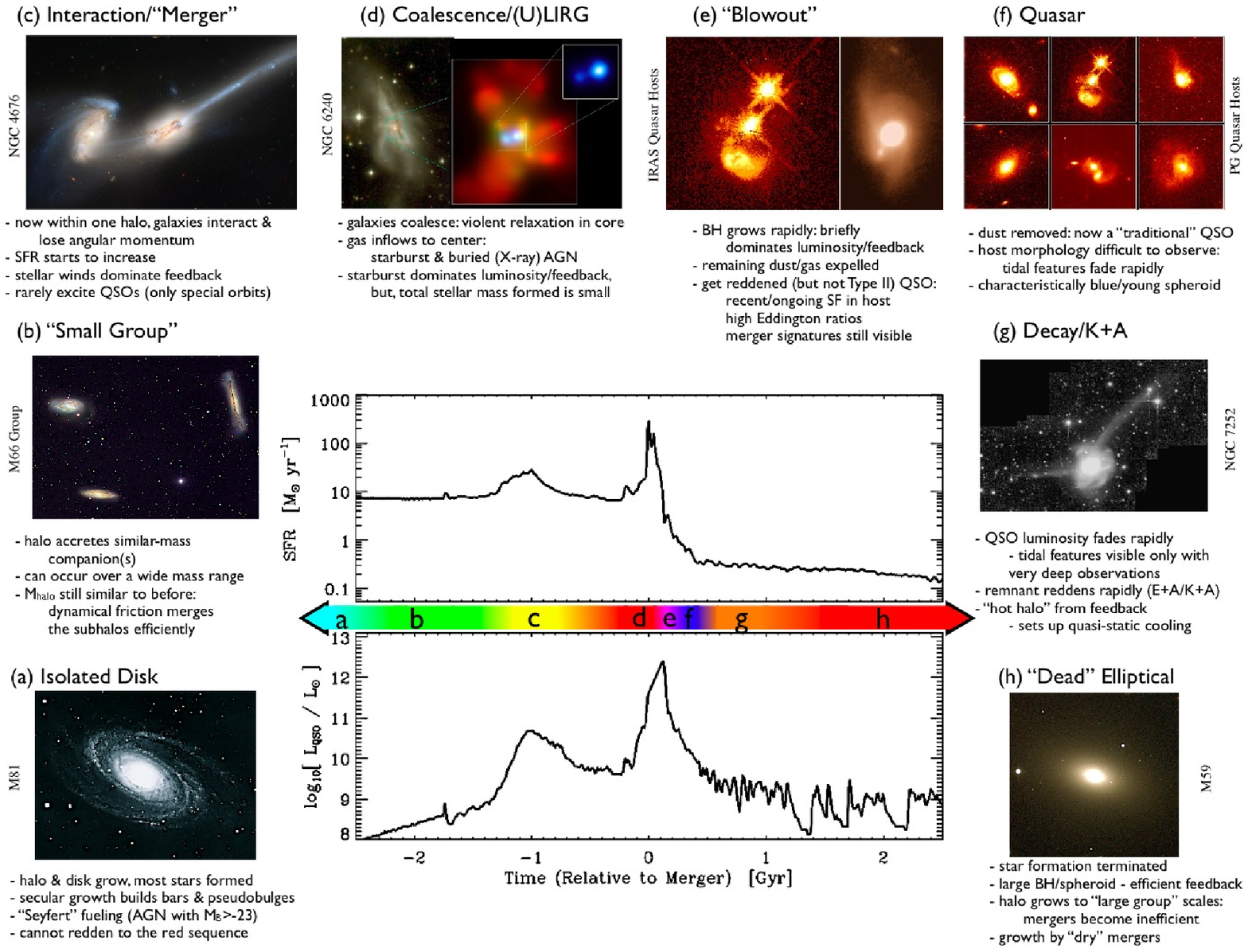}
    \caption{An schematic 
    outline of the phases of growth in a ``typical'' galaxy undergoing a 
    gas-rich major merger.
    {\em Image Credit:} (a) NOAO/AURA/NSF; (b) REU program/NOAO/AURA/NSF; 
    (c) NASA/STScI/ACS Science Team; (d) Optical (left): NASA/STScI/R.\ P.\ van 
    der Marel \&\ J.\ Gerssen; X-ray (right): NASA/CXC/MPE/S.\ Komossa et al.; (e) Left: 
    J.\ Bahcall/M.\ Disney/NASA; Right: Gemini Observatory/NSF/University of Hawaii 
    Institute for Astronomy; (f) J.\ Bahcall/M.\ Disney/NASA; (g) F.\ Schweizer (CIW/DTM); 
    (h) NOAO/AURA/NSF.
    \label{fig:outline}}
\xendsidefig

During the early stages of the merger,
tidal torques excite some enhanced star formation 
and black hole accretion, but the effect is relatively weak, and the combination 
of large galactic dust columns and relatively small nuclear black holes 
means that only in rare circumstances (involving particular initial 
orbits and/or bulge-to-disk ratios) will the pair be identified as Seyferts or quasars. 
Most observationally identified mergers (and essentially all merging pairs) 
will be in this stage, and numerical simulations suggest it is the last stage 
at which the distinct nuclei enable automated morphological selection 
criteria to efficiently 
identify the system as a merger \citep{lotz:gini-m20,lotz:merger.selection}. 
Care must therefore be taken with conclusions regarding the prevalence of 
starbursts and AGN in these samples, as the small observed 
incidence of quasar activity \citep{dasyra:mass.ratio.conditions,
myers:clustering.smallscale,straughn:tadpoles,alonso:agn.in.pairs} is actually expected. 

During the final coalescence of the galaxies, massive inflows of gas trigger
starbursts with strengths similar to those inferred for ULIRGs and
SMGs, although the actual mass in stars formed in these bursts is 
generally small compared to the stellar mass contributed by the merging disks. 
The high gas densities feed rapid black hole growth, but the
black holes are obscured at optical wavelengths by gas and dust 
and are initially small compared to the newly forming spheroid. However, 
by the final stages, high accretion rate, heavily obscured 
(and in some cases nearly Compton-thick) 
BH growth in a ULIRG stage (often with merging binary BHs) appears ubiquitous 
\citep{komossa:ngc6240,alexander:xray.smgs,borys:xray.ulirgs,brand:xray.ir.contrib}, and 
by high redshifts ($z\sim2$) may dominate the obscured luminous quasar 
population \citep{alexander:bh.growth,stevens:xray.qso.hosts,
martinez:host.obscured.qsos,brand:ulirg.qsos}.

Most of the nuclear gas is consumed by the starburst and eventually
feedback from supernovae and the black hole begins to disperse the
residual gas. This brief transition or ``blowout'' phase will be 
particularly associated with highly dust-reddened (as opposed to more 
highly obscured Type II) and/or IR-luminous 
quasars. As a relatively short phase, such objects 
constitute only $\sim20-40\%$ of the quasar population, similar to 
that observed \citep{gregg:red.qsos,white:red.qsos,richards:red.qsos,richards:seds,
hopkins:dust}. In fact, observational studies find 
that red quasar populations are related to mergers, 
with $\gtrsim75\%$ (and as high as $100\%$) showing clear evidence of 
recent/ongoing merging \citep{hutchings:redqso.lowz,hutchings:redqso.midz,
kawakatu:type1.ulirgs,guyon:qso.hosts.ir,urrutia:qso.hosts}, with young post-starburst stellar 
populations \citep{guyon:qso.hosts.ir}, much of the dust arising on 
scales of the galaxy \citep[in turbulent motions, inflow, and outflow;][]{urrutia:qso.hosts}, 
and extremely high Eddington ratios indicative of a 
still active period - making them (as opposed to most fully 
obscured quasars) a substantial contributor to the most luminous quasars 
in the Universe \citep{white:red.qsos,hutchings:redqso.midz,zakamska:qso.hosts}. 
As the dust is removed, the black hole is 
then visible as a traditional optical quasar (although very small-scale 
``torus'' obscuring structures may remain intact, allowing for 
some rare, bright Type II systems). 

Here, observations of the host morphology 
are more ambiguous \citep[see e.g.][]{bahcall:qso.hosts,canalizostockton01:postsb.qso.mergers,
floyd:qso.hosts,zakamska:qso.hosts,pierce:morphologies}, but this is expected, for two 
reasons. First, the point
spread function of the bright and unobscured optical quasar must be subtracted 
and host galaxy structure recovered, a difficult procedure. Second, 
by this time the merger is complete and the spheroid has formed, leaving only fading tidal 
tails as evidence for the recent merger. Mock observations constructed from the simulations 
\citep{krause:mock.qso.obs}
imply that, with the best presently attainable data, these features are difficult to 
observe even locally and (for now) nearly impossible to identify at the 
redshifts of greatest interest ($z\gtrsim1$). This appears to be borne out, as 
\citet{bennert:qso.hosts} have re-examined low-redshift quasars previously recognized from 
deep HST imaging as having relaxed spheroid hosts, and found (after 
considerably deeper integrations) that every such object shows clear evidence for 
a recent merger. These difficulties will lead us to consider a number of 
less direct, but more robust tests of the possible association between mergers and quasars. 

Finally, as the remnant relaxes, star formation and quasar activity decline as the
gas is consumed and dispersed, and the remaining galaxy resembles an
elliptical with a quiescent black hole satisfying observed correlations
between black hole and spheroid properties. During this intermediate $\sim$Gyr decay, 
depending on details of the 
merger and exact viewing time, the remnant may be classified as a low-luminosity 
(decaying) AGN in a massive (and relatively young) spheroid, or as a 
post-starburst (E+A/K+A) galaxy. Observationally, the link between 
K+A galaxies and mergers is well-established 
\citep[e.g.][and references therein]{yang:e+a.merger.ell,goto:e+a.merger.connection,hogg:e+a.env}, 
and there is a clear tendency for these galaxies to host low-luminosity 
AGN or LINERs \citep{yang:e+a.agn.connection,goto:e+a.agn.connection}. 
Again, for the reasons given
above, the situation is less clear for all low-luminosity AGN (and there 
will be, as noted above, many such sources driven by secular 
mechanisms in disks). But more importantly 
most objects seen in this stage are expected to have relaxed 
to resemble normal spheroids. 
The merger exhausts gas and star formation in an immediate sense very efficiently, 
so the remnant reddens rapidly onto the red sequence. If this is also associated 
with quenching of future star formation (see \papertwo), then the 
spheroid will evolve passively, growing largely by dry mergers. 

Individual simulations of mergers have enabled us to quantify the
duration of these stages of evolution and how this depends on
properties of the merging galaxies, such as their masses and gas
content and the mass ratio and orbit of the encounter.  In particular,
we used the results to suggest a physical interpretation of quasar
lifetimes \citep{hopkins:lifetimes.letter}, to examine how quasars 
\citep{hopkins:lifetimes.methods} 
and starbursts \citep{chakrabarti:SEDs} would evolve in
this scenario, and quantify structural properties of the remnant and
how they depend on e.g.\ the gas fractions of the merging galaxies 
\citep{cox:xray.gas,cox:kinematics,robertson:fp,robertson:msigma.evolution,
hopkins:bhfp}.

In addition to making predictions for individual systems, we would
also like to characterize how entire {\it populations} of objects
would evolve cosmologically in our picture to test the model against
the large body of observational data that exists from surveys of
galaxies, quasars, and starbursts.  Previously, we have adopted a 
semi-empirical approach to this problem, as follows.  In our
simulations, we can label the outcome by the final black hole mass in
the remnant, $M_{BH,f}$ or, equivalently, the peak bolometric
luminosity of the quasar, $L_{peak}$.  Our simulations predict a
regular behavior for the evolution of the different merger phases as a
function of $M_{BH,f}$ or $L_{peak}$ and also for the properties of
the remnant as a function of $M_{BH,f}$ or $L_{peak}$.  If we have an
estimate of the observed distribution of systems in one phase of the
evolution, we can then use our models to deconvolve the observations
to infer the implied birthrate of such objects as a function of
$M_{BH,f}$ or $L_{peak}$.  Given this, the time behavior of the
simulations provides a mapping between the different phases enabling
us to make independent predictions for other populations. 
For example, knowing the observed quasar luminosity function (QLF) 
at some redshift, 
our simulations allow us to predict how many quasar-producing mergers of a given 
mass must be occurring at the time, which can then be tested against the  
observed merger statistics. 

We exploited this approach to examine the relationship between the
abundance of quasars and other manifestations of quasar activity, and
showed that our model for quasar lifetimes and lightcurves yields 
a means to interpret the shape of the QLF
\citep{hopkins:lifetimes.interp}, 
provides a consistent explanation for observations of
the QLF at optical and X-ray frequencies \citep{hopkins:lifetimes.obscuration},
explains observed evolution in the faint-end slope of the QLF \citep{hopkins:faint.slope}, 
and can account for the spectral shape of the cosmic
X-ray background \citep{hopkins:qso.all,hopkins:bol.qlf}. 
Using this technique to map between different types of objects, we
demonstrated that the observed evolution and clustering of the quasar
population is consistent with observations of red galaxies 
\citep{hopkins:red.galaxies,hopkins:clustering,hopkins:old.age} and 
merging systems \citep{hopkins:transition.mass,hopkins:merger.lfs}, 
as well as the mass function of supermassive black holes
and its estimated evolution with redshift \citep{hopkins:qso.all,hopkins:bol.qlf}. 
In each case, we found
good agreement with observations provided that the mappings were based
on the lifetimes and lightcurves from our merger simulations and not
idealized ones that have typically been used in earlier theoretical
studies.  We further showed that our picture makes numerous
predictions \citep{hopkins:transition.mass,hopkins:qso.all} 
that can be used to test our
hypothesis, such as the luminosity dependence of quasar clustering
\citep{lidz:clustering}.  However, the cosmological context of our results
was not provided in an entirely theoretical manner because our
analysis relied on an empirical estimate of one of the connected
populations.

Obtaining a purely theoretical framework for our scenario is difficult
because cosmological simulations including gas dynamics currently lack
the resolution to describe the small-scale physics associated with
disk formation, galaxy mergers, star formation, and black hole growth.
Semi-analytic methods avoid
some of these limitations, but at the expense of parameterizing the
unresolved physics in a manner this is difficult to calibrate
independently of observational constraints.  For the time being,
neither approach is capable of making an entirely {\it ab initio}
prediction for how the various populations we are attempting to
model would evolve with time.

In this paper, we describe a strategy that enables us, for the first
time, to provide a purely theoretical framework for our picture.  Our
procedure is motivated by, but does not rely upon, observations
suggesting that there is a characteristic halo mass hosting bright
quasars.  This inference follows from measurements of the clustering
of quasars in the 2dF, SDSS, and other surveys
\citep{porciani2004,porciani:clustering,
wake:local.qso.clustering,croom:clustering,coil:agn.clustering,
myers:clustering,daangela:clustering,shen:clustering} and 
investigations of
the quasar proximity effect \citep{faucher:proximity,kim:proximity,guimaraes:proximity}.
By adopting simple models for the merger efficiency of galaxies as a
function of environment and mass ratio, we show that this
characteristic halo mass for quasars corresponds to the most favorable
environment for major mergers between gas-rich disks to occur, namely
the ``small group'' scale.  This finding argues for an intimate
link between such mergers and the triggering of quasar activity and
naturally leads to a method for determining the redshift evolution
of the quasar population from dark matter simulations of structure
formation in a $\Lambda{\rm CDM}$ Universe.

By combining previous estimates of the evolution of the halo mass
function with halo occupation models and our estimates for merger
timescales, we infer the statistics of mergers that excite quasar
activity.  We then graft onto this our modeling of quasar lightcurves
and lifetimes, obtained from our simulations of galaxy mergers that
include star formation and black hole growth to deduce, in an {\it ab
initio} manner, the redshift dependent
birthrate of quasars as a function of their peak
luminosities and the corresponding formation rate of black holes as a
function of mass.  Because our merger simulations relate starbursts,
quasars, and red galaxies as different phases of the same events, we
can then determine the cosmological formation rate of these various
populations and their evolution with redshift.  In particular, as we
demonstrate in what follows, the observed abundance of all these
objects is well-matched to our estimates, unlike for other theoretical
models, supporting our interpretation that mergers between gas-rich
galaxies represent the dominant production mechanism for quasars,
intense starbursts, supermassive black holes, and elliptical galaxies.

We investigate this in a pair of companion papers. Here (\paperone), 
we describe our model and use it to investigate the properties of 
mergers and merger-driven quasar activity. 
In the companion paper \citep[][henceforth \papertwo]{hopkins:groups.ell}, 
we extend our study to the properties of merger remnants and the 
formation of the early-type galaxy population.
Specifically, 
\S~\ref{sec:mergers} outlines our methodology, describing 
the physical criteria for and identification of major 
mergers (\S~\ref{sec:mergers:criteria}), the distribution of mergers 
across different scales and galaxy types (\S~\ref{sec:mergers:scales}), 
and the dependence of mergers on environmental properties 
(\S~\ref{sec:mergers:env}). We then examine the predicted merger 
mass functions, fractions, and clustering properties from this 
model, and compare with observations to verify that we are appropriately 
modeling the merger history of the Universe (\S~\ref{sec:mergers:populations}).
In \S~\ref{sec:quasars} we examine the consequences of 
a general model in which mergers trigger quasar activity. 
We present a number of robust predictions both independent of 
(\S~\ref{sec:quasars:mergers}) and including (\S~\ref{sec:quasars:qlf}) 
physical models for the quasar lightcurves and duty cycles in mergers. 
We contrast this with a ``secular'' model in which quasar activity 
is caused
by disk instabilities (\S~\ref{sec:quasars:secular}), and show 
that a variety of independent constraints suggest that such a mode cannot dominate 
the formation of bright, high redshift quasars. We discuss and summarize our 
conclusions in \S~\ref{sec:discussion}. 

Throughout, we adopt a WMAP3 
$(\Omega_{\rm M},\,\Omega_{\Lambda},\,h,\,\sigma_{8},\,n_{s})
=(0.268,\,0.732,\,0.704,\,0.776,\,0.947)$ cosmology 
\citep{spergel:wmap3}, and normalize all observations and models 
shown to these parameters.
Although the exact choice of 
cosmology may systematically 
shift the inferred bias and halo masses (primarily scaling with $\sigma_{8}$), 
our comparisons (i.e.\ relative biases) are for the most part unchanged, 
and repeating our calculations for 
a ``concordance'' $(0.3,\,0.7,\,0.7,\,0.9,\,1.0)$ cosmology or 
the WMAP1 $(0.27,\,0.73,\,0.71,\,0.84,\,0.96)$ results of \citet{spergel:wmap1}
has little effect on our conclusions. 
We also adopt a diet Salpeter IMF following \citet{bell:mfs}, and convert all stellar masses 
and mass-to-light ratios accordingly. Again, the choice of the IMF systematically 
shifts the normalization of stellar masses herein, but does not substantially change 
our comparisons. 
$UBV$ magnitudes are in the Vega system, and 
SDSS $ugriz$ magnitudes are AB.

\section{Mergers}
\label{sec:mergers}

\subsection{What Determines Whether Galaxies Merge}
\label{sec:mergers:criteria}

\subsubsection{Physical processes}
\label{sec:mergers:processes}

To begin, we postulate which 
mergers are relevant to our picture.
Minor mergers (mass ratios $\gg3:1$) will not trigger 
significant star formation or quasar activity
for most orbits, and consequently will neither exhaust a 
large fraction of the larger galaxy's gas supply nor be typically identified as mergers 
observationally. 
We are therefore specifically interested in major 
mergers, with mass ratios $\leq3:1$, but note that our conclusions are unchanged 
if, instead of this simple threshold, we include all mergers and adopt some 
mass-ratio dependent efficiency 
\citep[e.g.\ assuming the fractional BH/bulge growth scales with mass ratio $R$ in 
some power-law fashion, $\propto R^{-1}$, as suggested by numerical simulations;][]{younger:minor.mergers}. In this case, the decreasing efficiency of BH fueling in minor 
mergers leads (as expected) to the conclusion that they are only 
important at low masses/luminosities 
(similar to where secular activity may dominate quasar populations; see \S~\ref{sec:quasars:secular}), 
and our predictions for massive bulges and BHs are largely unaffected. 
If the timescale for two galaxies to merge 
is long compared to the Hubble time, they clearly will not have
merged in the actual Universe. However, the merger 
timescale must also be short compared to the time required to tidally strip or disrupt 
either of the galaxies -- if it is not, then by the time the galaxies finally
coalesce, the end result
will simply be tidal accretion of material at large radii. 

This defines two fundamental criteria for galaxy mergers to occur in the 
setting of a halo of mass $\mhalo$:
\begin{itemize}
\item The halo must host at least two galaxies of comparable mass $\sim\mgal$. Note that 
even for mergers of distinct host halos in the field, the halo-halo merger proceeds much 
faster than the merger of the galaxies, so there is some period where the two can 
be considered distinct substructures or distinct galaxies within a common host. 
\item The merger must be efficient -- i.e.\ occur in much less than a Hubble time. This requires 
that the mass of the galaxies and their associated (bound) dark matter subhalos 
be comparable to the mass of the parent halo (e.g.\ for the simplest dynamical 
friction arguments, requiring $\mhalo/\mgal \ll 30$).  
\end{itemize}

Together, these criteria naturally define a preferred 
mass scale for major mergers (host halo mass $\mhalo$) for 
galaxies of mass $\mgal$. A halo of mass $\langle\mhalo\rangle(\mgal)$ typically hosts a galaxy of mass 
$\mgal$. At smaller (relative) 
halo masses $\mhalo\ll\langle\mhalo\rangle$, the probability that the halo 
hosts a galaxy as large as $\mgal$ declines rapidly (and eventually must be zero or else violate 
limits from the cosmic baryon fraction). At larger $\mhalo\gg\langle\mhalo\rangle$, the 
probability that the halo will merge with or accrete another halo hosting a comparable $\sim\mgal$ 
galaxy increases, but the efficiency of the merger of these galaxies declines rapidly. Eventually the 
$\mgal$ galaxies are relatively small satellites in a large parent halo of mass 
$\mhalo\gg\langle\mhalo\rangle$, for which (satellite-satellite) mergers are extremely 
inefficient (given the high virial velocities of the host, and dynamical friction timescales 
$\gg \tH$). 

The preferred major-merger scale for galaxies of mass $\mgal$ is therefore only slightly 
larger (factor $\sim2$) than the average host halo mass for galaxies of this mass. 
We refer to this as the small group scale, and emphasize the term {\em small} in this name: 
the average halo of this mass still hosts only 1 galaxy of mass $\sim\mgal$, and 
the identifiable groups will only consist of $2-3$ members of similar mass 
(although there may of course be several much smaller systems in the group, 
which have little dynamical effect). This is very different from 
large group scales, easily identified observationally, which consist of $\gg3$ members.  

\begin{figure}
    \centering
    \figexpand
    \plotter{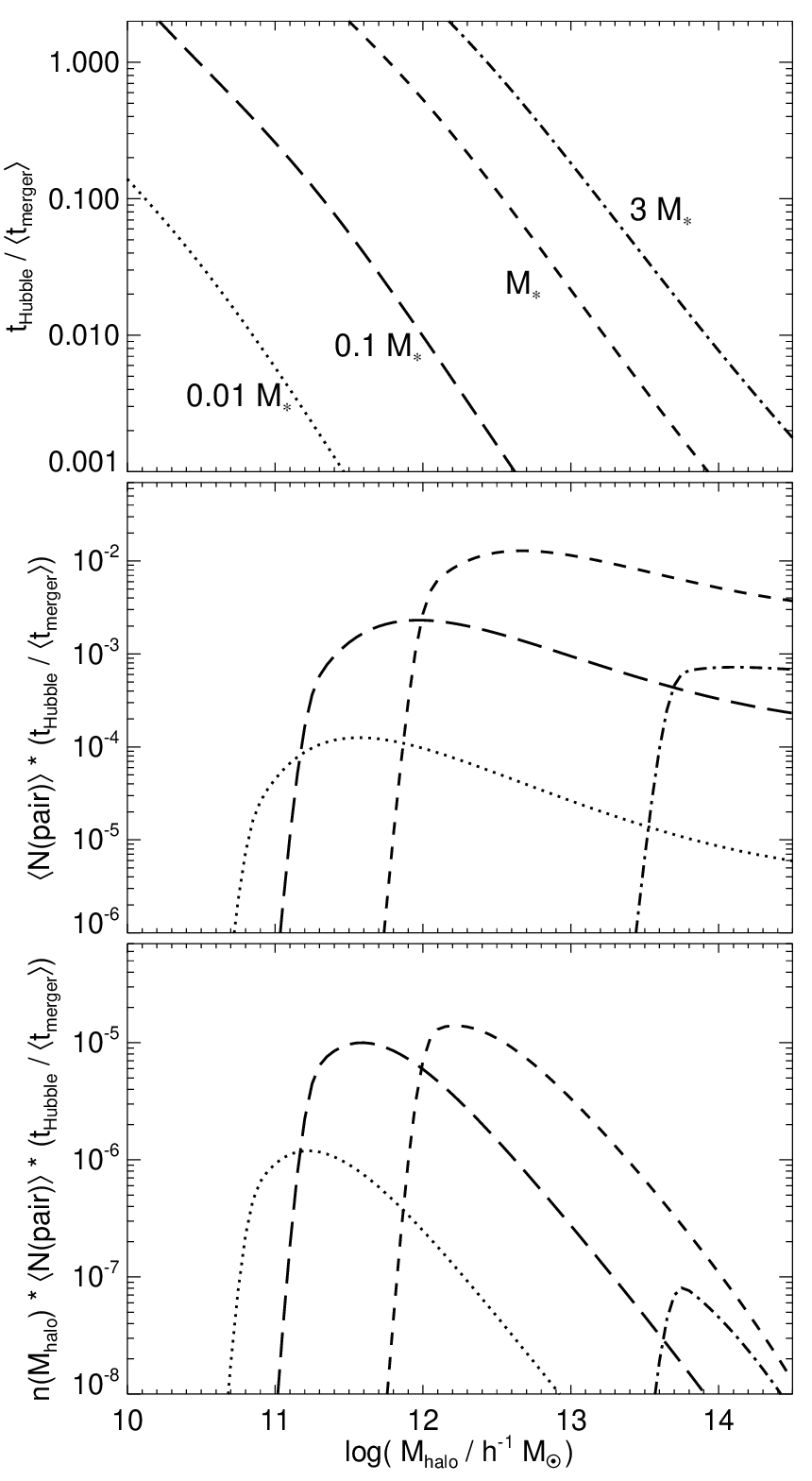}
    \caption{Efficiency of major galaxy mergers (of a certain galaxy mass relative to the 
    characteristic local Schechter-function $M_{\ast}$) as a function of host halo mass 
    (at $z=0$, but the results are qualitatively similar at all redshifts).
    {\em Top:} Merger timescale relative to the Hubble time (assuming a pair of galaxies of mass 
    $\mgal$ are hosted in a halo of mass $\mhalo$) -- mergers occur rapidly ($\tmerger\ll\tH$) 
    when the halo mass is small relative to the galaxy mass (we temporarily ignore 
    the obvious requirement that $\mgal<f_{\rm baryon}\,\mhalo$).
    {\em Middle:} Same, but now multiplied by the probability that the halo actually hosts a pair of 
    galaxies of the given mass (technically, within a mass ratio $3:1$), given the empirical 
    halo occupation model from \citet{wang:sdss.hod}. 
    Although mergers are most rapid in the lowest-mass 
    halos, these halos do not host relatively massive galaxies. 
    {\em Bottom:} Same, but further multiplied by the abundance of halos of a given mass -- 
    the fact that the halo mass function and merger efficiency are decreasing functions 
    of $\mhalo$ (for fixed $\mgal$) means that the 
    contribution to galaxy mergers of a given $\mgal$ will be dominated by the lowest-mass halos 
    in which there is a significant probability to accrete/host a pair of $\mgal$ galaxies -- 
    the small group scale. 
    \label{fig:merger.eff.demo}}
\end{figure}
Figure~\ref{fig:merger.eff.demo} illustrates several of these points. We adopt the merger 
timescales derived below and use the halo occupation fits from \citet{wang:sdss.hod} to 
determine the probability of a halo hosting a pair of galaxies of a given mass: 
the details of the formalism are described below and used throughout, but we wish to illustrate 
the key qualitative points. The merger timescale for galaxies of a given mass is shortest 
when they are large relative to their host halo mass, as expected from dynamical friction 
considerations. However, the probability of a pair being hosted cuts off sharply at low 
halo masses. Moreover, the contribution to mergers of galaxies of mass $\mgal$ 
from larger halos is further suppressed by the simple fact that there are fewer halos of 
larger masses. 

Modern, high-resolution dark matter-only cosmological simulations \citep[e.g.][]{springel:millenium} 
have made it possible to track the merger histories of galaxy halos over large 
ranges in cosmic time and halo mass. For our purposes, the critical information 
is contained in the subhalo mass function, which has been quantified in great detail  
directly from such simulations \citep{kravtsov:subhalo.mfs,gao:subhalo.mf,nurmi:subhalo.mf}
and from extended Press-Schechter theory and semi-analytic approaches 
\citep{taylor:substructure.evolution,zentner:substructure.sam.hod,vandenbosch:subhalo.mf} 
calibrated against numerical simulations. 

When a halo (containing a galaxy and its own 
subhalo populations) is accreted, the accretion process is relatively rapid -- the 
accreted halo will always be identifiable for {\em some} period of time 
as a substructure in the larger halo. Although the new subhalo may lose mass to tidal stripping, 
there will still be some dark matter subhalo associated with the accreted galaxy, which 
will remain until the substructure merges with the central galaxy via dynamical friction 
or (much more rarely) another satellite substructure. Therefore, knowing the 
subhalo populations of all halos at a given instant, the calculation of the rate and distribution of 
{\em galaxy} mergers depends only on calculating the efficiency 
of the subhalo/galaxy mergers within these halos. This is a great advantage -- we 
do not need to calculate halo-halo merger rates, which are not well-defined 
(even when extracted directly from cosmological simulations) and depend 
sensitively on a number of definitions \citep[see, e.g.][]{gottlober:merger.rate.vs.env,
maller:sph.merger.rates}, but instead work 
from the robust (and well-defined) subhalo mass function 
\citep[see][and references therein]{gao:subhalo.mf}. 

This is similar to many 
of the most recent semi-analytic models, which adopt a hybrid approach to 
determine galaxy mergers, 
in which galaxies survive independently so long as their host halo remains a distinct 
substructure, after which point a dynamical friction ``clock'' is started and the galaxy merges 
with the central galaxy in its parent halo at the end of the dynamical friction time. 
Fortunately, for our purposes we are only interested in major mergers with mass 
ratios $\lesssim 3:1$. In these cases, dynamical friction acts quickly on the subhalos 
(infall time $\lesssim \tH /3$ at all redshifts), and the primary ambiguity will be 
the {\em galaxy} merger time in their merged or merging subhalos. 

To perform this calculation, we need to know the properties of the merging galaxies. 
For now, we only want to calculate where and when galaxies are merging, not 
how they evolved to their present state in the first place. This is our primary reason for 
not constructing a full semi-analytic model: rather than introduce a large number of 
uncertainties, theoretical prescriptions which we are not attempting to test here, 
and tunable parameters in order to predict that e.g.\ a $10^{11}\,\msun$ halo 
typically hosts a $\sim10^{10}\,\msun$ star-forming galaxy, we can adopt 
the established empirical fact that this is so. In detail, we populate subhalos according 
to an empirical halo occupation model \citep[e.g.,][]{tinker:hod,conroy:monotonic.hod,
valeostriker:monotonic.hod,vandenbosch:concordance.hod,wang:sdss.hod};
i.e.\ matching the observed statistics of 
where galaxies of a given type live (accounting for different occupations for 
different galaxy types/colors, and the scatter in galaxies hosted in 
halos of a given mass). 

This is sufficient for most of our predictions. We do not necessarily need to know 
exactly how long it will take for these mergers to occur, only that they are 
occurring at a given redshift -- i.e.\ that the objects will merge and that the 
merger time is shorter than the Hubble time (which for the mass ratios of interest 
is essentially guaranteed). For example, predicting the clustering of galaxy mergers 
does not require knowledge of how rapidly they occur, only {\em where} they occur. 
Even predicting the observed merger mass function does not rely 
sensitively on this information, 
since the duration over which the merger is visible will be comparable (albeit 
not exactly equal)
to the duration over which the merger occurs (such that a fixed fraction $\sim1$ of 
all merging systems are observable). 

However, for the cases where it is necessary, 
we estimate the timescales for the galaxies to merge and 
to be identified as mergers. This is the most uncertain element in our model. 
Part of this uncertainty owes to the large parameter space of mergers (e.g.\ differences 
in orbital parameters, relative inclinations, etc.). 
These uncertainties are fundamental, but can at least be controlled by 
comparison to large suites of hydrodynamic simulations which sample these 
parameter spaces \citep{robertson:fp} and allow us to quantify the 
expected range of merger properties owing to these (essentially random) differences. 
The more difficult question is how appropriate any analytic merger timescale or 
cross section can be. To address this, we will throughout this paper consider 
a few representative models: 

{\em Dynamical Friction:} The simplest approximation is that the 
galaxies are point masses, and (once their subhalos merge) they fall 
together on the 
dynamical friction timescale. This is what is adopted in most semi-analytic 
models. In fact, this is only an appropriate description when the galaxies are small 
relative to the enclosed halo mass, and are both 
moving to the center of the potential well -- which is often not the case at these 
late stages. While unlikely to be incorrect by orders of magnitude, 
this approximation begins to break down when the galaxies are relatively 
large compared to their halos (common in $\lesssim10^{12}\,\msun$ halos) 
and when the galaxies are very close (and could e.g.\ enter a stable orbit). What 
finally causes galaxies to merge is not, in fact, 
simple dynamical friction, but dissipation of angular momentum via a resonance 
between the internal and orbital frequencies.

{\em Group Capture (Collisional):} On small scales, 
in satellite-satellite mergers, or in the merger 
of two small field halos, it is more appropriate to consider galaxy mergers 
as a collisional process in which there is some effective gravitational cross section. 
In other words, galaxy mergers proceed once the galaxies pass at sufficiently 
small distances with sufficiently low relative velocity. There have been a number of 
theoretical estimates of these cross sections -- we adopt here the fitting 
formulae from \citet{krivitsky.kontorovich}, who calibrate the appropriate 
cross-sections from a set of numerical simulations of different encounters and 
group environments. This compares well with other calculations \citep[][and 
references therein]{white:cross.section,makino:merger.cross.sections,mamon:groups.review}, 
and we find little difference using these alternative estimations. For large 
mass ratios and separations, 
the expressions appropriately reduce to the dynamical friction case.

{\em Angular Momentum:} \citet{binneytremaine} consider this problem 
from the perspective of the angular momentum-space in which 
galaxy mergers are allowed. This approach is similar to the capture estimates 
above, but accounting for capture into orbits as well. Whether or not such 
orbits will merge is, of course, somewhat ambiguous -- it is likely that 
some significant fraction are stable, and will not merge, while others 
decay rapidly owing to resonance between the disk circular frequencies and 
the orbital frequency. Nevertheless, this serves to bracket the range of 
likely merger configurations.

\subsubsection{Synopsis of model and uncertainties}
\label{sec:mergers:synopsis}

Thus, to summarize our approach: at a given redshift, we calculate the 
halo mass function $n(\mhalo)$ for our adopted cosmology following 
\citet{shethtormen}. For each halo, we calculate the 
(weakly mass and redshift dependent) subhalo mass function (or distribution of 
subhalos, $P[N_{\rm subhalo}\, | \, M_{\rm subhalo},\ \mhalo]$)
following \citet{zentner:substructure.sam.hod} 
and \citet{kravtsov:subhalo.mfs}. Alternatively, we 
have adopted it directly from \citet{gao:subhalo.mf,nurmi:subhalo.mf} or 
calculated it following \citet{vandenbosch:subhalo.mf,valeostriker:monotonic.hod}, and 
obtain similar results. Note that the subhalo masses are 
defined as the masses upon accretion by the parent halo, which 
makes them a good proxy for the hosted galaxy mass \citep{conroy:monotonic.hod} 
and removes the uncertainties owing to tidal mass stripping. 

Mergers are identified by the basic criteria described above. 
We populate these halos and subhalos 
with galaxies following the empirical halo occupation models 
of \citet{conroy:monotonic.hod} \citep[see also][]{valeostriker:monotonic.hod} normalized directly 
with group observations following \citet{wang:sdss.hod} at $z=0$ 
\citep[considering instead the occupation fits in][makes little difference]{yang:clf,
cooray:highz,cooray:hod.clf,zheng:hod,vandenbosch:concordance.hod}. 
This determines both the mean stellar mass and dispersion in stellar masses of 
galaxies hosted by a given halo/subhalo mass $P(\mgal\,|\,M_{\rm subhalo})$, 
which (optionally) can be broken down 
separately for blue and red galaxy types. 
 
\begin{figure}
    \centering
    \figexpand
    \plotter{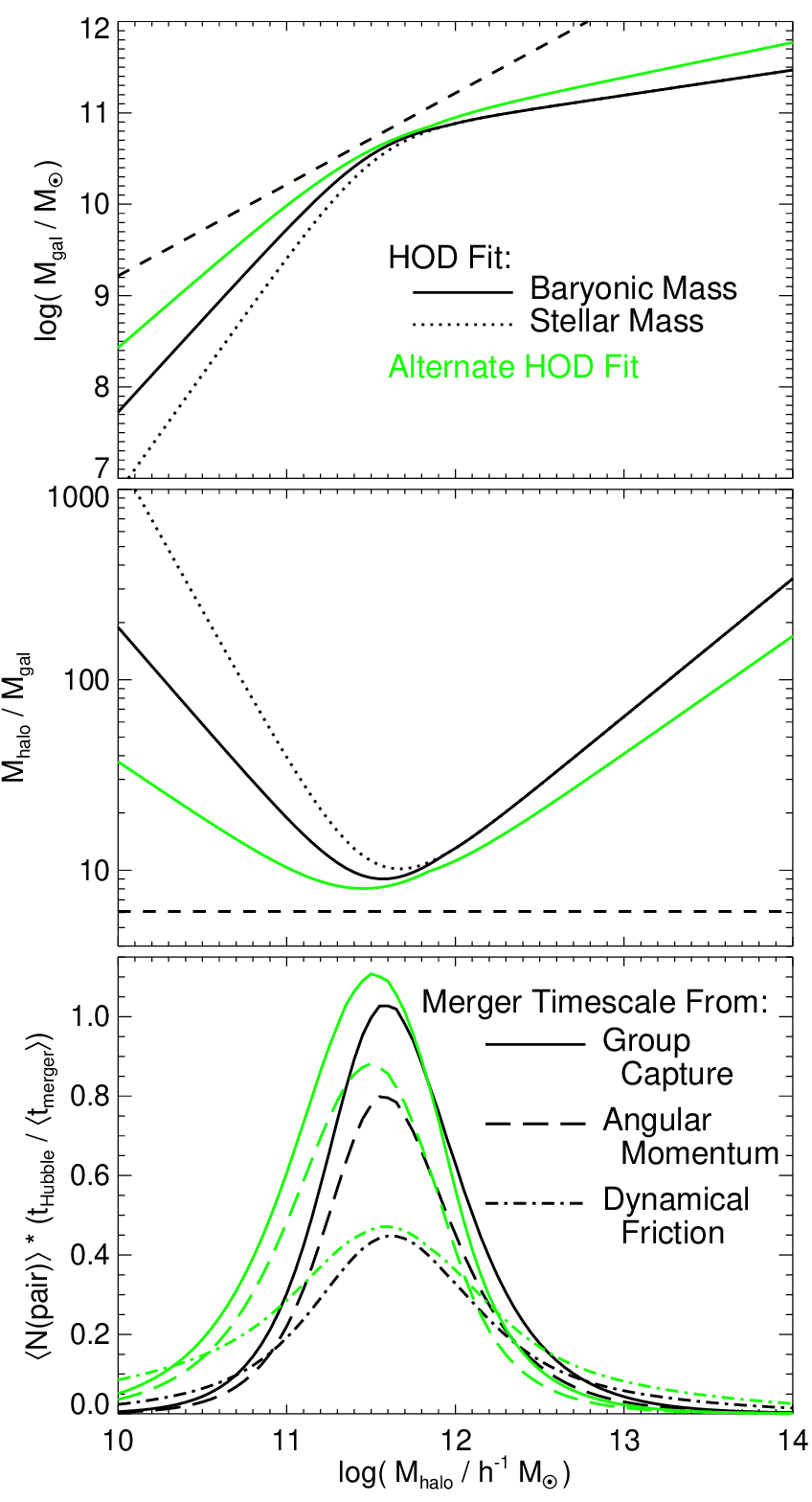}
    \caption{Illustration of basic elements of importance to where 
    galaxy-galaxy mergers occur. {\em Top:} Average central galaxy 
    stellar (dotted) and baryonic (solid) mass as a function of host 
    halo mass, in our typically adopted halo occupation 
    model \citep[][black]{conroy:monotonic.hod,valeostriker:monotonic.hod}, 
    and the alternate halo occupation model from
    \citet[][green; only baryonic mass shown]{yang:clf}
    {\em Middle:} Corresponding halo-to-galaxy mass ratio. 
    {\em Bottom:} Average major merger timescale/efficiency (calculated as 
    in the middle panel of Figure~\ref{fig:merger.eff.demo}, but for the 
    appropriate mean $\mgal(\mhalo)$). Timescales are determined
    as described in the text, from dynamical friction (dot-dashed), 
    group capture (solid), or angular momentum (long dashed) considerations. 
    \label{fig:merger.eff.mean}}
\end{figure}

Figure~\ref{fig:merger.eff.mean} shows the mean galaxy mass as a function of 
halo mass from this model at $z=0$. Since the halo occupation models 
consider stellar mass or luminosity, we use the baryonic and stellar mass 
Tully-Fisher relations calibrated by \citet{belldejong:tf} to convert between the two. 
(We have also compared the global baryonic mass function estimated in this manner with 
that observationally inferred in \citet{bell:baryonic.mf} and find good agreement). 
If necessary, we calculate the galaxy-galaxy merger efficiency/timescale 
using the different estimators described above. Figure~\ref{fig:merger.eff.mean} 
also shows the expected merger efficiency as a function of halo mass 
for these mean values (i.e.\ probability of hosting a subhalo within the appropriate 
mass range convolved with the calculated merger timescale). The qualitative 
features are as expected from Figure~\ref{fig:merger.eff.demo}. 
The different merger timescale estimators agree well at large halo masses, 
with the dynamical friction treatment yielding a somewhat longer 
(factor $\lesssim$ a few) timescale at intermediate masses (but this is near the regime 
of low $\mhalo/\mgal$ where the dynamical friction approximation is 
least accurate). 

The main elements and their uncertainties in our model are: 

{\bf 1.\ Halo Mass Function:} We begin by computing the overall halo mass function. 
There is very little ambiguity in this calculation at all redshifts and masses 
of interest \citep[$z\lesssim6$; see e.g.][]{reed:halo.mfs}, and 
we do not consider it a significant source of 
uncertainty. 

{\bf 2.\ Subhalo Mass Function:} The subhalo mass function of each halo is 
then calculated. Although numerical simulations and semi-analytic 
calculations generally give 
very similar results \citep[especially for the major-merger mass ratios of interest 
in this paper, as opposed to very small subhalo populations; see][]{vandenbosch:subhalo.mf}, 
there is still some (typical factor $<2$) disagreement between different estimates. 
We therefore repeat most of our calculations adopting both 
our ``default'' subhalo mass function calculation 
\citep{zentner:substructure.sam.hod,kravtsov:subhalo.mfs} and an alternative 
subhalo mass function calculation \citep{vandenbosch:subhalo.mf} 
\citep[normalized to match cosmological simulations 
as in][]{shaw:cluster.subhalo.statistics}, which bracket the range 
of a number of different estimates \citep[e.g.,][]{springel:cluster.subhalos,
tormen:cluster.subhalos,delucia:subhalos,gao:subhalo.mf,nurmi:subhalo.mf} 
and demonstrate the uncertainty 
owing to this choice. The difference is ultimately negligible 
at $\mgal\gtrsim10^{10}\,\msun$ at all redshifts, and rises to only a factor $\sim2$ at 
$\mgal\lesssim10^{10}\,\msun$ (probably owing to differences in the 
numerical resolution of different estimates at low halo masses). 

{\bf 3.\ Halo Occupation Model:} We then populate the 
central galaxies and ``major'' subhalos with an empirical halo occupation model. 
Although such models are constrained, by definition, to reproduce the mean 
properties of the halos occupied by galaxies of a given mass/luminosity, there 
are known degeneracies between parameterizations that give rise to 
(typical factor $\sim2$) differences between models. We therefore again 
repeat all our calculations for our ``default'' model 
\citep{conroy:monotonic.hod} \citep[see also][]{valeostriker:monotonic.hod} and 
an alternate halo occupation model \citep{yang:clf} \citep[see also][]{yan:clf.evolution,zheng:hod}, which 
bracket the range of a number of calculations 
\citep[e.g.,][]{cooray:highz,cooray:hod.clf,zheng:hod,vandenbosch:concordance.hod}. 
Again, we find this
yields negligible differences 
at $\mgal\gtrsim10^{10}\,\msun$ (as the clustering and abundances 
of massive galaxies are reasonably well-constrained, and most of these 
galaxies are central halo galaxies), and even at low masses the 
typical discrepancy rises to only $\sim0.2\,$dex. 

We note that we have also considered a variety of prescriptions for the 
redshift evolution of the halo occupation model: including that 
directly prescribed by the quoted models, a complete re-derivation 
of the HOD models of \citet{conroy:monotonic.hod} and 
\citet{valeostriker:monotonic.hod} 
at different redshifts from the observed mass functions of 
\citet{fontana:highz.mfs,bundy:mfs,borch:mfs,blanton:lfs} (see \S~\ref{sec:quasars:mergers}), 
or simply assuming no evolution (in terms of galaxy mass
distributions at fixed halo mass; for either all galaxies or 
star-forming galaxies). We find that the resulting differences are 
small (at least at $z\lesssim3$), comparable to 
those inherent in the choice of halo occupation model. 
This is not surprising, as a number of recent 
studies suggest that there is very little evolution in halo occupation 
parameters (in terms of mass, or relative to $L_{\ast}$) with 
redshift \citep{yan:clf.evolution,cooray:highz,
conroy:monotonic.hod}, or equivalently that the masses of galaxies hosted in a 
halo of a given mass are primarily a function of that halo mass, not 
of redshift \citep{heymans:mhalo-mgal.evol,
conroy:mhalo-mgal.evol}. This appears to be especially true for 
star-forming and $\sim L_{\ast}$ galaxies \citep[of greatest importance for 
our conclusions;][]{conroy:mhalo-mgal.evol}, unsurprising 
given that ``quenching'' is not strongly operating in those systems to change 
their mass-to-light ratios. 

{\bf 4.\ Merger Timescale:} Having populated a given halo and its subhalos 
with galaxies, we then calculate the timescale for mergers between major galaxy 
pairs. This is ultimately the largest source of uncertainty in our calculations, 
at all redshifts and masses. 
Again, we emphasize that some of our calculations are completely 
independent of these timescales. However, where adopted, we illustrate  
this uncertainty by presenting all of our predictions for three estimates of 
the merger timescale: a simple dynamical friction formula, a 
group capture or collisional cross section estimate, and an angular 
momentum (orbital cross section) capture estimate, all
as described above. At large masses 
and redshifts $z\lesssim2.5$, this is a surprisingly weak source of 
uncertainty, but the estimated merger rates/timescales 
can be very different at low masses $\mgal\lesssim 10^{10}\,\msun$ 
and the highest redshifts $z\sim3-6$. 

At low masses, this owes 
to a variety of effects, including the substantial difference 
between infall or merger timescales and the timescale for 
morphological disturbances to be excited (different in e.g.\ an 
impact approximation as opposed to the circular orbit decay 
assumed by dynamical friction). 

The difference in redshift 
evolution is easily understood: at fixed mass ratio, the 
dynamical friction timescale scales as 
$t_{\rm df}\propto \tH\propto \rho^{-1/2}$, 
but a ``capture'' timescale will scale with fixed cross section as 
$t\propto 1/(n\,\langle\sigma\,v \rangle)\propto \rho^{-1}$, 
so that (while the details of the cross-sections and dependence 
of halo concentration on redshift make the 
difference not quite as extreme as this simple scaling) the very large
densities at 
high redshift make collisional merging increase rapidly in efficiency. 
The true solution is probably some effective 
combination of these two estimates, and the 
``more appropriate'' approximation 
depends largely on the initial orbital parameters of the subhalos. 
At present, we therefore must recognize this as an inherent 
uncertainty, but one that serves to bracket the likely range of 
possibilities at high redshifts.

\subsection{Where Mergers Occur}
\label{sec:mergers:scales}

\begin{figure}
    \centering
    \figexpand
    \plotter{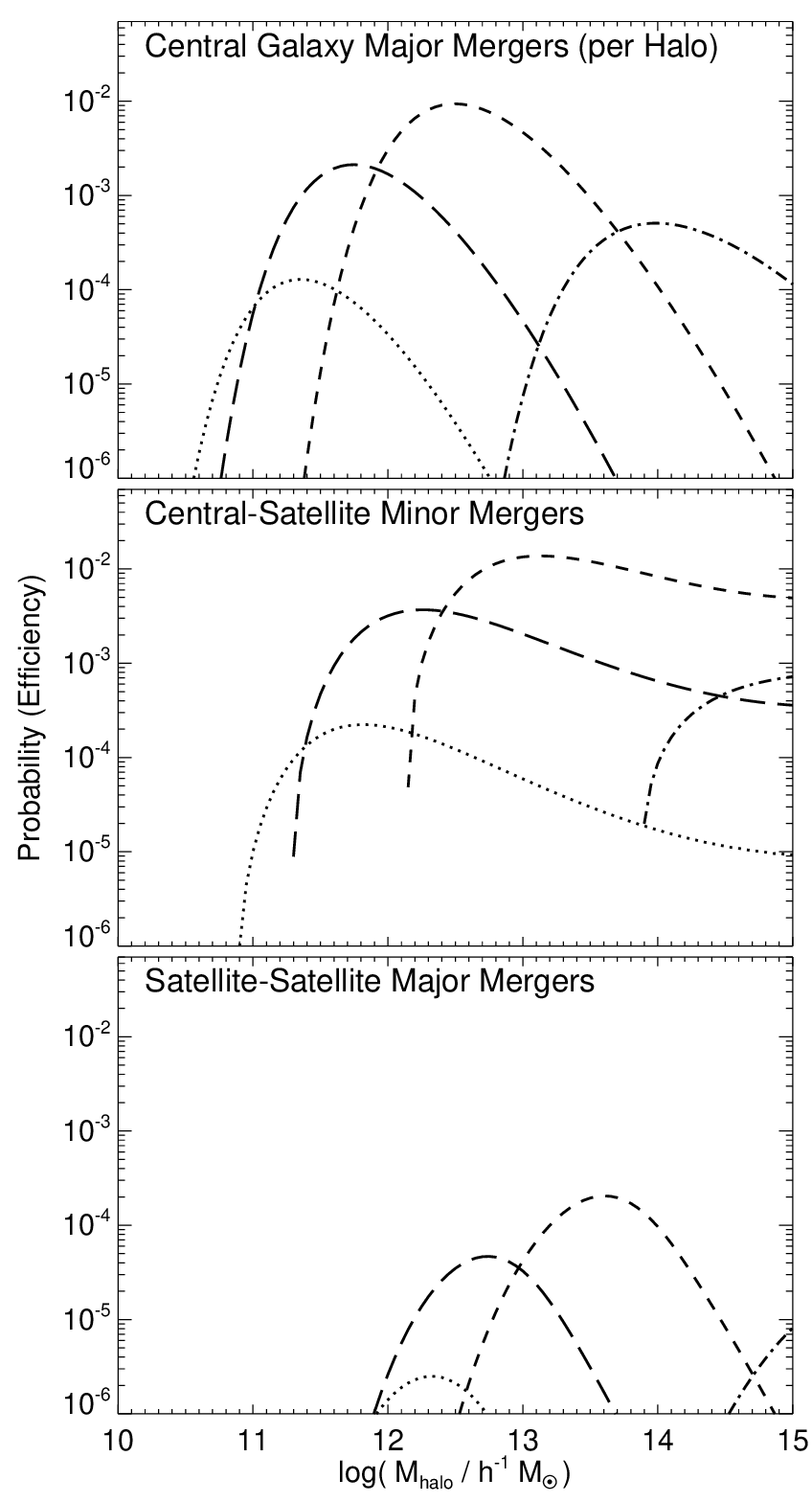}
    \caption{Merger efficiency (arbitrary units; defined in the same manner as the lower panel 
    of Figure~\ref{fig:merger.eff.demo}, with different linestyles in the same style for various mass 
    galaxies) for different classes of mergers. Using the subhalo mass functions and halo 
    occupation models, we can separate major mergers onto the 
    central galaxy in a halo ({\em top}), 
    minor (mass ratio $>3:1$ but $<10:1$) mergers onto the central galaxy ({\em middle}), 
    and satellite-satellite mergers ({\em bottom}). Major mergers occur efficiently in central galaxies 
    near the small group scale for each $\mgal$. When galaxies live in very massive halos, they 
    experience a large number of minor mergers from the satellite population. Satellite-satellite 
    mergers are a relatively small effect at all galaxy and halo masses. 
    \label{fig:merger.eff.centralsat}}
\end{figure}

We are now in a position to predict the statistics of mergers. First, we illustrate some 
important qualitative features. Figure~\ref{fig:merger.eff.centralsat} shows the 
merger efficiency (as in Figure~\ref{fig:merger.eff.demo}) for different classes of 
mergers: major mergers with the central galaxy in a halo, minor mergers with the 
central galaxy, and major mergers of two satellite galaxies in the halo. We show 
the results for our ``default'' model, adopting the dynamical friction merger 
timescale, but the qualitative results are independent of these choices.
The key features 
are expected: major mergers are efficient at small group scales (halo 
masses) comparable to or just larger than the average host halo mass for a given 
$\mgal$. At larger $\mhalo$, major mergers become more rare for the reasons in 
\S~\ref{sec:mergers:criteria}. 
However, although dynamical friction times increase, the rapidly increasing 
number of satellite systems in massive halos means that minor merger accretion onto 
the central galaxy proceeds with a relatively constant efficiency. This will not 
trigger substantial quasar or starburst activity or morphological transformation, but 
may be important for overall mass growth in large cD galaxies, although 
recent cosmological simulations \citep{maller:sph.merger.rates} suggest that 
major mergers dominate minor mergers in the assembly of massive galaxies 
(although their simulation does not extend to the largest cD galaxies). 

Satellite-satellite 
minor mergers are a small effect at all masses, as expected (by the time a halo is sufficiently massive 
to host a large number of satellites of a given $\mgal$, the orbital velocity of the 
galaxies about the halo is much larger than their individual internal velocities). 
In what follows, we will generally ignore satellite-satellite mergers. Including them 
is a very small correction (generally $\ll10\%$), and their dynamics are 
uncertain. Moreover, their 
colors and star formation histories are probably affected by processes 
such as tidal stripping, harassment, and ram-pressure stripping, which we 
are neither attempting to model nor test. We have however checked that there 
are no significant or qualitative changes to our predictions if we 
(naively) include the satellite-satellite term.

\begin{figure}
    \centering
    \figexpand
    \plotter{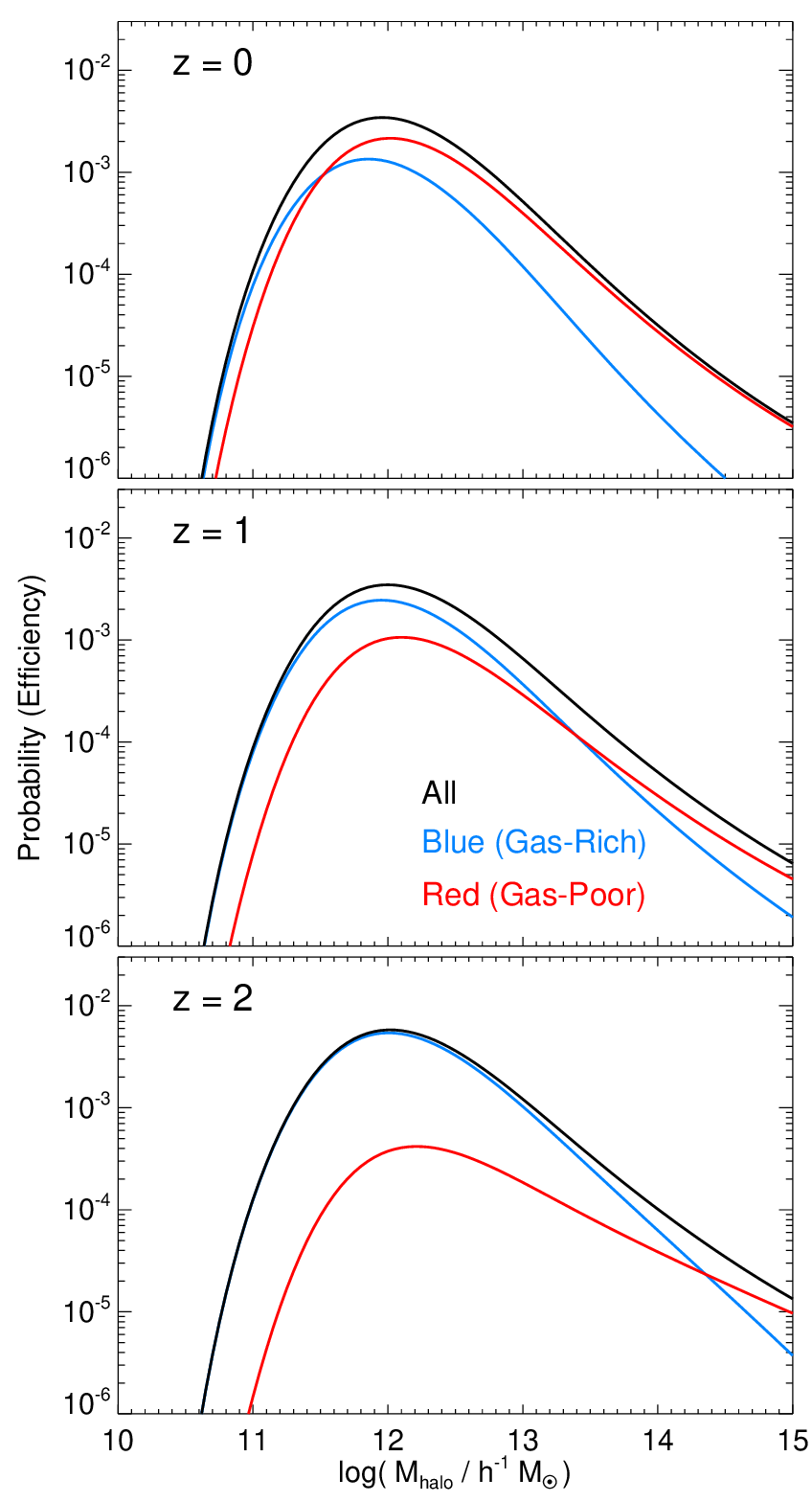}
    \caption{Merger efficiency (arbitrary units; 
    calculated as in Figure~\ref{fig:merger.eff.demo}) 
    as a function of halo mass (adopting the mean $\mgal(\mhalo)$ from 
    Figure~\ref{fig:merger.eff.mean}). Using the type-separated 
    galaxy mass functions from 
    \citet{bell:mfs,borch:mfs,fontana:mfs} at $z=0,\,1,\,2$, respectively, 
    we show the fraction of galaxies 
    at each mass expected to be gas-rich and gas-poor, at each of 
    three redshifts. At high redshifts, all but the most massive merging galaxies 
    will be gas-rich, whereas at low masses the gas-poor population dominates 
    at most masses where mergers are efficient. 
    \label{fig:merger.redblue}}
\end{figure}

Although the consequences of the merger will be very different, 
the efficiency with which
two galaxies merge does not depend strongly on whether they 
are star-forming or red/passive (all else being equal). It is therefore a consequence that, 
at low redshifts, gas-rich mergers are generally relegated to low stellar masses and field 
environments where such galaxies are common. Figure~\ref{fig:merger.redblue} 
illustrates this. We plot the mean efficiency of major, central galaxy mergers 
(as in Figure~\ref{fig:merger.eff.centralsat}, but for the mean $\mgal$ at each $\mhalo$) 
as a function of halo mass at each of three redshifts. At each redshift, we divide this into  
the observed fraction of red and blue galaxies at the given galaxy/halo mass, 
using the appropriate observed, type-separated galaxy mass functions. The efficiency of 
mergers at a given halo and galaxy mass 
does not evolve (note that this is {\em not} a statement that the overall 
merger rates will not change, but rather a statement that the same galaxies in 
the same halos will merge at the same rate). However, at low redshifts, red galaxies 
dominate the mass budget, whereas at high redshifts, most galaxies are 
still blue (star-forming) in all but the most massive halos. We will discuss 
the possibility that mergers themselves drive this change in the 
blue and red fractions in \papertwo, but for now illustrate that 
the locations of gas-rich and dry mergers reflect where 
gas-rich and gas-poor galaxies dominate the population, respectively, 
which is empirically determined at the redshifts of interest here. We note 
that our halo occupation models do not explicitly model a dependence of 
halo populations on central galaxy properties; i.e.\ the tentative 
observational suggestion that, at fixed halo and galaxy mass, 
red central galaxies are preferentially 
surrounded by red (as opposed to blue) satellites \citep{weinmann:obs.hod}. If real, 
the effect of such a trend is to make the transition plotted in 
Figure~\ref{fig:merger.eff.centralsat} somewhat sharper -- this has 
little effect on our conclusions, but does somewhat lower 
the predicted gas-rich merger rates (and corresponding predicted 
quasar luminosity density) at $z\lesssim0.5$ (since a red central 
galaxy would have a lower probability of an infalling, gas-rich system). 

\begin{figure*}
    \centering
    \figexpand
    \plotone{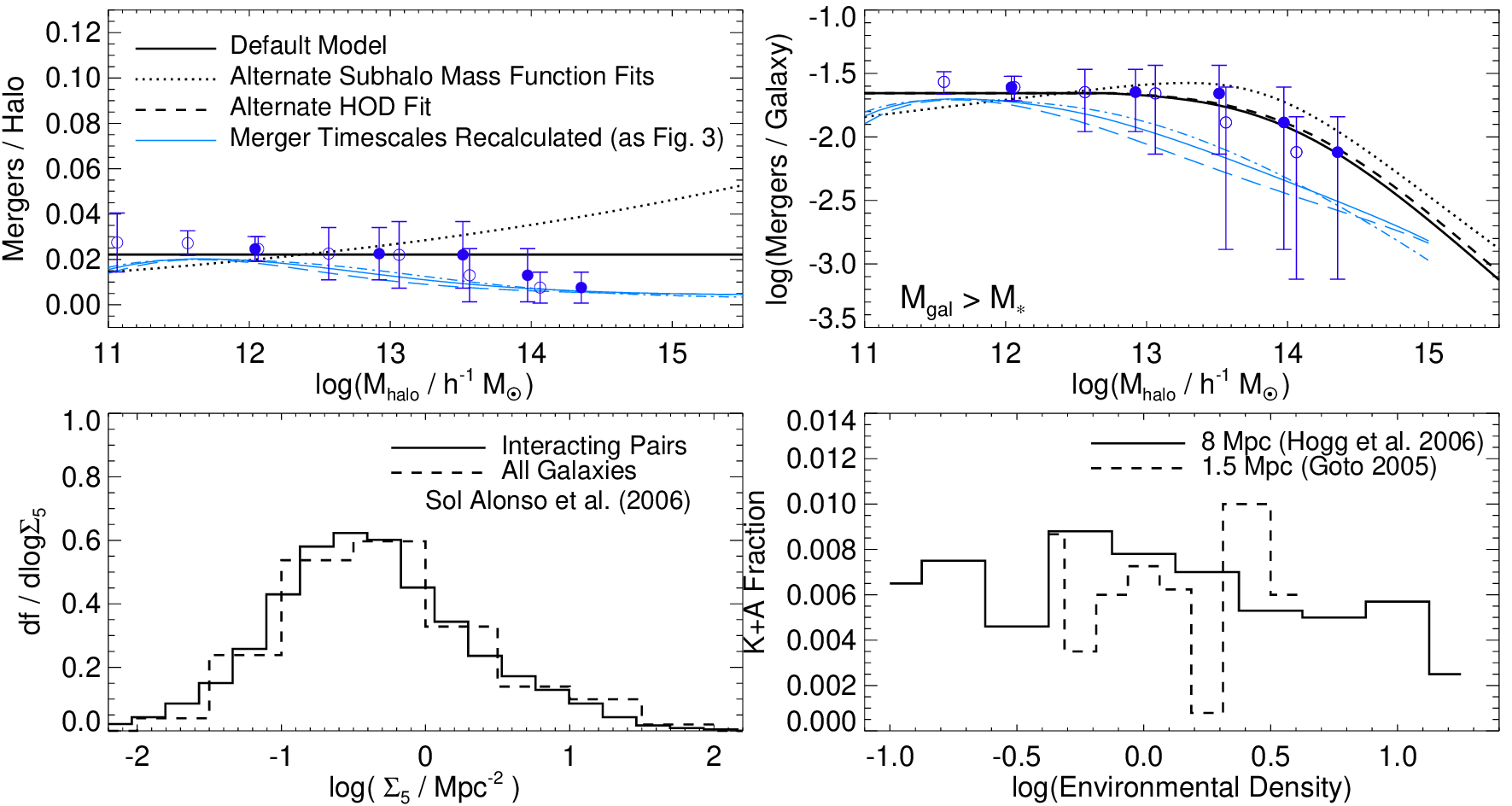}
    \caption{{\em Top:} Merger fraction as a function of host halo mass. The 
    fraction of all halos (groups) predicted to host at least one major merger of 
    galaxy mass $\gtrsim10^{10}\,\msun$ is plotted ({\em left}), 
    as is the fraction of all galaxies in halos of a given $\mhalo$ which are 
    merging ({\em right}). We show the predictions for several variations of 
    our standard model (described in the text) used to identify all merging systems
    (black lines, as labeled), 
    and adding a more detailed calculation of the actual 
    timescale for the physical galaxy mergers (blue lines, as labeled) and 
    ability to morphologically identify them. 
    Both are compared with observed merger fractions 
    (points) from \citet[purple circles][]{alonso:groups} \citep[we convert 
    their measured intermediate-scale densities to average halo masses 
    following][shown as open and filled points, 
    respectively]{baldry06:redfrac.vs.m.env,kauffmann:sf.vs.env}. 
    {\em Bottom:} The observed distributions (fraction of objects per logarithmic interval in 
    galaxy surface density) of merger and normal galaxy 
    environments, from the group catalogues of \citet{alonso:groups} ({\em left}), 
    and the fraction of recent merger remnant (post-starburst, K+A) galaxies 
    as a function of galaxy surface density averaged 
    on intermediate ($1.5\,{\rm Mpc}$) 
    and large ($8$\,Mpc) scales ({\em right}). Mergers occur on all scales 
    and in halos of all masses, without a strong feature at a particular scale.
    \label{fig:merger.fraction.mhalo}}
\end{figure*}

Integrating over the appropriate galaxy 
populations, Figure~\ref{fig:merger.fraction.mhalo} compares the predicted $z=0$ 
merger fraction as a function of 
halo mass from this model with that observed. The agreement is good over a wide 
dynamic range. Although there is a significant (factor $\sim2$) systematic difference 
based on how this fraction is calculated, this is within the range of present 
observational uncertainty. It is also important to distinguish the merger fraction of 
parent halos (i.e.\ fraction of groups which contain a merger) and that of 
galaxies (i.e.\ fraction of all galaxies at a given $\mgal$ or $\mhalo$ which 
are merging), as at large halo masses the rate of mergers onto the central galaxy 
could remain constant (giving a constant merger rate per halo), but the inefficient 
merging of the increasingly large number of satellites will cause the 
galaxy merger fraction to fall rapidly. 

We also show the distribution of mergers (interacting pairs) and all galaxies 
in environmental density (local projected surface density 
$\Sigma_{5}=5/(\pi\,d_{5}^{2})$, where $d_{5}$ is the distance to the 
fifth nearest-neighbor) from the local group catalogues of \citet{alonso:groups} 
-- we compare this data set directly to our prediction by converting 
$\Sigma_{5}$ to $\mhalo$ using the mean relation from \citet{croton:sam}, 
as in \citet[][]{baldry06:redfrac.vs.m.env} (although as they note, the relation has considerable scatter).
Similarly, we show the post-starburst (generally merger remnant) 
fraction from \citet{hogg:e+a.env} and \citet{goto:e+a.merger.connection}, as a function of 
surface density on large scales. 

Our predictions and the observations 
emphasize that galaxy mergers occur on all scales (in halos of all masses), 
and in all environments. In a global sense, there is no preferred merger scale. 
That is not to say that mergers of galaxies of a particular mass do not 
have a preferred scale (indeed, in our modeling, this is explicitly the 
small group scale), but rather because this scale is a function of galaxy mass, 
mergers of {\em some} mass occur in all halo masses and environments. 
It is clear that it is a mistake to think that mergers would not occur in field 
(or even void) environments, a fact which is very important to the formation of 
spheroids and quasars in these locations.

\subsection{How Mergers Are Influenced By Environment}
\label{sec:mergers:env}

Figure~\ref{fig:merger.fraction.mhalo} demonstrates that, 
all else being equal, mergers do not depend on the large scale 
environment. This is conventional wisdom, of course, because 
mergers are an essentially {\em local} process. However, there 
is one sense in which the merger rate should depend on environment. 
If the local density of galaxies (supply of systems for major mergers) 
is enhanced by some factor $1+\delta$, then the probability (or rate) 
of major mergers should be enhanced by the same factor. 

In detail, our adopted model for the merger/capture cross section 
of galaxies (\S~\ref{sec:mergers:criteria})
allows us to calculate the differential probability that 
some halo/subhalo or galaxy population at a given distance $r$ 
will merge with the central galaxy in a time $<\tH$. Given the observed 
galaxy-galaxy correlation function as a function of 
stellar mass \citep{li:clustering}, we can trivially calculate the mean number density of 
galaxies (possible fuel for major mergers) in a shell $dr$ at $r$, 
and combining this with the merger rate/cross section calculation 
determines the differential contribution to the total merger 
rate of galaxies of that mass, from pairs at the separation $dr$. 
This can be thought of as either a capture process from 
halo/subhalo orbits, or a global inflow rate from 
dynamical friction and gravitational motions; the results are 
the same, modulo the absolute merger rate normalization 
\citep{binneytremaine,masjedi:merger.rates}. 
Next, assume that the density of these companions is 
multiplied, at this radius, by a factor $1+\delta_{r}$ (relative to the 
mean $\langle(1+\delta_{r})\rangle$ expected 
at that $r$ for the given central halo mass). Integrating over all 
radii, we obtain the total merger rate/probability, with the 
appropriate enhancement. 

\begin{figure}
    \centering
    \figexpand
    \plotone{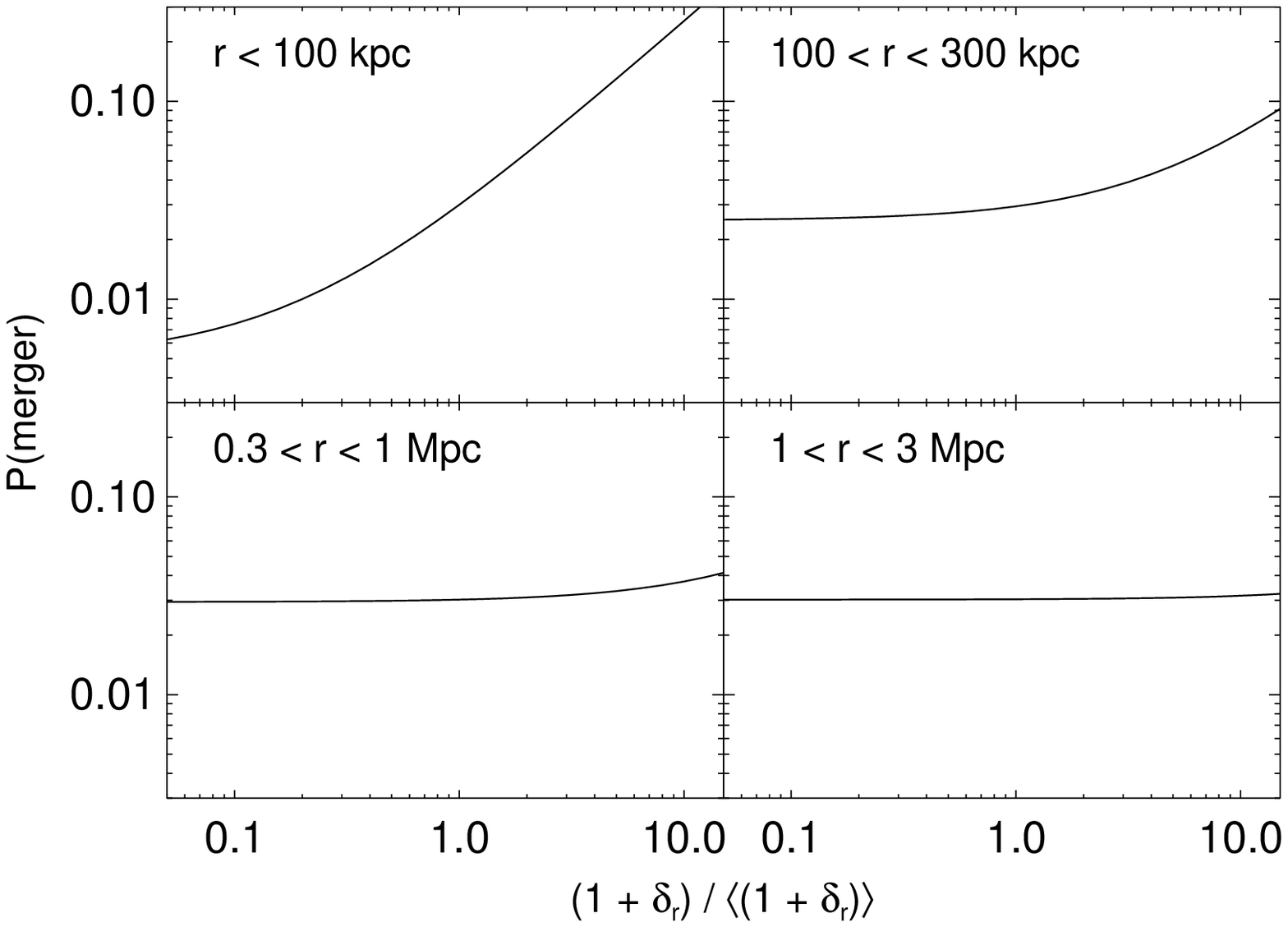}
    \caption{Dependence of the merger rate/probability on environmental density 
    decrement/enhancement
    within a given radius $r$; i.e.\ galaxy 
    overdensity $(1+\delta_{r})/\langle(1+\delta_{r})\rangle$ 
    at a fixed galaxy and 
    host halo mass (absolute units are arbitrary here, and depend on these quantities). 
    On scales less than the typical virial radii of interest, the merger rate 
    increases with overdensity (linearly at $\delta_{r}\gg1$), but it is independent 
    (for a fixed halo mass) of large-scale environment. 
    \label{fig:merger.density.dept}}
\end{figure}

Figure~\ref{fig:merger.density.dept} illustrates this, 
calculated in several 
radial shells using our gravitational capture cross sections 
to estimate the enhancement (the other cross sections yield 
similar results). The absolute value of the 
probability shown will be a function of galaxy mass, halo mass,
and redshift, but the qualitative behavior is similar. Unsurprisingly, 
density enhancements on small scales ($r\lesssim100\,$kpc, where 
most systems will merge) linearly increase the merger rate 
accordingly. Note that density decrements decrease the merger 
rate only to a point -- this is because even for a galaxy with no companions 
within a $100\,$kpc radius, there is of course some non-zero probability that 
companions will be accreted or captured from initially larger radii and 
merge in $t\ll\tH$. 

At larger radii, the enhancement is less pronounced. 
A galaxy in the center of a 
halo of a given mass in a $\sim3\,$Mpc overdensity is not substantially 
more likely to experience a major merger, because there is little contribution 
to its merger rate from those large radii (at least on short timescales; of course, 
over $t\sim\tH$ subhalos may be accreted from these radii, but by then the 
density structure will change and the merger rate will reflect that). 
Naturally, an overdensity at the $\sim3\,$Mpc scale implies an enhanced 
density within that scale. However, we are considering this for 
galaxies and halos of a specific mass, for which the virial radii are generally much smaller 
than these scales, so the increased density in this annulus does not necessarily 
imply an enhanced galaxy density within the halos themselves 
(for that $\mhalo$), although it may affect the overall abundance of the halos. As a 
general rule, merger rates will scale with environmental density on scales less than the 
virial radii of the masses of interest, and be independent of density on larger scales. 

If the merger rate increases in regions with small-scale overdensities, then 
mergers themselves should be biased to such regions. To the extent 
that the small-scale galaxy overdensity around a merger traces this overdensity 
(which we caution is not {\em necessarily} true, as one of the initial galaxies 
in this overdensity is, by definition, consumed in the merger), 
this implies that mergers and merger remnants should preferentially exhibit 
small-scale density excesses. The magnitude of this excess is straightforward 
to determine: for a given galaxy/halo mass, the distribution of 
environments (densities ($1+\delta_{r}$) on a given scale $r$) is 
known. Then, for each scale $r$, the calculation in Figure~\ref{fig:merger.density.dept} 
gives the relative probability of a merger as a function of overdensity. 
Convolving the probability of any object being in given overdensity with the probability of a 
merger in that overdensity gives the mean overdensity of mergers at that scale, i.e.\ 
\begin{equation} 
\frac{\langle x_{\rm merger}\rangle}{\langle x_{\rm all}\rangle}
=\frac{\int{x\,P_{\rm merger}(x)\,P(x\,|\,\mhalo)\,
{\rm d}x}}{\int{x\,P(x\,|\,\mhalo)\,
{\rm d}x}}, 
\end{equation}
where $x\equiv(1+\delta_{r})$.

\begin{figure}
    \centering
    \figexpand
    \plotter{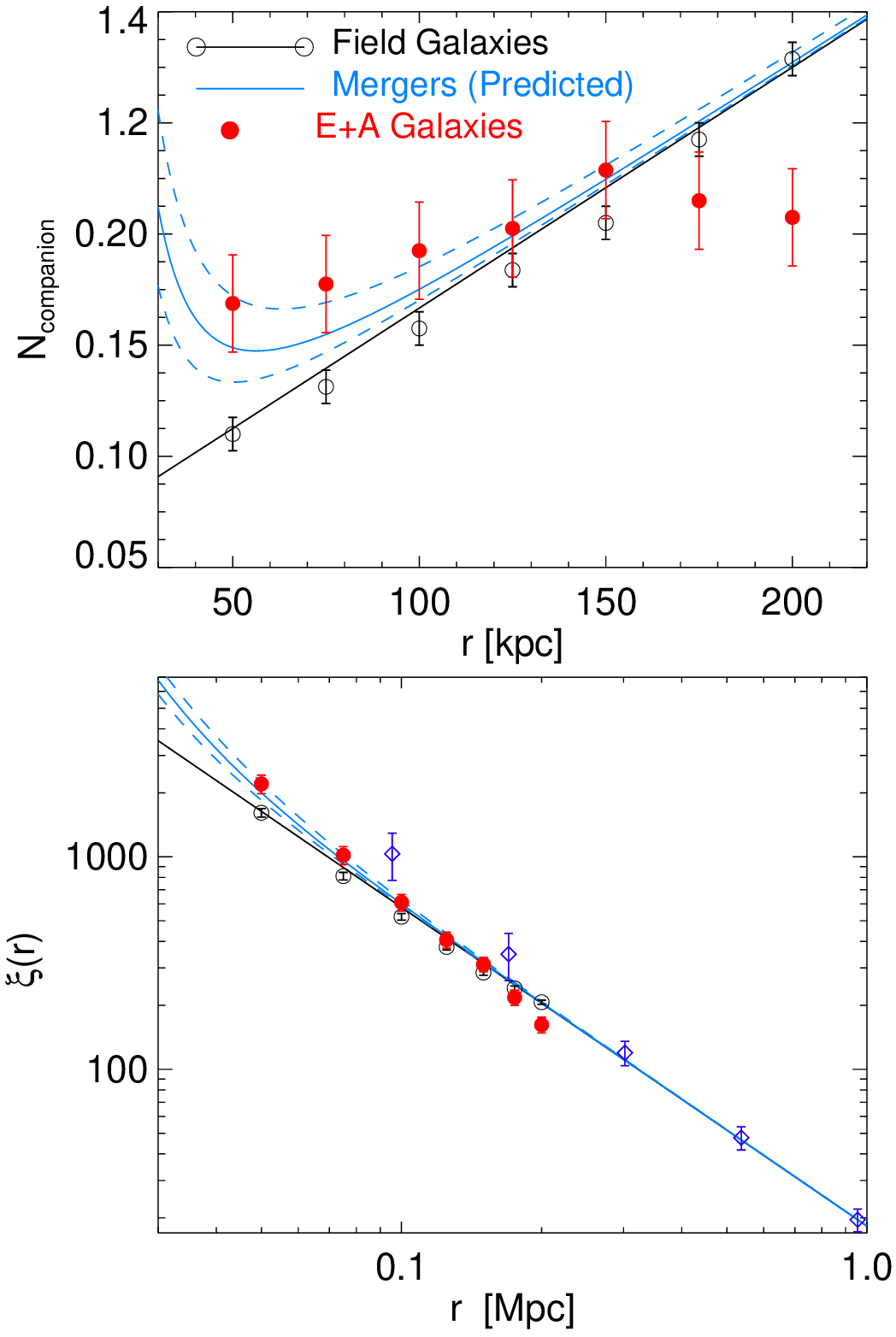}
    \caption{Excess galaxy overdensity on small scales predicted for 
    mergers from our model. Because mergers are more likely when there is a 
    galaxy overdensity on small scales (Figure~\ref{fig:merger.density.dept}), 
    mergers will, on average, occur in regions with slightly enhanced small-scale densities. 
    We show the real-space correlation function ({\em bottom}; technically 
    the merger-galaxy cross correlation function) and corresponding 
    number of companions within a given radius ({\em top}) of all field galaxies
    \citep{goto:e+a.merger.connection}, 
    and then this multiplied by the predicted excess on small scales 
    for mergers (essentially integrating over the probability bias to large overdensity 
    on small scales in Figure~\ref{fig:merger.density.dept}). Dashed blue lines 
    indicate the errors in our estimate from the combination of uncertainties in the field 
    galaxy correlation function, the range of galaxy masses considered (which 
    slightly shifts the physical scale on which the effect is important), and 
    the inclusion/exclusion of Poisson noise in the distribution of overdensities for a 
    given halo mass. The observed number of companions and clustering 
    of post-starburst (likely merger remnant) galaxies is shown for 
    comparison, from \citet[][red circles]{goto:e+a.merger.connection} 
    and \citet[][purple diamonds]{hogg:e+a.env}.
    \label{fig:excess.clustering.mergers}}
\end{figure}

It is straightforward in extended Press-Schechter theory to calculate of 
the probability of forming a halo of a given mass in a given overdensity 
on a particular scale \citep{mowhite:bias}. However, 
since we are calculating a galaxy overdensity in radii about the 
merger candidate, Poisson noise is 
dominant on small scales where the average number of companions is 
$\lesssim1$ -- nevertheless it is again straightforward to calculate the probability of 
a given overdensity. In any case we account for both effects, and show the 
results in Figure~\ref{fig:excess.clustering.mergers}. 
Specifically, we show the average number of companions within a radius of a 
given $r$ about a merger, for all field galaxies. We then 
multiply the field curve by the calculated 
overdensity of mergers as a function of $r$. The exercise can then be trivially repeated 
for the correlation function $\xi(r)$. We compare with observed post-starburst 
populations (E+A/K+A) galaxies, and find that they display a similar excess on small scales. 
As before, the difference on large scales is negligible -- 
unsurprisingly, the density excess becomes important at $r\lesssim r_{\rm vir}$ for 
the typical galaxies of interest. 

Finally, we stress that the excess of companions 
on small scales does {\em not}, in this model, stem from those galaxies themselves having 
any interaction with the central merger (remnant), but reflects a genuine small-scale 
overdensity (as in small groups), in which mergers will be more likely. 

\subsection{Integrated Merger Populations Over Time}
\label{sec:mergers:populations}

At a given redshift, we use our model to predict the mass function of mergers. 
For clarity, we take the mass of a merger to be the total stellar mass of the 
remnant galaxy (roughly the total baryonic mass of the merger 
progenitors). This avoids ambiguity in merger mass ratios, tends to be 
observationally representative (since mergers are generally labeled 
by total luminosity/stellar mass), and has been shown in simulations to 
be a better proxy for the merger behavior than the initial mass of 
either progenitor \citep[as long as it is still a major merger;][]{hopkins:qso.all}. 

\begin{figure}
    \centering
    \figexpand
    \plotone{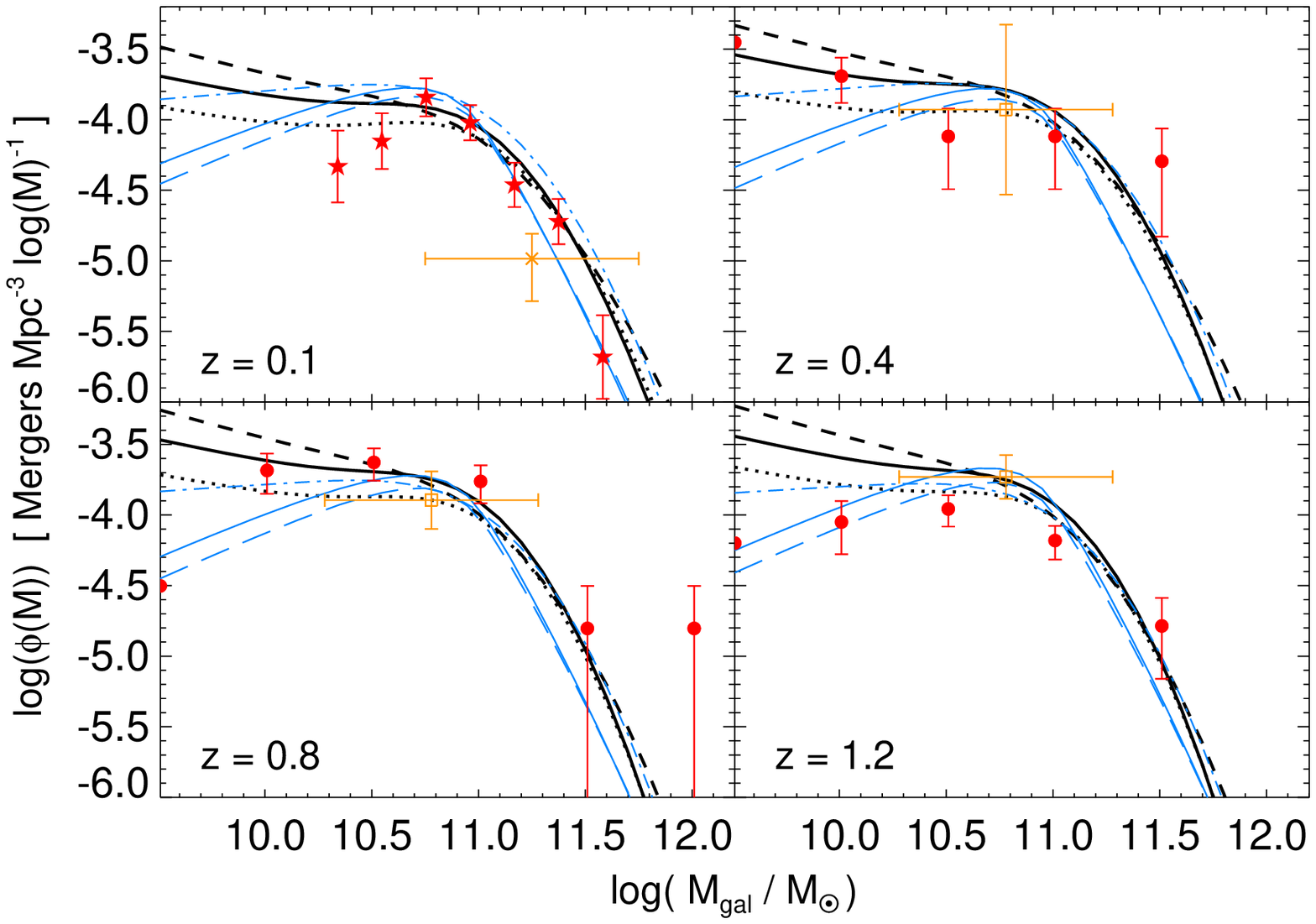}
    \caption{Mass functions (in terms of the remnant stellar mass) of 
    ongoing mergers at each of several redshifts (labeled). Observed mass functions  
    (solid red points) are shown from \citet[][stars]{xu:merger.mf} and \citet[][circles]{bundy:mfs}
    \citep[for a detailed analysis of the mass functions, see][]{hopkins:transition.mass}.
    Error bars do {\em not} include cosmic variance. Observed merger 
    fractions (open orange points), converted to a mass function estimate 
    over the mass range sampled (horizontal errors) are shown 
    from \citet[][cross]{bell:merger.fraction} 
    and \citet[][squares]{lotz:merger.fraction}, with errors including cosmic variance. 
    We compare the prediction of our default model (thick solid black line), for the abundance of 
    mergers and merging pairs. Dotted line employs a different halo occupation 
    model, and dashed line adopts a different 
    fit to the subhalo mass functions (see Figure~\ref{fig:merger.fraction.mhalo} 
    and \S~\ref{sec:mergers:synopsis}).
    We also show the predictions 
    for morphologically identified mergers (thin blue lines), which requires estimating the merger 
    timescale/capture efficiency and duration of morphological disturbances
    (see \S~\ref{sec:mergers:criteria}). 
    We estimate these using a group capture/collisional model (solid), 
    angular momentum capture cross-sections (long dashed), and simple dynamical friction 
    considerations (dotted), calibrating the duration of disturbances from numerical 
    simulations \citep{lotz:merger.selection}. At masses $\gtrsim10^{10}\,\msun$, there is 
    little difference owing to methodology. At very low masses, simulations suggest that the 
    merger timescale (i.e.\ orbital or crossing time after first passage) is considerably longer 
    than the time period over which strong disturbances are excited; however, this is below the 
    mass scales of interest for most of our predictions.     
    \label{fig:merger.mfs}}
\end{figure}

Figure~\ref{fig:merger.mfs} shows the mass functions of ongoing 
mergers at each of several redshifts. We first consider 
the mass function of ``all'' objects which will merge efficiently -- i.e.\ the mass function of 
merging pairs. This requires no knowledge of the 
actual timescale of the merger or e.g.\ lifetime of tidal disturbances. 
The results agree well with the mass functions and merger fractions 
estimated at all $z\lesssim1.5$, suggesting that our model does 
indeed reasonably describe the true nature of galaxy mergers. For comparison, we 
show the results obtained using 
a different halo occupation model to associate galaxies and 
halos, or using a different set of simulations/models to estimate the subhalo 
mass functions. As noted in \S~\ref{sec:mergers:criteria}, these choices make very little difference 
(considerably smaller than e.g.\ the systematics in the observations). 

It is not always clear, however, that observations capture all merging pairs 
(or that our definition of ``all'' is appropriate as, for some mergers, 
$t_{\rm merger}\rightarrow\tH$). Often, 
systems are identified as mergers on the basis of tidal disturbances and other 
clear morphological signatures. We therefore calculate the mass function of systems 
observed in this manner. This requires that we adopt one of the 
models in \S~\ref{sec:mergers:criteria} 
for the merger timescale, which tells us how long it will characteristically take for a given 
merger to reach the interaction cross section where tidal disturbances will be excited. 
Then, using numerical simulations to estimate the typical duration of those features 
\citep[in which they will be identified by typical morphological classification schemes, see][]
{lotz:merger.selection}, we obtain the observed ``disturbed morphology'' mass functions. 
We perform this calculation using each of the methods for calculating the merger 
timescale described in \S~\ref{sec:mergers:criteria}. Note that the number of systems 
according to this convention
can exceed that in our ``all pairs'' definition if the timescale on which 
disturbances are visible is longer than the ``infall'' timescale or timescale on 
which the subhalo survives (the case for very efficient infall/capture). 

At high masses, the difference between samples of merging pairs and 
those of disturbed systems is small, as is the difference between our choice of 
methodology in calculating the merger abundances and/or timescales. This is 
because high-mass systems merge more quickly, excite morphological 
disturbances more easily on first passage, and are brighter (making 
faint morphological features easier to identify). At very low masses 
$\mgal\lesssim10^{10}\,\msun$, our predictions do diverge -- this is because the 
overall infall or merger timescale can become substantially longer than 
the timescale over which morphological disturbances are excited (in these cases, 
this occurs closer to the final coalescence). Although this conclusion 
merits more detailed numerical 
investigation in future work, it has little effect on any 
of our predictions -- for example, the total merger fraction (especially at high redshift) 
is restricted to larger-mass $\mgal\gtrsim\mstar$ systems, where the predictions 
agree well, and the overall merger mass density is nearly identical regardless of 
the methodology. Furthermore, quasar and galaxy formation processes are 
probably influenced (or even dominated) by other mechanisms 
(such as secular disk instabilities and quenching via infall as a 
satellite galaxy) at these low masses, which we do not attempt to model.

\begin{figure*}
    \centering
    \figexpand
    \plotone{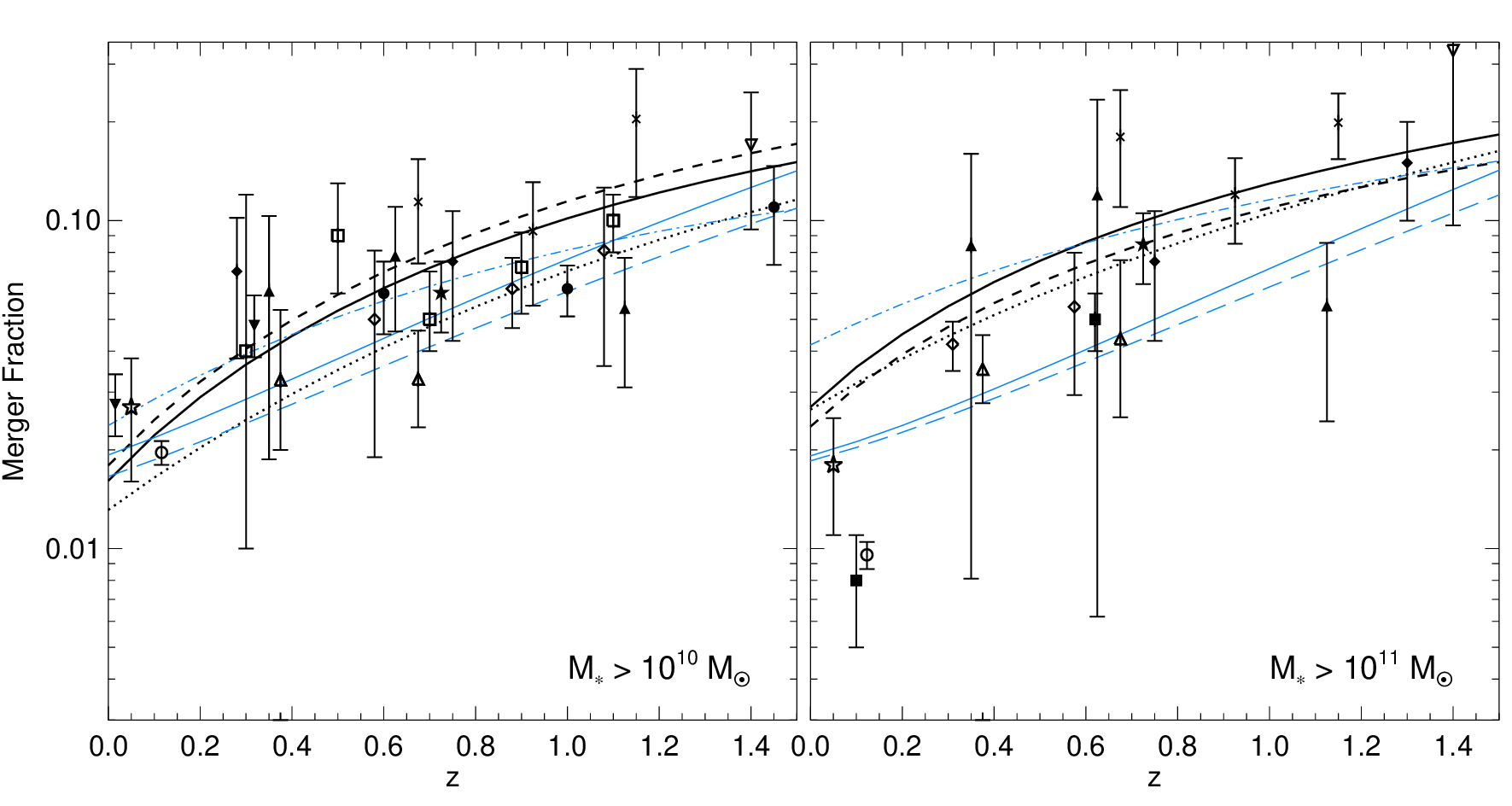}
    \caption{Predicted merger fraction as a function of redshift (lines, 
    same style as Figure~\ref{fig:merger.mfs}), above two approximate mass 
    limits. Observations (points) are shown from 
    \citet[][filled inverted triangles]{patton:merger.fraction}, \citet[][filled circles]{conselice:merger.fraction}, 
    \citet[][filled triangles]{bundy:merger.fraction}, 
    \citet[][open diamonds]{lin:merger.fraction}, \citet[][open stars]{xu:merger.mf}, 
    \citet[][open circles]{depropris:merger.fraction}, \citet[][filled diamonds]{cassata:merger.fraction}, 
    \citet[][filled stars]{wolf:merger.mf}, \citet[][open triangles]{bundy:mfs},     
    \citet[][open inverted triangles]{lotz:morphology.evol}, \citet[][open squares]{lotz:merger.fraction},
    \citet[][filled squares]{bell:merger.fraction}, and 
    \citet[][$\times$'s]{bridge:merger.fractions}.
    Note that the mass limit 
    is only approximate in several of these cases, as they are selected by optical luminosity. 
    The predicted merger fractions agree well, especially for the deeper case which 
    resolves $\mstar$ galaxies. 
    \label{fig:merger.fraction}}
\end{figure*}

We next integrate the mass functions in Figure~\ref{fig:merger.mfs} above a given 
mass limit to predict the merger fraction as a function of redshift, shown in 
Figure~\ref{fig:merger.fraction}. The fraction is determined relative to the mass 
functions in \citet{fontana:highz.mfs}, who provide a continuous fit over the range of 
interest. But we note that since this is an integrated quantity, the difference 
adopting other mass function estimates \citep[e.g.][]{borch:mfs} is small
(at least at $z\lesssim1.5$). Comparing 
this to a range of observations, the agreement is good, especially 
for the deeper mass limit. For high mass 
mergers ($\mgal\gtrsim10^{11}\,\msun$) there is greater scatter in the observations, 
which most likely owes to cosmic variance (especially at $z\lesssim0.2$).  In both 
cases, however, the merger fraction is not an especially steep function of 
redshift. In fact, between $z= 0.3-1.5$, the fraction increases by 
only a factor $\sim3-4$, 
consistent with most observations finding a relatively flat merger fraction in this 
range \citep[e.g.][]{lin:merger.fraction,lotz:merger.fraction} and 
recent cosmological simulations \citep{maller:sph.merger.rates}. 
Further, although halos may be merging more frequently at high redshift, they 
are also merging more rapidly, meaning that the fraction merging at any instant 
can be relatively flat. 

Finally, given our model for the halos hosting mergers, it is straightforward to calculate 
the predicted clustering properties of those mergers. Specifically, we have 
already predicted a number density of mergers as a function of halo mass, galaxy mass, and 
redshift; i.e.\ some $n_{\rm merger}(\mgal\,|\,\mhalo,\,z)$. 
Knowing the clustering amplitude or bias of each host halo $b(\mhalo\,|\,z)$, it is straightforward 
to predict the clustering of the merging galaxies, in the same manner by which halo 
occupation models construct the clustering of a given population: 
\begin{equation}
b(\mgal) = \frac{\int{b(\mhalo)\,n_{\rm merger}(\mgal\,|\,\mhalo)\,{\rm d}{\mhalo}}}
{\int{n_{\rm merger}(\mgal\,|\,\mhalo)\,{\rm d}{\mhalo}}}. 
\end{equation}
We calculate $b(\mhalo)$ following \citet{mowhite:bias} as updated 
by \citet{shethtormen} to agree with the results of numerical simulations. 

\begin{figure}
    \centering
    \figexpand
    \plotone{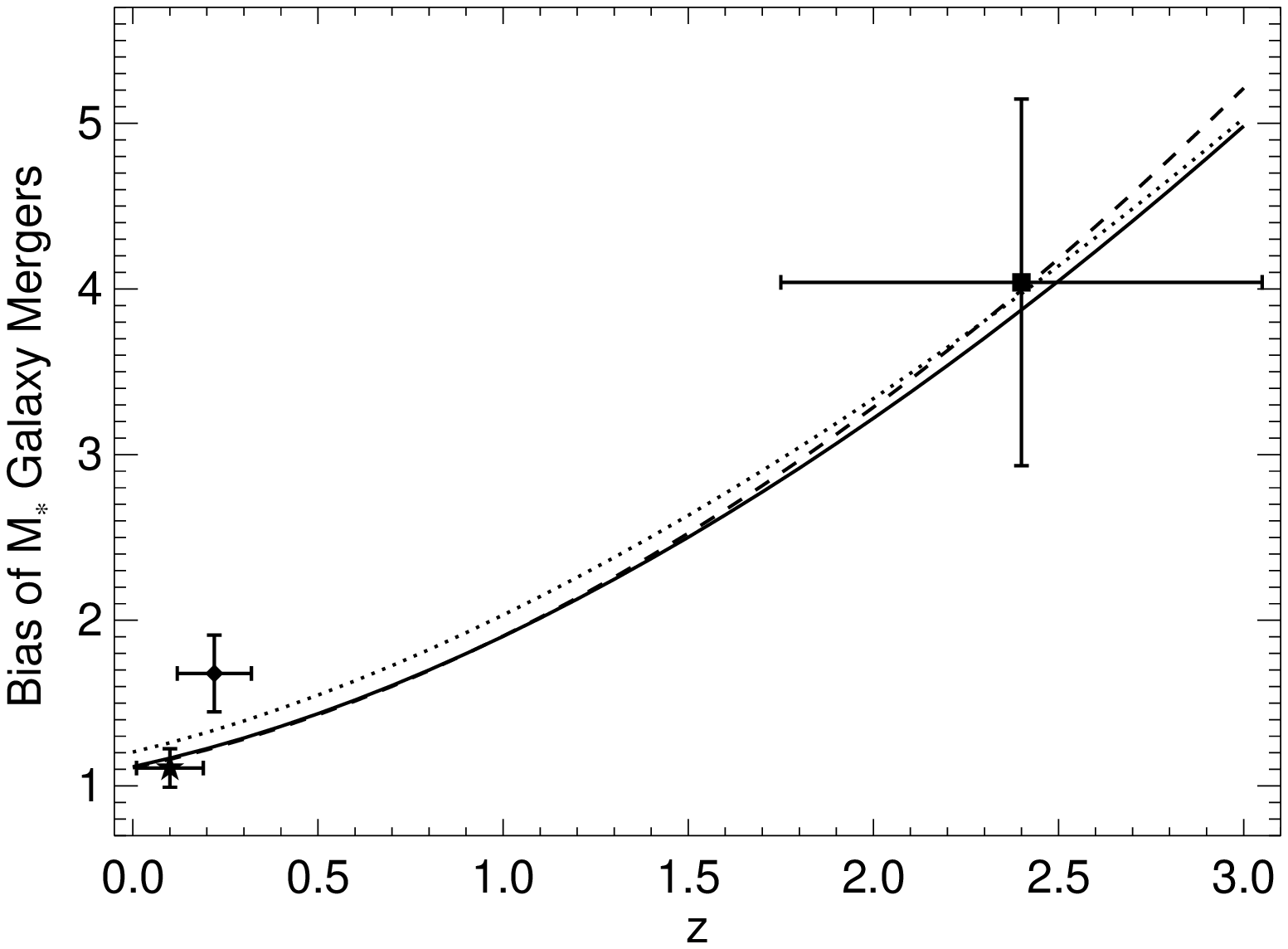}
    \caption{Comparing our predicted clustering of $\sim\mstar$ major mergers (lines; 
    style as in Figure~\ref{fig:merger.mfs}) 
    as a function of redshift to that various populations usually associated with 
    galaxy mergers (points): post-starburst (E+A/K+A) galaxies 
    \citep[][star]{blake:e+a.clustering}, 
    close galaxy pairs \citep[][diamond]{infante:pair.clustering}, and 
    sub-millimeter galaxies \citep[][square]{blain:smg.clustering}. 
    \label{fig:merger.clustering}}
\end{figure}

Figure~\ref{fig:merger.clustering} shows this as a function of redshift. Since 
observations generally sample near $\mgal\sim\mstar$, we plot this for 
$\mgal=\mstar(z=0)\approx10^{11}\,\msun$. We compare with available 
clustering measurements for 
likely major-merger populations. At low redshifts, \citet{blake:e+a.clustering} have 
measured the clustering of a large, uniformly selected 
sample of post-starburst (E+A/K+A) galaxies in the 2dF. 
\citet{infante:pair.clustering} have also measured the 
large-scale clustering of close galaxy pairs selected from the SDSS at 
low redshift. At high redshift, no such samples exist, but \citet{blain:smg.clustering} 
have estimated the clustering of a moderately large sample of 
spectroscopically identified sub-millimeter galaxies at $z\sim2-3$, 
which as discussed in \S~\ref{sec:intro} are believed to originate in major mergers. 
Our prediction is consistent with these constraints -- 
however, given the very limited nature of the data and the lack of 
a uniform selection criteria for ongoing or recent mergers at different 
redshifts, we cannot draw any strong conclusions. 

One caution should be added: 
recent higher-resolution simulations suggest that the approximation here 
(and in many -- but not all -- halo occupation models), that bias is a function only of 
halo mass at a given redshift, may not be accurate
\citep[e.g.,][]{gao:assembly.bias,harker:marked.correlation.function,wechsler:assembly.bias}. 
In particular, because mergers 
have particularly recent halo assembly times for their post-merger masses, 
they may represent especially biased regions of the density distribution. 
Unfortunately, it is not clear how to treat this in detail, as there remains considerable 
disagreement in the literature as to whether or not a significant ``merger bias'' exists 
\citep[see, e.g.][]{kauffmann:qso.clustering,percival:merger.bias,furlanetto:merger.bias,
lidz:merger.bias}. 
Furthermore the distinction between galaxy-galaxy and 
halo-halo mergers (with the considerably longer timescale for most galaxy mergers) 
means that it is not even clear whether or not, after the galaxy merger, there would be a 
significant age bias. 

In any case, most studies suggest the effect is quite small: using 
the fitting formulae from \citet{wechsler:concentration,wechsler:assembly.bias}, 
we find that even in extreme cases 
(e.g.\ a $\mhalo\gg\mstar$ halo merging at $z=0$ as opposed to an average 
assembly redshift $z_{f}\approx6$) the result is that the standard EPS formalism 
underestimates the bias by $\approx30\%$. For the estimated 
characteristic quasar host halo masses 
and redshifts of interest here, the maximal effect is $\lesssim 10\%$ at all $z=0-3$, 
much smaller than other systematic effects we have considered (and 
generally within the range of 
our plotted variant calculations in Figure~\ref{fig:merger.clustering}). 
This is consistent with \citet{gao:assembly.bias} and \citet{croton:assembly.bias} 
who find that assembly bias is only important 
(beyond the $10\%$ level) for the most extreme halos or galaxies in their simulations, 
where for example the clustering 
of small halos which are destined to be 
accreted as substructure in clusters ($\gtrsim 10^{15}\,h^{-1}\,M_{\sun}$) will be 
very different from the clustering of similar-mass halos in field or void environments. 
Indeed, our own calculation in Figure~\ref{fig:excess.clustering.mergers} suggests 
that merger bias applies only on small scales, and that mergers show no preference 
for excess densities on the large scales for which the linear bias description is 
meaningful. 
The effect may grow with redshift, however, so care should be taken in extrapolating 
the predictions in Figure~\ref{fig:merger.clustering} to higher redshifts. 
For further discussion of the effects on the data and predictions shown here, 
we refer to \citet{hopkins:clustering}. 

\begin{figure}
    \centering
    \figexpand
    \plotter{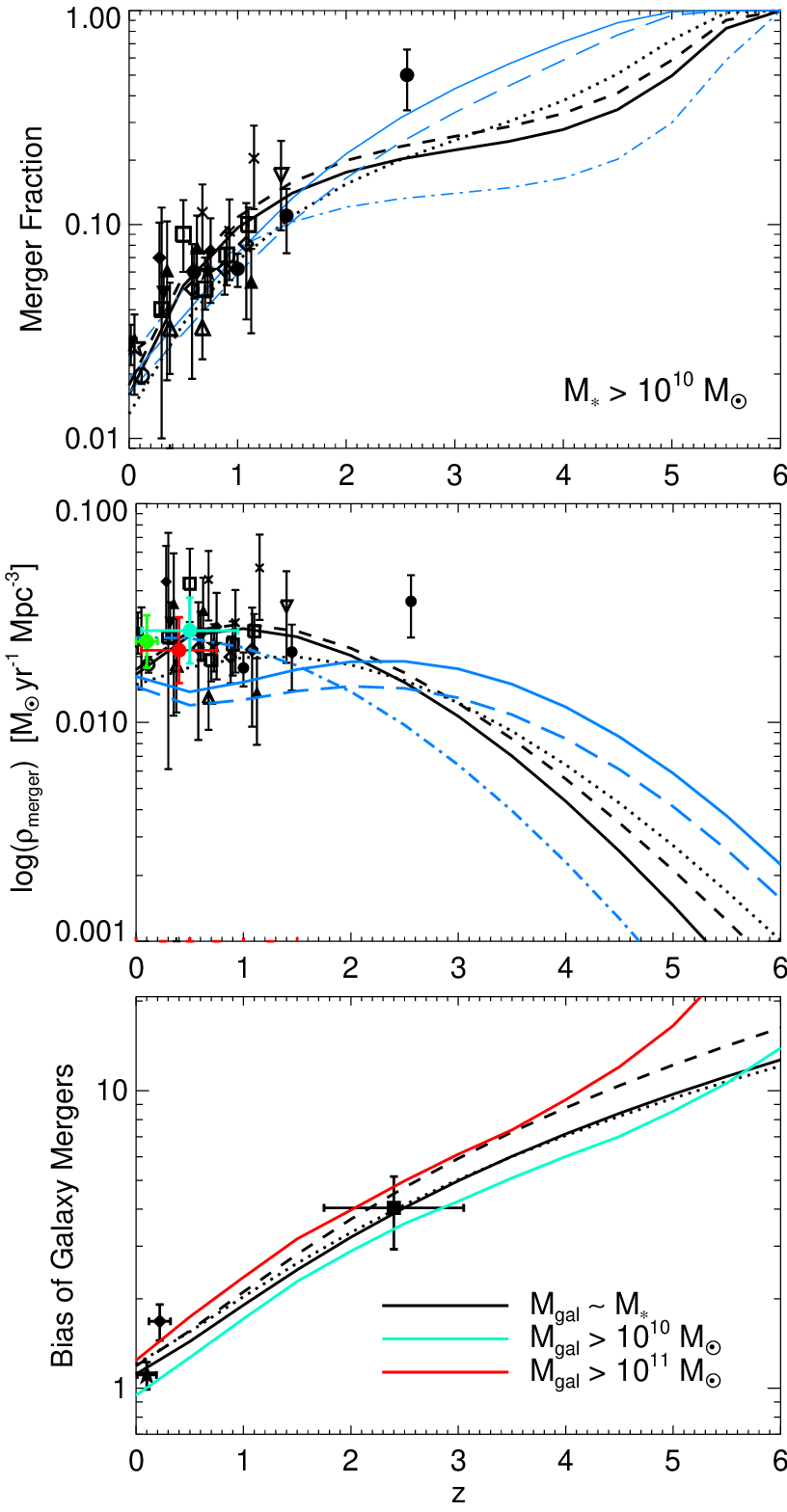}
    \caption{{\em Top:} As Figure~\ref{fig:merger.fraction}, but 
    extending our predicted merger fractions to high redshift.
    {\em Middle:} Mass flux through mergers (i.e.\ total rate of stellar mass 
    merging). Black points are observed merger fractions converted to an 
    estimated mass flux rate following \citet{hopkins:transition.mass}. 
    Green, red, and blue circle show the observationally inferred 
    mass flux through the ``green valley'' (i.e.\ from blue cloud to red sequence), 
    rate of growth of the red sequence, and rate of mass loss off the 
    blue cloud (respectively), from $z\sim0-1$ \citep{martin:mass.flux} 
    (see \papertwo\ for a more detailed comparison).
    {\em Bottom:} As Figure~\ref{fig:merger.clustering}, but 
    extended to higher redshift. Blue and red lines show the clustering of 
    mergers above the given mass thresholds. 
    \label{fig:merger.highz}}
\end{figure}

For the sake of future comparison, we show in Figure~\ref{fig:merger.highz} 
our predictions for the merger fractions and clustering of 
mergers (Figure~\ref{fig:merger.fraction} \&\ \ref{fig:merger.clustering}, 
respectively) at all redshifts $z=0-6$. We note the caveat that 
our merger fraction is defined relative to the mass functions in 
\citet{fontana:highz.mfs}, which become uncertain at high redshifts, 
although this uncertainty is comparable to the differences between 
the methods of calculating the merger timescale (as discussed in 
\S~\ref{sec:mergers:synopsis}). It is also less clear 
what the observable consequences of mergers at 
the highest redshifts may be -- if merger 
rates are sufficiently high, there may be a large number of multiple 
mergers (as in \citet{li:z6.quasar}),
or systems may effectively be so gas rich that merging 
preserves disks and operates as a means of ``clumpy accretion'' 
\citep[e.g.][]{robertson:disk.formation}.

Although the estimates differ at the highest redshifts, we stress that their 
integrated consequences at low redshifts $z\lesssim3$ are
similar, as this is where most merging activity and spheroid/BH mass 
buildup occurs. 
We also note that high-redshift mergers are likely to be the most 
massive $\mgal\gg M_{\ast}$ systems, so we show 
our predictions for the clustering of mergers assuming different 
mass limits (as opposed to strictly at $\mgal=M_{\ast}$). We 
also plot the mass flux in mergers, i.e.\ the 
integrated rate at which galaxy baryonic/stellar mass is merged, 
$\int \mgal\,\dot{n}(\mgal)\,{\rm d}\log{\mgal}$. This compares favorably 
with the observationally inferred rates at which mass is moved 
off the blue cloud, through the ``green valley,'' and onto 
the red sequence \citep[from the evolution in galaxy mass functions 
and color-magnitude relations; see][]{martin:mass.flux}, as expected 
in a model where mergers drive such a transition (for details, see 
\papertwo). Future observations of these quantities at high redshift 
will improve the constraints on our halo occupation and 
merger timescale estimates, allowing for more accurate calculations
of e.g.\ quasar triggering and spheroid formation rates at these
epochs.

\section{Quasars}
\label{sec:quasars}

\subsection{Consequences of Merger-Driven Fueling: 
What Determines Where and When Quasars Live}
\label{sec:quasars:mergers}

Having developed in \S~\ref{sec:mergers} 
a physically-motivated model of merger rates as a function of 
galaxy and halo mass, environment, and redshift (and tested that this 
model is consistent 
with the existing body of merger observations), we can now extend 
our application. As discussed in \S~\ref{sec:intro}, the argument for an 
association between mergers and quasars has a long history. We therefore 
make the simple ansatz: {\em Every major merger of star-forming/gas-rich galaxies 
triggers a quasar}. 

\begin{figure*}
    \centering
    \figexpand
    \plotone{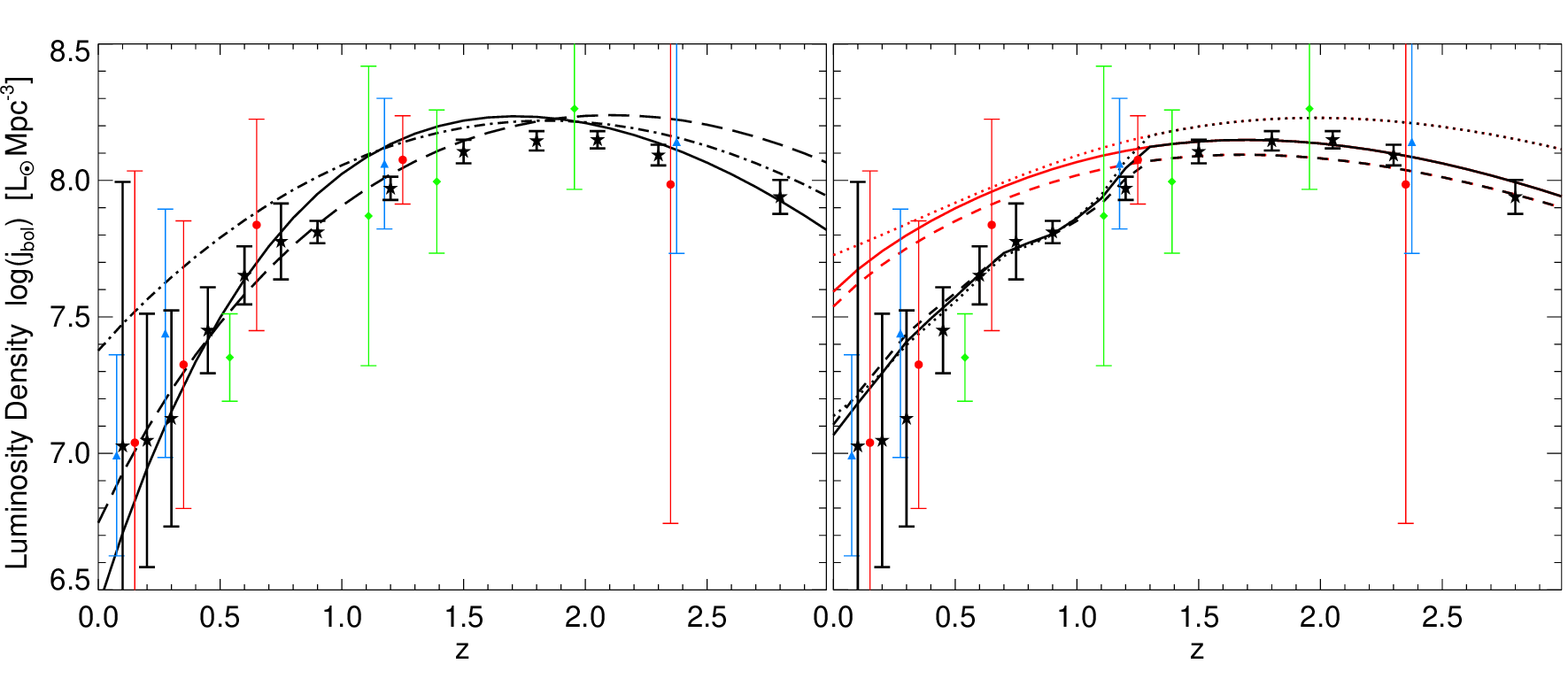}
    \caption{Predicted quasar luminosity density, if quasars are triggered in mergers, 
    as a function of redshift. {\em Left:} Prediction from a simplified toy model 
    in which all halos hosting $\sim\lstar$ galaxies undergo major mergers near their 
    characteristic small group mass scale, and build a BH which obeys the appropriate 
    $\mbh-\mhalo$ relation for that redshift
    \citep[estimated $\mbh-\mhalo$ as a function of redshift from][corresponding to 
    solid, long dashed, 
    and dot-dashed lines, respectively]{hopkins:clustering,fine:mbh-mhalo.clustering,
    hopkins:bhfp}. 
    Points show observational estimates from 
    the measured QLFs of \citet[][red circles]{ueda03:qlf}, \citet[][blue triangles]{hasinger05:qlf}, 
    \citet[][green diamonds]{richards05:2slaq.qlf}, and the large compilation of 
    multiwavelength QLF data in \citet[][black stars]{hopkins:bol.qlf}. The observations 
    from specific bands are converted to a bolometric luminosity density using the 
    bolometric corrections calibrated in \citet{hopkins:bol.qlf}. 
    {\em Right:} Same, but the predicted luminosity density is calculated properly accounting for all 
    galaxy and halo masses from the merger rate functions determined in \S~\ref{sec:mergers}, and 
    adopting the observed ratio of BH to host galaxy spheroid mass as a function of redshift 
    \citep[e.g.][]{peng:magorrian.evolution}. Linestyles correspond to different 
    means of estimating the exact merger rates, as in Figure~\ref{fig:merger.mfs}. Red lines 
    assume all mergers will trigger quasars, black (lower) lines assume only gas-rich (``wet'') mergers 
    can trigger bright quasar activity (adopting the observed fraction of 
    gas-rich/star-forming/blue galaxies as a function of $\mgal$ and 
    $\mhalo$ as in Figure~\ref{fig:merger.redblue}). 
    A merger-driven model naturally predicts both the rise and fall of the global quasar luminosity density 
    to high precision.
    \label{fig:lum.density}}
\end{figure*}
From this statement, we can make a number of robust predictions. In \S~\ref{sec:mergers} 
we derived the characteristic host halo mass for mergers of $\sim\mstar$ galaxies. 
To the extent that these are gas-rich systems, this should therefore also 
represent the characteristic host halo mass of quasars, and (since the mass density of 
the Universe is dominated by systems near $\sim\mstar$) dominate the buildup of black 
hole mass. 

From the \citet{soltan82} argument, the black hole mass density of the Universe 
must be dominated by growth in typical, bright quasar phases with canonical radiative 
efficiency $\epsilon_{r}\sim0.1$. Let us construct the simplest possible model: 
mergers (of $\mstar$ galaxies) characteristically occur at a host halo mass $\sim \mmerger$. 
From the halo mass function, it is straightforward to calculate the rate at which halo mass 
crosses this mass threshold, 
\begin{equation}
\dot{\rho}_{\rm halo} = \bar{\rho}\,\frac{{\rm d}F(>\mhalo)}{{\rm d}t}, 
\end{equation}
where $F(>\mhalo,\,z)$ is the fraction of mass in halos of mass greater than 
$\mhalo$, determined from the Press-Schechter formalism revised following 
\citet{shethtormen}. 
Assume that every such halo undergoes a merger approximately 
upon crossing this mass threshold, which transforms its galaxy from disk to spheroid. The 
hosted BH mass therefore grows from some arbitrarily small amount to the expected mass 
given the BH-host mass relations, which we can write as $\mbh=\nu(z)\,\mhalo$ 
(we distinguish this from $\mbh=\mu(z)\,\mgal$). 
The ratio $\nu(z)$ is determined to $z\sim3$ from the clustering of active BHs 
of a given mass at each redshift 
\citep[see e.g.,][]{daangela:clustering,fine:mbh-mhalo.clustering,
hopkins:clustering,hopkins:bhfp}, and indirectly from determinations of the 
BH host galaxy masses \citep{peng:magorrian.evolution}. 
The total rate at which BH mass is built up is then
\begin{equation}
\dot{\rho}_{\rm BH} = \nu(z)\,\dot{\rho}_{\rm halo} = \nu(z)\,\bar{\rho}\,\frac{{\rm d}F(>\mhalo)}{{\rm d}t}, 
\end{equation}
and the bolometric luminosity density is $j_{\rm bol}=\epsilon_{r}\,\dot{\rho}_{\rm BH}\,c^{2}$. 
Figure~\ref{fig:lum.density} compares this simple estimate with the observed bolometric 
quasar luminosity density as a function of redshift. 

The agreement is striking, which suggests that this toy model, such that
the bulk of the 
assembly of BH mass occurs near the transition halo mass, is reasonable. This also 
naturally explains the rise and fall of the quasar luminosity density with time. However, 
this is ultimately just a simple approximation -- we can consider this in greater detail adopting 
our previous estimate of the merger rate as a function of stellar mass and redshift, $\dot{n}(\mgal\,|\,z)$, from \S~\ref{sec:mergers}. Each major merger transforms disks to spheroids, building a BH of 
average mass $\mbh = \mu(z)\,\mgal$. We should properly only consider mergers of 
gas-rich or star-forming systems, as dry mergers will, by definition, not be able to trigger 
quasar activity and form new BH mass. Therefore, we empirically adopt the fraction of 
red and blue galaxies at each $\mgal,\,\mhalo$ (as in \S~\ref{sec:mergers}) to restrict 
only to mergers of blue galaxies. 
Again, $\mu(z)$ has been directly determined from 
observations \citep{peng:magorrian.evolution}, and estimated from theoretical arguments 
\citep{hopkins:bhfp}. For convenience, we adopt the numerical best-fit estimate of 
$\mu(z)$ from \citet{hopkins:bhfp}. A good approximation to this 
numerical function is 
\begin{equation}
\mu(z) \approx 0.0012\,{\Bigl(}\frac{1+z^{5/2}}{1+(z/1.775)^{5/2}}{\Bigr)}, 
\end{equation}
which matches the asymptotic observed values at low and high redshift 
\citep{haringrix,walter04:z6.msigma.evolution}, and captures the observed weak evolution 
to $z\sim1$ and rapid evolution between $z=1-3$ \citep{shields03:msigma.evolution,
peng:magorrian.evolution,salviander:msigma.evolution}. 
Given the merger rate $\dot{n}(\mgal\,|\,z)$, we can then convert this to a cosmic 
rate of formation or build-up of BHs in merger-driven quasars, 
\begin{equation}
\dot{n}(\mbh\,|\,z) = \int{P(\mbh\,|\,\mgal)\,\dot{n}(\mgal\,|\,z)\,{\rm d}\log{\mgal}}. 
\end{equation}
The intrinsic dispersion about the mean BH-host mass relation appears, at all redshifts, to be 
roughly lognormal with width $\approx0.27\,$dex, so we model $P(\mbh\,|\,\mgal)$ as such.  
Once the total rate of formation of BH mass is calculated, the same conversion above 
yields the quasar luminosity density. 

Figure~\ref{fig:lum.density} shows the results of this 
more detailed calculation. They are similar to the results from our extremely simplified model -- 
which reflects the fact that most of the mass/luminosity density is contained near 
$\mstar$ or $\lstar$. Note that 
considering all mergers (i.e.\ including dry mergers) overpredicts the quasar luminosity 
density at low redshifts. This demonstrates that the decrease in the quasar luminosity density 
at low redshifts is, in part, driven by the fact that an increasing fraction of massive systems have 
already been transformed to ``red and dead'' systems at late times, and are no longer available to fuel 
quasars, even if they undergo subsequent dry mergers. By $z\sim0$, for example, 
a large fraction ($\sim50\%$) of the mass density in $>M_{\ast}$ systems has already 
been gas-exhausted (discussed in detail in \papertwo), 
and therefore such mergers are no longer a viable fuel supply 
for quasar activity. As discussed in \S~\ref{sec:mergers:scales}, the predicted gas-rich merger 
mass density (and corresponding quasar luminosity density) at $z\lesssim0.5$ will be slightly 
lower if these gas-exhausted systems are preferentially surrounding by 
gas-exhausted satellites (compared to gas-rich central galaxies of the same mass in 
similar halos), but it is clear in Figure~\ref{fig:lum.density} that this is completely 
consistent with the observations (especially if secular processes contribute significantly 
to the quasar luminosity density at low redshifts and luminosities, as we expect from 
our comparisons in \S~\ref{sec:quasars:secular}).

\begin{figure}
    \centering
    \figexpand
    \plotone{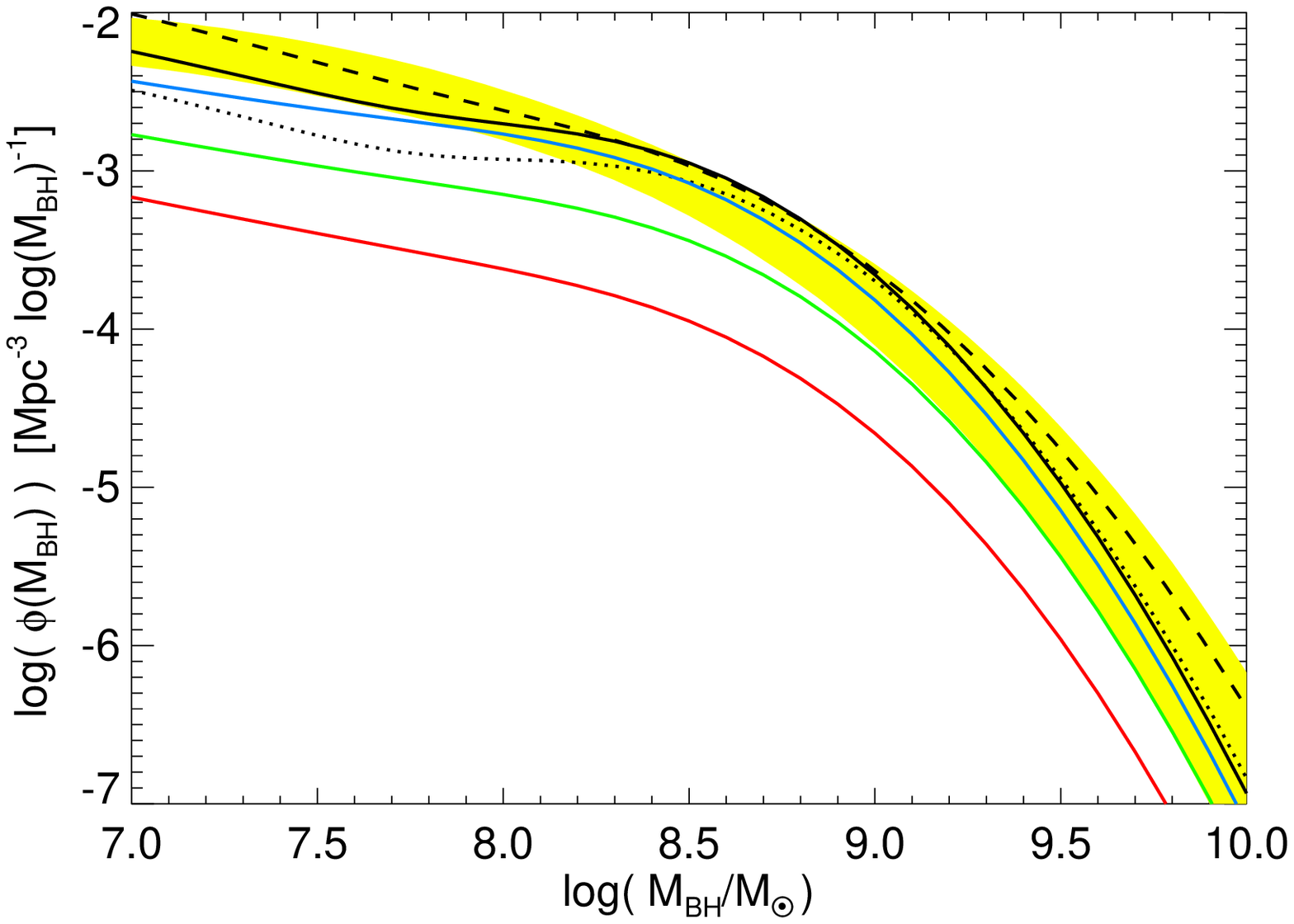}
    \caption{Predicted BH mass function (BHMF) from gas-rich merger-driven quasar/BH formation 
    (Figure~\ref{fig:lum.density}, right). Results are shown at $z=0$ (black lines; linestyles 
    correspond to different calculations of the merger rates, as in Figure~\ref{fig:merger.mfs}), 
    and $z=1,\,2,\,3$ (blue, green, and red, respectively; for clarity, only our fiducial calculation 
    -- solid line -- is shown, but relative evolution with redshift for each calculation is similar). 
    Yellow (shaded) range shows the $z=0$ observational estimate of the BHMF 
    in \citet{marconi:bhmf}. Integrating forward the 
    merger mass functions as a function of redshift yields a good match to the local BHMF. 
    The effect of dry mergers is included, but is small. 
    \label{fig:bhmf}}
\end{figure}
Having calculated the rate of BH formation as a function of the remnant BH mass, 
$\dot{n}(\mbh\,|\,z)$, it is trivial to integrate this forward and predict the BH mass 
function (BHMF) at any time. Figure~\ref{fig:bhmf} shows the result of this calculation at 
$z=0$, compared to the observationally estimated BHMF. The two agree well at all masses, 
even at very large $\mbh\sim10^{10}\,\msun$. We also show the BHMF at several other redshifts. 
Interestingly, there is a downsizing behavior, where a large 
fraction of the most massive 
BHs are in place by $z=2$, while less massive BHs form later \citep[essentially required by the 
fact that few $\sim10^{9}\,\msun$ BHs are active at low redshift, while a very high fraction are 
active at $z\sim2$, see][]{mclure.dunlop:mbh,kollmeier:eddington.ratios,fine:mbh-mhalo.clustering}. 
If we were to ignore dry mergers at low redshifts, this effect would be even more pronounced, 
but at $z\lesssim1$ their effect is to move some of the BH mass density from lower-mass systems 
into higher mass $\gtrsim10^{9}\,\msun$ systems (at higher redshifts, the effects are negligible). 
It is not obvious, however, that this translates to 
downsizing in galaxy mass assembly, since the ratio of BH to galaxy mass $\mu(z)$ evolves 
with redshift. We will return to this question in \papertwo. 

\begin{figure}
    \centering
    \figexpand
    \plotone{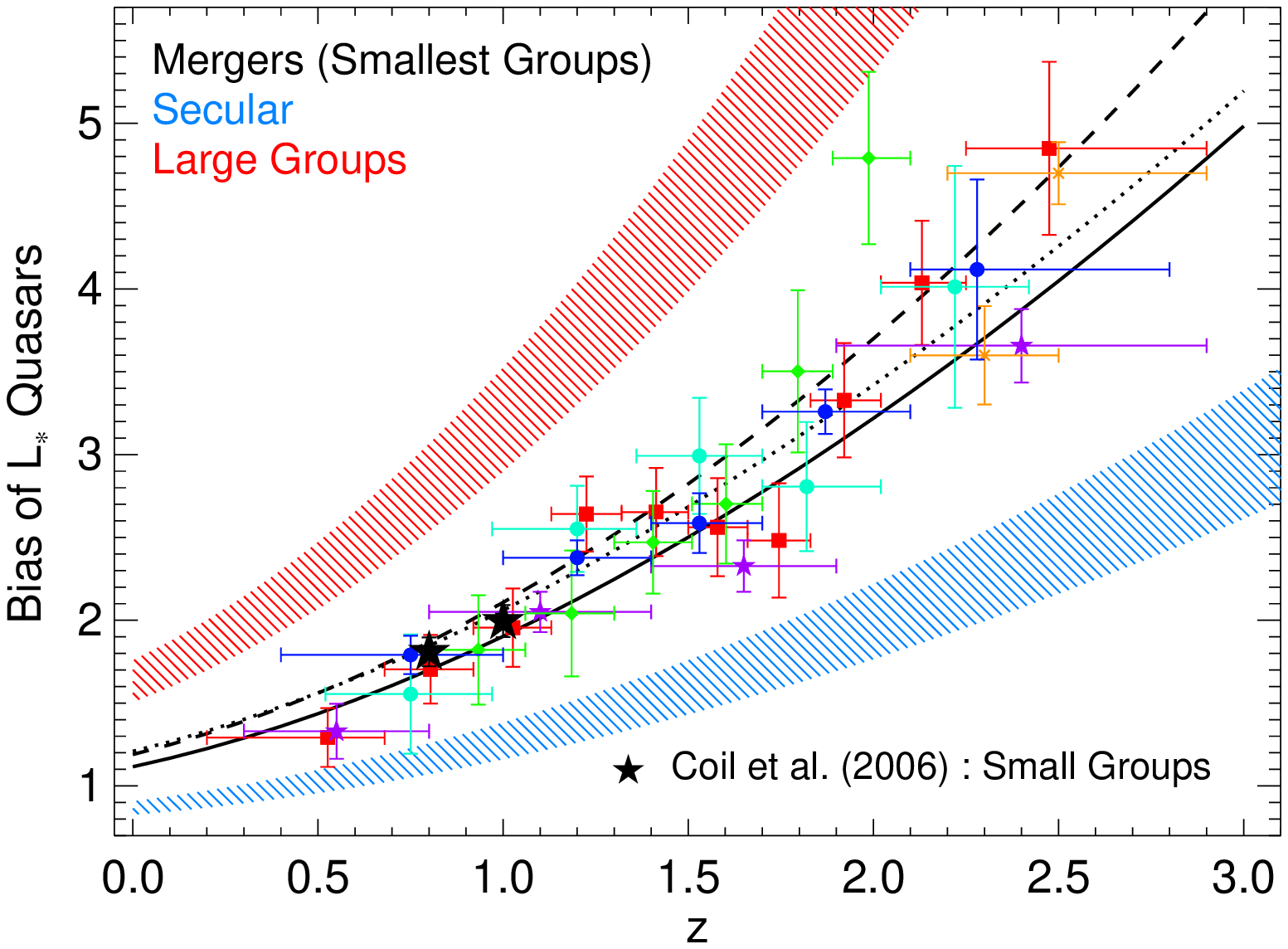}
    \caption{Predicted quasar clustering as a function of redshift, assuming merger-triggering 
    (black lines, as in Figure~\ref{fig:merger.mfs}), corresponding to the small group scale of 
    $\sim\mstar$ galaxies. Red (upper) shaded range show the prediction if quasars were associated 
    with large group scales, blue (lower) range show the prediction from a secular model in 
    which quasar clustering traces that of star-forming galaxies observed at each redshift (lines show 
    $\pm1\,\sigma$ range estimated from the 
    compiled observations in \citet{hopkins:clustering}, from \citet{shepherd:clustering.by.type,
    giavalisco:lbg.clustering,norberg:clustering.by.lum.type,coil:prelim.clustering,zehavi:local.clustering,
    adelberger:lbg.clustering,allen:lum.dep.lbg.clustering,
    phleps:midz.clustering,meneux:clustering.vs.z,lee:lbg.clustering}). Points show 
    quasar clustering measurements from \citet[][red squares]{croom:clustering}, 
    \citet[][green diamonds]{porciani:clustering}, 
    \citet[][cyan and blue circles]{myers:clustering,myers:clustering.old}, 
    and \citet[][violet stars]{daangela:clustering}. 
    Large black stars 
    show the observed clustering of $z\sim1$ small groups (of $\sim\lstar$ galaxies) 
    from \citet{coil:clustering.groups}, corresponding to the most efficient scales for 
    major $\sim\lstar$ galaxy mergers. Quasar clustering measurements are consistent 
    with the small group scale in which mergers proceed efficiently. 
    \label{fig:quasar.bias}}
\end{figure}
Since we begin our calculation with the halos hosting quasars, we should 
be able to predict the bias of quasars as a function of redshift. As in Figure~\ref{fig:merger.clustering}, 
we use the known clustering of the halos hosting mergers to calculate the 
clustering of those mergers as a function of redshift. Assuming each merger produces a 
quasar of the appropriate mass, this yields the expected clustering of quasars 
as a function of redshift. Figure~\ref{fig:quasar.bias} compares this prediction 
to observed quasar clustering as a function of redshift. Technically, we adopt 
the quasar lightcurve models from \S~\ref{sec:quasars:qlf} below to determine the clustering 
specifically of $\lstar$ quasars (i.e.\ determining the relative contribution to $\lstar$ from 
different host masses and their clustering as in Figure~\ref{fig:merger.clustering}), 
but the result is nearly identical 
to assuming that $\lstar$ quasars trace $\mstar$ mergers (Figure~\ref{fig:merger.clustering}). 
This should 
be true in any model, as long as the quasar lifetime is a smooth function of luminosity or 
host mass. We also compare with the directly observed clustering of 
small groups similar to our definition. 

The agreement is quite good at all 
$z\lesssim2$. At higher redshifts, the observations show considerably larger scatter, perhaps 
owing to their no longer being complete near the QLF $\lstar$ -- future observations, sufficiently 
deep to clearly resolve $\lstar$,
are needed to test this in greater detail. We also consider 
the predicted clustering if $\lstar$ quasars were associated with the large group scale of 
$\mstar$ galaxies (for simplicity we take this to be halo masses $\gtrsim5-10$ times larger than 
the small group scale, where 
our halo occupation model predicts of order $\gtrsim3$ satellite $\sim M_{\ast}$ galaxies), 
and the expectation from a secular model, in which quasar clustering 
traces the observed clustering of star-forming galaxies \citep[taken from the observations 
collected in][]{hopkins:clustering}
-- neither agrees with the observations. Note that these estimates may not be 
applicable to the highest-redshift quasar clustering measurements, where flux limits 
allow only the most massive $L\gg L_{\ast}$ systems to be observed 
(but see Figure~\ref{fig:merger.highz} for how the clustering amplitude varies with 
merger masses). 

\begin{figure}
    \centering
    \figexpand
    \plotter{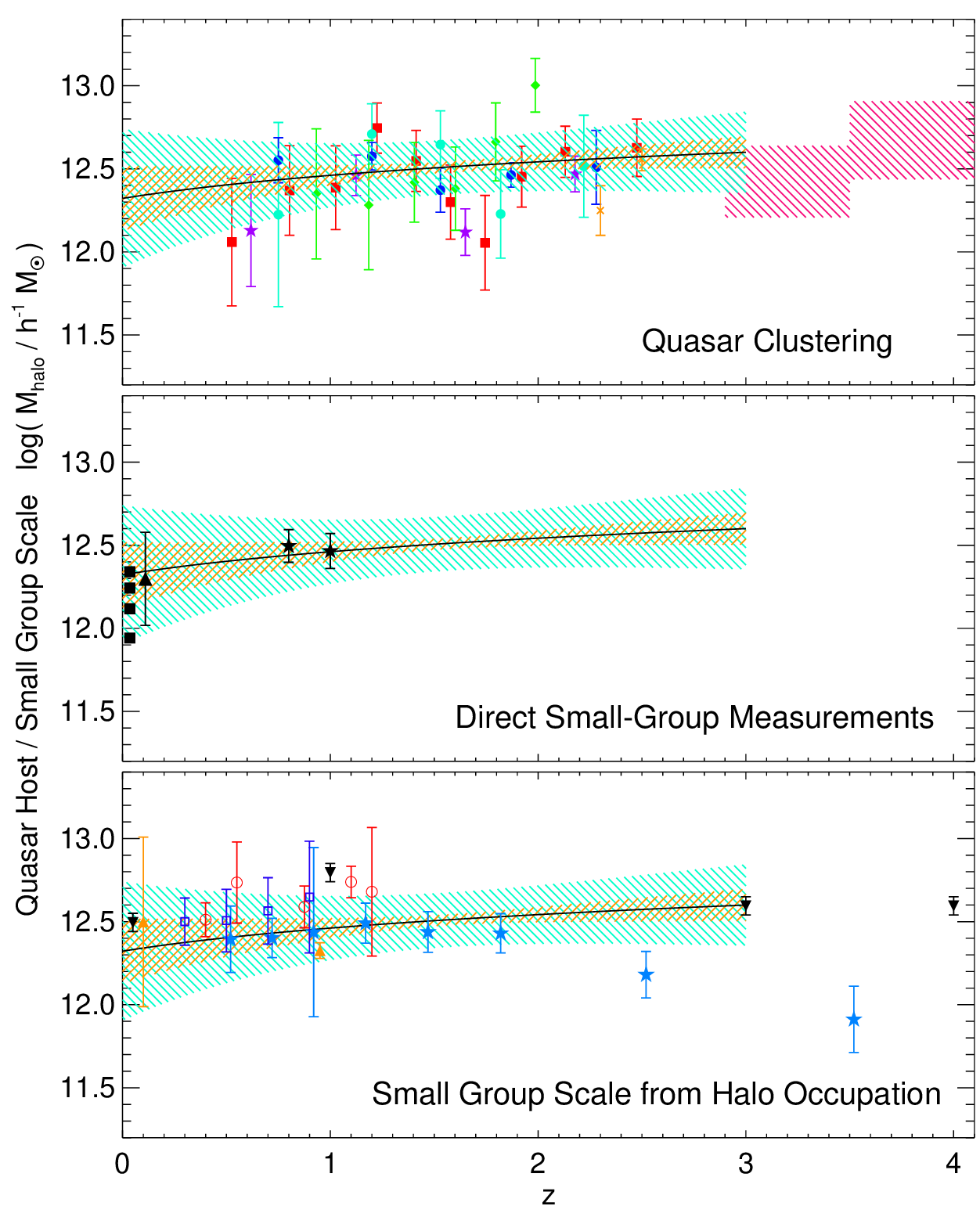}
    \caption{{\em Top:} Characteristic halo mass implied by quasar clustering measurements. 
    Points show the $1\sigma$ allowed range in host halo mass $\mhalo$ corresponding to the quasar
    bias measurements in Figure~\ref{fig:quasar.bias} (in the same style). Shaded magenta 
    regions show the range of halo masses for the corresponding redshift bins in the SDSS 
    \citep{shen:clustering}. The solid line shows 
    the best-fit $\mhalo(z)$ to all observations, with the $1\sigma$ ($2\sigma$) allowed range 
    shaded orange (cyan). {\em Middle:} Shaded range again 
    shows the characteristic host halo mass 
    implied by quasar clustering. Points show the halo mass scale 
    implied by direct measurements of 
    observationally identified small groups (velocity dispersions $\lesssim200\,{\rm km\,s^{-1}}$), 
    from \citet{brough:group.dynamics} at $z\approx0$ (squares), 
    and from from clustering 
    measurements of groups from \citet[][triangles]{eke:groups} 
    and \citet[][stars]{coil:clustering.groups}. 
    {\em Bottom:} Same, but showing the small group halo mass 
    estimated indirectly from the empirically determined halo occupation distribution (HOD). 
    Black inverted triangles adopt the best-fit HOD from \citet{conroy:monotonic.hod} (our 
    default model), other points adopt the 
    methodology of \citet{valeostriker:monotonic.hod} to construct the 
    HOD from various measured 
    galaxy stellar mass functions in 
    \citet[][blue stars]{fontana:highz.mfs}, \citet[][purple squares]{borch:mfs}, 
    \citet[][red circles]{bundy:mfs,bundy:mtrans}, and 
    \citet[][orange triangles]{blanton:lfs}. The characteristic 
    scale of $\sim\lstar$ quasar hosts appears to robustly trace the characteristic small group scale of 
    $\sim\lstar$ galaxies; i.e.\ the mass scale at which galaxy mergers are most efficient. 
    \label{fig:quasar.small.groups}}
\end{figure}
We can invert this, and compare the empirically determined scales of quasar host 
systems with the small group scale which should dominate gas-rich $\sim\lstar$ galaxy mergers. 
Figure~\ref{fig:quasar.small.groups} shows the mean host mass $\mhalo$ which corresponds to 
various quasar clustering measurements (i.e.\ range of $\mhalo$ for which the expected 
quasar bias agrees with the observed $\pm1\,\sigma$ range). We compare this with direct measurements of the halo masses corresponding to small groups of $\sim\mstar$ galaxies, 
determined from both clustering measurements and velocity dispersion measurements of 
observationally identified groups with dispersions $\sigma\lesssim200\,{\rm km\,s^{-1}}$. 
We can also estimate the appropriate small group scale from the halo occupation 
formalism. 

Specifically, following the formalism of \citet{conroy:monotonic.hod}, if galaxy luminosity/mass 
is monotonic with subhalo mass (at the time of subhalo accretion), then we can take any 
galaxy mass function, monotonically rank it and match to our halo+subhalo mass functions, 
and obtain a new halo occupation model which predicts a small group scale -- i.e.\ the range 
of halo masses at which satellites of mass $\sim\lstar$ first appear. As discussed in 
\S~\ref{sec:mergers:criteria}, 
the choice of mass functions and how the HOD is constructed makes little difference 
(factor $<2$) to our predictions, so (unsurprisingly) these all yield a similar estimate of the 
small group scale to our default model predictions.

At all observed redshifts, 
the scale of $\sim\lstar$ quasars appears to trace the small group scale -- i.e.\ whatever 
mechanism triggers $\sim\lstar$ quasars operates preferentially at 
the characteristic small group scale for $\sim\lstar$ galaxies, where mergers are expected 
to be most efficient. 


\begin{figure*}
    \centering
    \figexpand
    \plotone{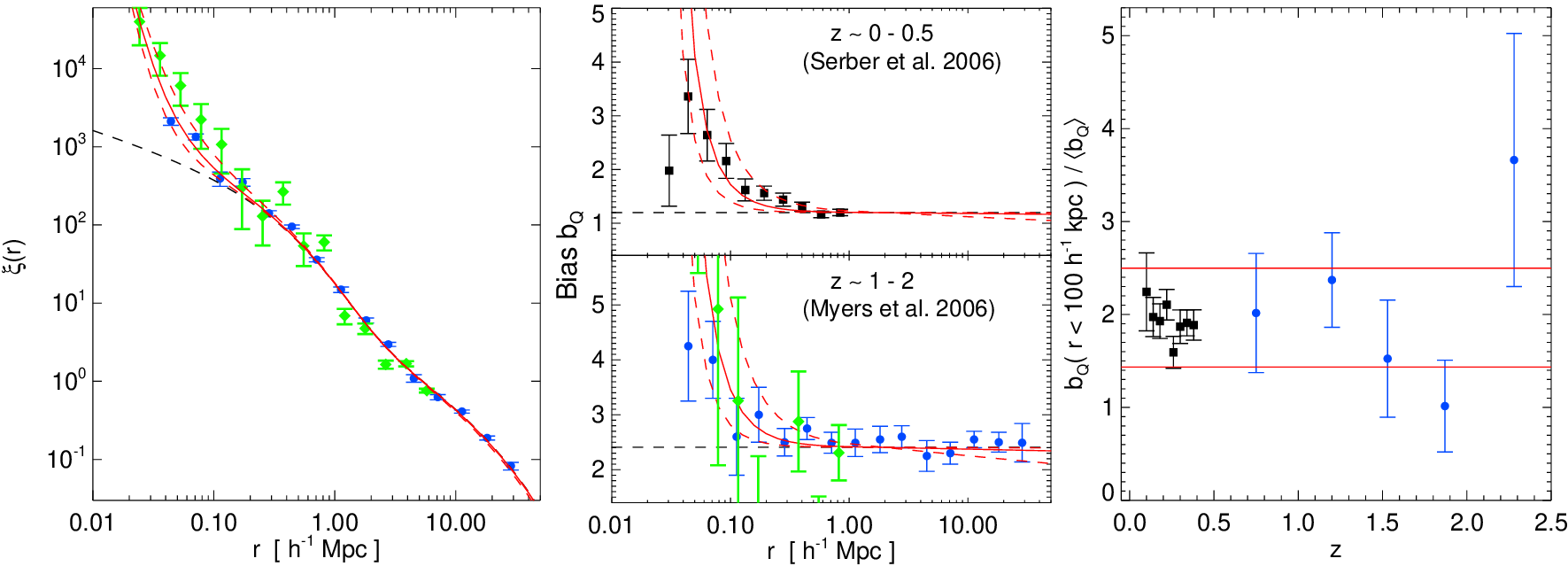}
    \caption{Excess small-scale clustering of quasars expected if they are triggered in 
    mergers, as in Figure~\ref{fig:excess.clustering.mergers}. {\em Left:} Observed 
    correlation functions from \citet[][blue circles]{myers:clustering.smallscale} and 
    \citet[][green diamonds]{hennawi:excess.clustering}, measured for $\sim\lstar$ quasars over the 
    redshift ranges $z\sim1-2$. Dashed black line shows 
    the expected correlation function (nonlinear dark matter clustering from 
    \citet{smith:correlation.function}, multiplied by the appropriate constant large-scale bias factor) 
    without a small-scale excess. 
    Red lines multiply this by the predicted additional bias as a function of scale 
    from \S~\ref{sec:mergers:env}, namely the fact that small-scale overdensities increase 
    the probability of mergers. Solid line shows our mean prediction, dashed 
    the approximate $\sim1\sigma$ range, as in Figure~\ref{fig:excess.clustering.mergers}. 
    {\em Center:} Same, but dividing out the best-fit large-scale correlation 
    function (i.e.\ bias as a function of scale). Black squares in upper panel show the 
    measurement for true optical quasars ($-23.3 > M_{i} > -24.2$) from 
    \citet{serber:qso.small.scale.env} at $z\sim0.1-0.5$. {\em Right:} Ratio of the mean 
    bias at small radii ($r < 100\,h^{-1}\,{\rm kpc}$) to that at large radii (the asymptotic values in 
    the center panel), at all redshifts where this has been observed. 
    Lines show the predicted excess from 
    the previous panels (lower line averages down to a minimum radius 
    $r=50\,h^{-1}\,{\rm kpc}$, upper line to a -- potentially unphysical -- minimum 
    $r=10\,h^{-1}\,{\rm kpc}$). 
    \label{fig:excess.clustering.qso}}
\end{figure*}
In \S~\ref{sec:mergers:env}, we demonstrated that the increased probability of mergers 
in regions with 
excess {\em small scale} overdensities means that the typical merger is more likely to 
exhibit an excess of clustering on small scales, relative to average systems of the same halo mass.  
If quasars are triggered in mergers, this should be true as well. We therefore apply 
the identical methodology from Figure~\ref{fig:excess.clustering.mergers} to calculate the 
excess clustering signal expected in active quasars. Figure~\ref{fig:excess.clustering.qso} shows 
the results of this exercise. We adopt the large-scale mean clustering expected from 
\citet{myers:clustering.smallscale}, specifically using the formulae of \citet{smith:correlation.function} 
to model the expected nonlinear correlation function in the absence of any bias, then apply the 
formalism from Figure~\ref{fig:excess.clustering.mergers} to estimate the additional bias as a 
function of scale. Comparing this to observations, the measurements clearly favor an excess 
bias on small scales \citep[$r\lesssim100-200\,h^{-1}\,{\rm kpc}$;][]{hennawi:excess.clustering}, 
similar to our prediction, 
over a constant bias at all scales. This appears to be true at all observed redshifts; the excess 
relative bias we predict at small scales is simply a consequence of how the probability of 
a merger scales with local density, so it does not vary substantially as a function of redshift. 

It should be noted that the excess of quasar clustering on small scales might 
also reflect an excess of merging binary quasars, i.e.\ merging systems in which 
the interaction has triggered quasars in each merging counterpart. For the 
reasons given in \S~\ref{sec:intro}, this 
situation is expected to be relatively rare (even if all quasars 
are initially triggered by galaxy mergers), but \citet{myers:clustering.smallscale} 
note that only a small fraction of merging pairs need to excite quasar activity in 
both members in order to explain the observed clustering excess. 
Figure~\ref{fig:excess.clustering.qso} demonstrates that a similar excess 
is observed in both the quasar-quasar autocorrelation function 
\citep{hennawi:excess.clustering,myers:clustering.smallscale} 
and the quasar-galaxy cross-correlation function 
\citep{serber:qso.small.scale.env}, 
arguing that it primarily reflects a genuine preference for 
quasar activity in small-scale overdensities. In any case, 
however, the excess on small scales is a general feature of a merger-driven 
model for quasar activity. Indeed, the predicted excess is also seen in 
high-resolution cosmological simulations \citep{thacker:qso.turnover.small.scale.excess}, 
if quasars are specifically identified with (``attached to'') major mergers.
Secular (bar or disk instability) fueling mechanisms, on 
the other hand, should (by definition) show no clustering excess relative 
to median disk galaxies of the same mass and properties, in contrast to what is 
observed (although in agreement with what is seen for low-luminosity Seyfert galaxies, 
see \S~\ref{sec:quasars:secular}).

\subsection{Model-Dependent Predictions: Additional Consequences of Quasar Light Curves}
\label{sec:quasars:qlf}

To proceed further, we must 
adopt some estimates for quasar lightcurves and/or lifetimes.  Following
the methodology developed by \citet{springel:multiphase} and 
\citet{springel:models},
\citet{hopkins:qso.all,hopkins:faint.slope} use a large set of several hundred 
hydrodynamical simulations \citep[see][]{robertson:fp} of galaxy mergers, 
varying the relevant physics, galaxy properties, orbits, and system masses, 
to quantify the quasar lifetime (and related statistics) as a
function of the quasar luminosity. They define the quantity $t_{Q}(L\,|\,M_{\rm BH})$, 
i.e.\ the time a quasar of a given BH mass $\mbh$ (equivalently, peak quasar 
luminosity $L_{\rm peak}$) will be observed at a given luminosity $L$. They 
further demonstrate that this quantity is robust across the wide range of varied 
physics and merger properties; for example, to the extent that the final 
BH mass is the same, any major 
merger of sufficient mass ratio (less than $\sim3:1$) will produce an identical effect. 
We adopt these estimates in what follows, and note that while there is still 
considerable uncertainty in a purely empirical determination of the quasar lifetime, 
the model lightcurves are consistent with the present observational constraints 
from variability studies \citep[][and references therein]{martini04}, clustering 
\citep{croom:clustering,adelbergersteidel:lifetimes,
porciani:clustering,myers:clustering,daangela:clustering,shen:clustering}, 
Eddington ratio measurements \citep{mclure.dunlop:mbh,kollmeier:eddington.ratios}, 
active BH mass functions \citep{vestergaard:mbh,fine:mbh-mhalo.clustering,greene:active.mf}, 
and cosmic background measurements \citep{volonteri:xray.counts,hopkins:bol.qlf}. 

\begin{figure*}
    \centering
    \figexpand
    \plotone{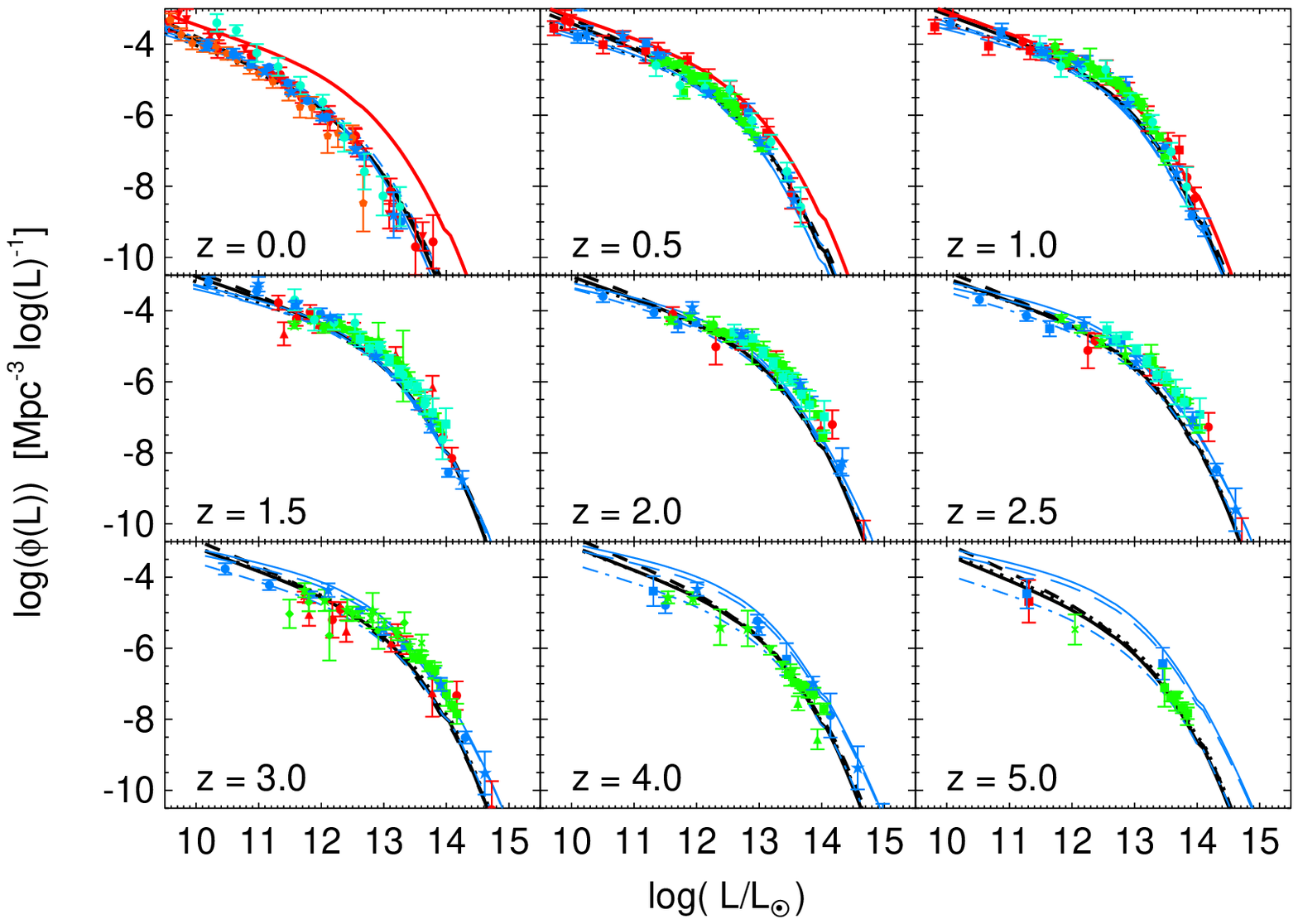}
    \caption{Predicted quasar luminosity functions, convolving 
    our predicted merger rate functions (Figure~\ref{fig:merger.mfs}; same line styles) with 
    quasar lightcurves from simulations \citep{hopkins:qso.all}. Red lines allow 
    dry mergers to trigger quasar activity as well (leading to an overestimate 
    at low redshifts, as in Figure~\ref{fig:lum.density}). Points show observed 
    bolometric luminosity functions at each redshift, from the compilation of 
    observations in \citet{hopkins:bol.qlf}. QLF measurements derived from 
    observations in the optical, soft X-ray, hard X-ray, mid-IR, and 
    narrow emission lines are shown as green, blue, red, cyan, and orange points, 
    respectively. The merger-driven model naturally 
    predicts the observed shape and evolution of the QLF at all redshifts. 
    \label{fig:qlf}}
\end{figure*}
The quasar luminosity function $\phi(L)$ is given by the convolution over the merger rate (rate of 
formation of BHs of final mass $\mbh$ in mergers) and quasar lifetime (differential 
time spent at luminosity $L$ by a BH of final mass $\mbh$):
\begin{equation}
\label{eqn:qlf.convolution}
\phi(L) = \int t_{Q}(L\,|\,\mbh)\,\dot{n}(\mbh\,|\,z) \,{\rm d}\log{\mbh}. 
\label{eqn:qlf}
\end{equation}
Note this technically assumes $t_{Q}\ll \tH$, but this is true for all 
luminosities and redshifts of interest here. Figure~\ref{fig:qlf} shows this 
prediction at a number of redshifts, compared to the large compilation of 
QLF measurements from \citet{hopkins:bol.qlf}. The agreement 
is surprisingly good at all redshifts. At the most extreme luminosities 
$L_{\rm bol} > 3\times10^{14}\,L_{\sun}$ at each redshift, our predictions 
may begin to fall short of the observed QLF, but this somewhat expected, as these 
luminosities naively imply $>10^{10}\,\msun$ BHs accreting at the 
Eddington limit. It is therefore likely that a full resolution at the most extreme 
luminosities involves either revising the estimate of these bolometric 
luminosities (i.e.\ the bolometric corrections adopted may not be appropriate 
for the most extreme objects, or there may be beaming effects) or 
including processes beyond the scope of our current investigation (e.g.\ 
super-Eddington accretion or multiple mergers in massive BCGs). Nevertheless, 
our simple merger-driven scenario appears to accurately predict the distribution 
and evolution of most quasar activity. 

\begin{figure}
    \centering
    \figexpand
    \plotter{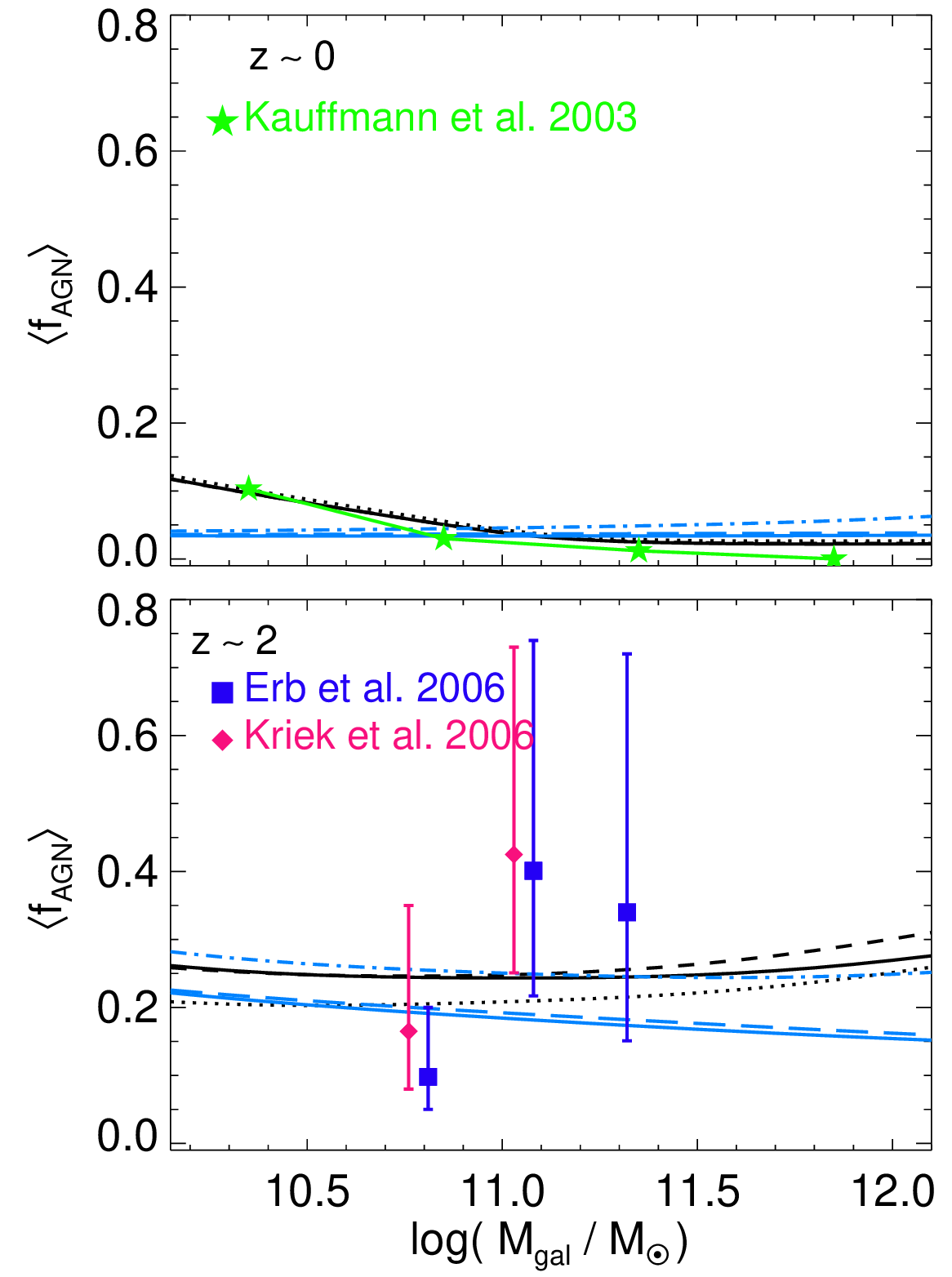}
    \caption{Predicted AGN fraction as a function of host properties. 
    {\em Top:} Low-redshift 
    quasar fraction (defined here by Eddington ratios $\dot{m}>0.1$) 
    as a function of galaxy mass. Black lines show 
    the prediction of our merger-driven model, in the style of 
    Figure~\ref{fig:merger.mfs}. Observed fractions are shown down to 
    (roughly) their completeness limit, from \citet{kauffmann:qso.hosts}.
    {\em Bottom:} Same, but at $z\approx2$, with the AGN fraction determined observationally in 
    LBG \citep{erb:lbg.masses} and $K$-selected \citep{kriek:qso.frac} samples. 
    Some caution should be applied at $\mgal\lesssim10^{10}\,\msun$, as 
    the AGN luminosities become sufficiently low that even moderate star formation 
    will dominate the observed luminosity and systems may not be classified as AGN. 
    \label{fig:active.fraction}}
\end{figure}
Integrating the QLF over the appropriate range, we trivially obtain the active fraction, and 
can calculate this separately for each host mass $\mgal$ or BH mass $\mbh$ in 
Equation~(\ref{eqn:qlf}).
Figure~\ref{fig:active.fraction} compares this to 
observations at both low and high redshift, for 
systems with $\dot{m}\equiv L/L_{\rm Edd} > 0.1$, 
representative of typical Seyfert and quasar populations 
\citep[e.g.][]{mclure.dunlop:mbh}. Note that the quasar lifetime 
integrated above this threshold is close to a constant value
$\lesssim10^{8}$\,yr, similar to observational estimates \citep{martini04}.
At very low masses/levels of activity, other fueling 
mechanisms may be dominant -- for comparison
with e.g.\ the active fractions in \citet{hao:local.lf} of typical $\lesssim10^{7}\,\msun$ BHs 
($\lesssim10^{10}\,\msun$ hosts), we refer to secular and/or ``stochastic'' accretion 
models in disks \citep[e.g.][]{hopkins:seyferts} and old ellipticals \citep{martini:ell.center.dust}.
Furthermore, at the lowest masses plotted, the typical AGN luminosities 
become extremely faint (typical $M_{B}\gtrsim-18$ in $\mgal\lesssim10^{10}\,\msun$ hosts), 
and so such systems may be more often classified as non-AGN or typical star-forming systems
\citep[e.g.][]{rodighiero:obscured.agn}. 
At high levels of accretion, however, the merger-driven prediction agrees well with 
observations at low and high redshift, and predicts a downsizing trend similar to that 
seen -- namely that from $z=2$ to $z=0$, quasar activity has been particularly suppressed 
in the most massive systems (although it has been suppressed to some extent at all host 
masses), presumably owing to the conversion of these systems to ``red and dead'' spheroids 
without cold gas supplies (see \papertwo).

\begin{figure*}
    \centering
    \figexpand
    \plotone{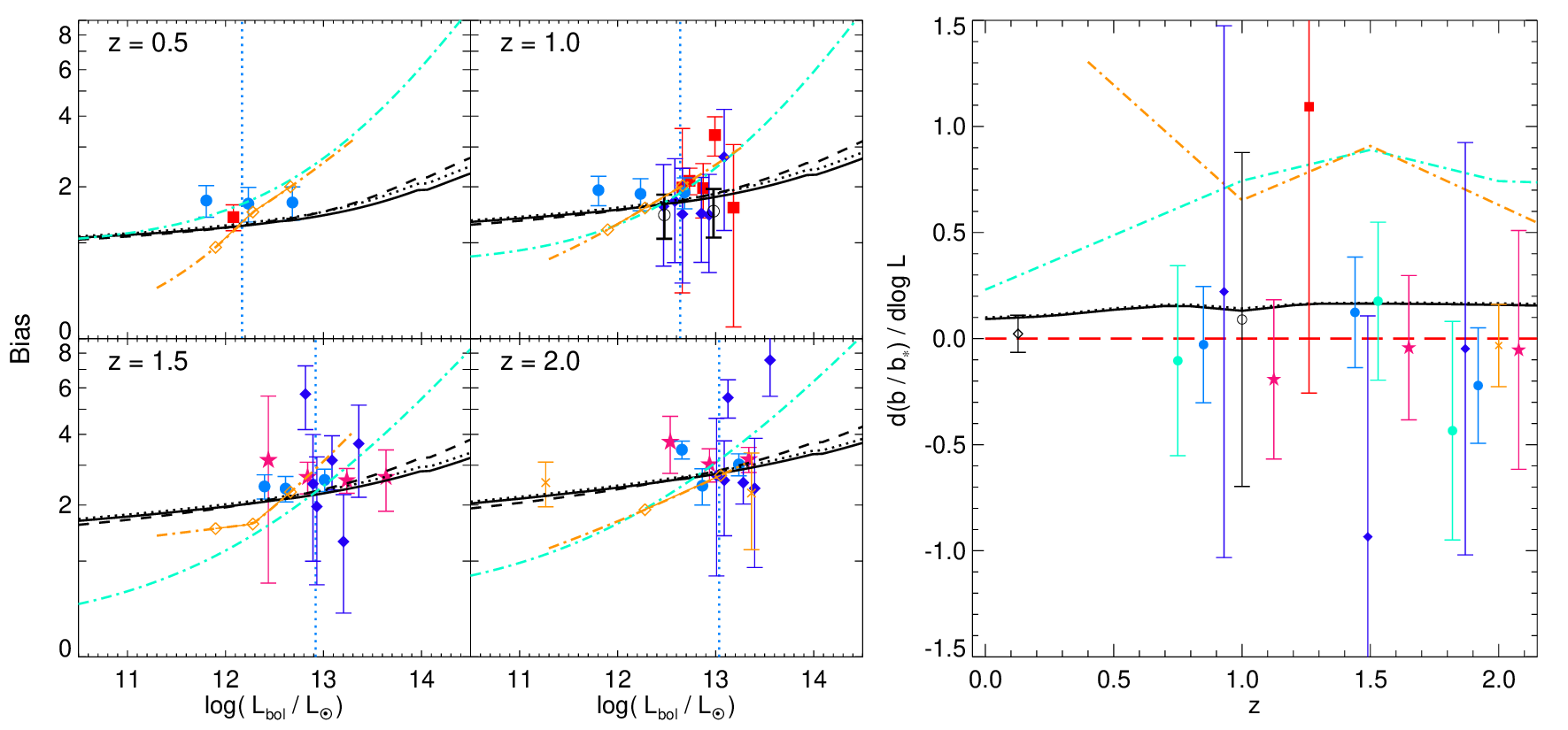}
    \caption{{\em Left:} Predicted bias as a function of quasar luminosity 
    from our merger-driven model (black lines, style as in Figure~\ref{fig:merger.mfs}). 
    To contrast, the expected bias $b(L)$ from the semi-analytic models 
    of \citet[][cyan]{wyitheloeb:sam} and \citet[][orange with diamonds]{kh00} are 
    plotted (dot-dashed lines); these adopt simplified (constant or exponential 
    ``on/off'') quasar lightcurves. Points are measurements from 
    \citet[][red squares]{croom:clustering}, \citet[][orange crosses]{adelbergersteidel:lifetimes}, 
    \citet[][purple diamonds]{porciani:clustering},  
    \citet[][blue circles]{myers:clustering},  
    \citet[][magenta stars]{daangela:clustering}, and 
    \citet[][black open circles]{coil:agn.clustering}. For ease of comparison, all luminosities are 
    converted to bolometric luminosities using the corrections from \citet{hopkins:bol.qlf}. 
    Vertical blue dotted lines show $\lstar$ in 
    the QLF at each redshift, from \citet{hopkins:bol.qlf}.
    {\em Right:} 
    The best-fit slope of the dependence of bias on luminosity at the QLF $\lstar$, i.e.\ 
    ${\rm d}(b/b_{\ast}) / {\rm d}\log{(L/L_{\ast})}$, where $b_{\ast}\equiv b(L_{\ast})$. 
    Points are determined from the observations at left, with the observations 
    from \citet[][cyan circles]{myers:clustering} and 
    \citet[][black open diamond]{grazian:local.qso.clustering,wake:local.qso.clustering}
    added. Lines are in the style of the left panel, with the red dashed line showing 
    no dependence of bias on luminosity. 
    Adopting an a priori model for merger-triggered quasar activity reproduces the 
    empirical prediction from \citet{lidz:clustering}, that quasar bias should depend 
    weakly on quasar luminosity. 
    \label{fig:bias.vs.l}}
\end{figure*}
We next follow \citet{lidz:clustering}, and extend 
Equation~(\ref{eqn:qlf.convolution}) to convolve over the expected 
bias of the active systems at each quasar luminosity $L$, 
\begin{equation}
b(L) = \frac{1}{\phi(L)}\,\int b(\mbh)\,t_{Q}(L\,|\,\mbh)\,\dot{n}(\mbh\,|\,z) \,{\rm d}\log{\mbh},
\end{equation}
where $b(\mbh)$ is determined just as $b(\mgal)$ in 
\S~\ref{sec:mergers:populations}, by convolving 
over the contributions to each merging range in $\mbh$ from all $\mhalo$. 
Figure~\ref{fig:bias.vs.l} plots the expected bias as a function of luminosity 
at each of several redshifts. As originally demonstrated in \citet{lidz:clustering}, 
our model for quasar lightcurves and the underlying triggering rate of quasars 
predicts a relatively weak dependence of clustering on quasar luminosity. 
Here, we essentially re-derive this result with an {\em a priori} prediction of 
these triggering rates, as opposed to the purely empirical (fitted to the QLF) rates 
from \citet{lidz:clustering}, and find that the conclusion is robust. However, this 
prediction is not necessarily a consequence of merger-driven models 
(nor is it unique to them) -- 
we show the predictions from the semi-analytic models of \citet{wyitheloeb:sam} and 
\citet{kh00}, who adopt simplified ``lightbulb''-like quasar lightcurves 
\citep[for a detailed discussion of these differences, see][]{hopkins:clustering}. 

The reason for the weak dependence of quasar clustering on luminosity in 
Figure~\ref{fig:bias.vs.l} is, in fact, the nature of the quasar lightcurve. Quasars grow 
rapidly in mergers to a peak quasar phase at the final stages of the merger, 
which exhausts and expels the remaining gas, after which the quasar 
decays to lower luminosities. This decay moves objects of the same host 
properties to fainter luminosities in the QLF, making the clustering properties 
flat as a function of luminosity. Thus, while an important test of our modeling 
(that the correct halos and galaxies host quasars of the appropriate luminosities), 
this is not a unique prediction of merger-driven models.

We can also use our model to estimate the infrared luminosity 
functions of various populations versus redshift. By construction, our 
assumed halo occupation model reproduces the observed star-forming (blue) 
galaxy mass function at each redshift. Using the corresponding fitted star-formation 
histories as a function of baryonic mass from \citet{noeske:sfh} (which fit the observations locally 
and their evolution at least to $z\sim1.5$), we immediately obtain an 
estimate of the star formation rate function in ``quiescent'' (non-merging) galaxies 
at each redshift. We include a scatter of $\sim0.25\,$dex in SFR at fixed 
stellar mass, comparable to that observed (in blue galaxies), but this makes relatively 
little difference, as the most extreme SFR populations are dominated by mergers. 
We then adopt the standard conversion from \citet{kennicutt98} to 
transform this to an infrared luminosity function (where we refer to the total IR 
$8-1000\,\mu{\rm m}$ luminosity). 

Our model also yields the mass function of 
gas-rich mergers, for which we can estimate their distribution of star formation rates. 
In \citet{hopkins:merger.lfs}, we quantify the distribution of star formation rates as a 
function of galaxy properties from the same large suite of simulations 
used to estimate the quasar lifetime. Essentially, this quantifies the ``lifetime'' above a 
given SFR in a merger, which can be reasonably approximated as a simple function of 
galaxy mass and (pre-merger) gas fraction, 
\begin{equation}
t(>\dot{M}_{\ast}) = t_{\ast}\,\Gamma{\Bigl(}0,\,\frac{\dot{M}_{\ast}}{M_{f}\,\fgas\,/t_{\ast}}{\Bigr)}, 
\label{eqn:t.sfr}
\end{equation}
where $M_{f}$ is the post-merger galaxy mass (i.e.\ our $\mgal$) and $t_{\ast}\approx0.3\,$Gyr 
is a fitted characteristic time. This functional form simply amounts to the statement that there is a 
mean characteristic timescale $t_{\ast}$ in which most of the gas mass of the merger 
($M_{f}\,\fgas$) is converted into stars, which we find is (unsurprisingly) similar to the 
dynamical time of the merger and to observational estimates of the 
characteristic star formation timescale in starbursts and ULIRGs \citep{kennicutt98}. 
Since the fitted star-formation histories of \citet{noeske:sfh} implicitly define a 
gas fraction as a function of time (or can be used in combination with the 
Schmidt-Kennicutt star formation law to infer the gas fraction), we simply adopt these 
for the pre-merger galaxies \citep[but we have checked 
that they correctly reproduce observed gas fractions as a function of mass 
at $z=0,\,1,\,2$; see][]{hopkins:bhfp}. It is worth noting that, with this estimate, 
the explicit dependence on $\fgas$ 
can be completely factored out in Equation~(\ref{eqn:t.sfr}), and we can write it as 
an estimate of the amount of time a system spends above a given enhancement in 
SFR (basically a merger enhances the $\tau$-model SFR by $\sim \tau/t_{\ast}$), 
relative to the pre-merger SFR.
Using the same SFR to $L_{\rm IR}$ conversion, we obtain a rough estimate of the 
IR luminosity function of mergers. 

Finally, adopting the empirically calculated obscured fraction as a function of 
quasar luminosity from \citet{gilli:obscured.fractions}, and assuming that 
the obscured bolometric luminosity is re-radiated in the IR, we convert 
our predicted bolometric QLF to an IR QLF of obscured quasars. 
Technically, not all of the luminosity will be obscured, of course, 
but we find that e.g.\ using the full distribution of column densities as a function 
of quasar luminosity from \citet{ueda03:qlf} to attenuate a template AGN SED yields 
a very similar answer \citep[see also][]{franceschini:faint.xr.qsos}, as does using a mean 
X-ray to IR bolometric correction of obscured AGN \citep{elvis:atlas,
zakamska:multiwavelength.type.2.quasars,polletta:obscured.qsos}. 
Including the IR contribution from un-obscured quasars 
is a negligible correction.   

\begin{figure*}
    \centering
    \figexpand
    \plotone{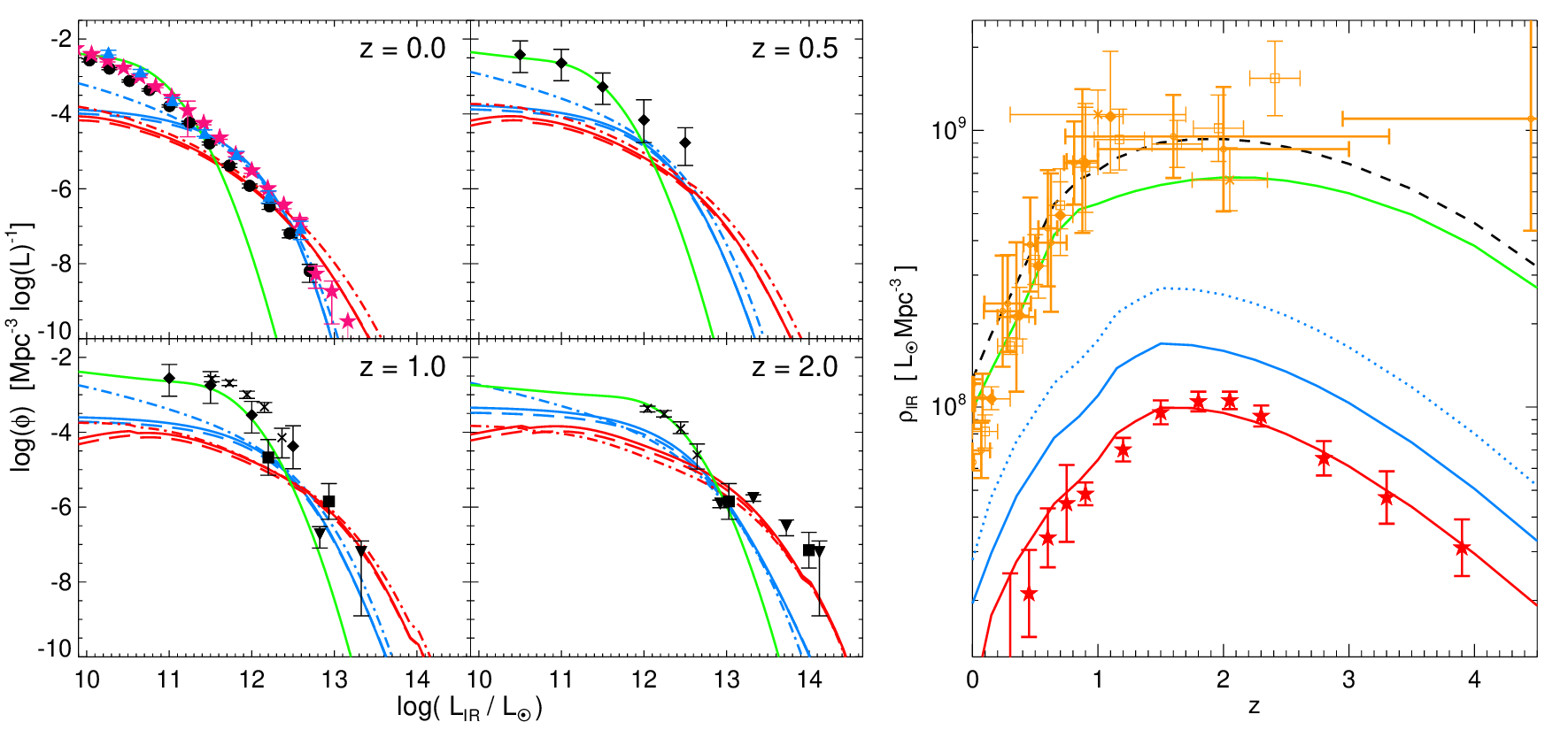}
    \caption{{\em Left:} Predicted total IR ($8-1000\,\mu{\rm m}$) luminosity functions at different 
    redshifts (as labeled). Green, blue, and red lines shows the estimated contribution from 
    non-merging systems, star formation in mergers, and obscured AGN in mergers, respectively. 
    Linestyles are as in Figure~\ref{fig:merger.mfs}, for the variants of the merger calculations.    
    Points show observational estimates from 
    \citet[][magenta stars]{saunders:ir.lfs}, 
    \citet[][blue triangles]{soifer:60m.lfs}, \citet[][black circles]{yun:60m.lfs}, 
    \citet[][black diamonds]{lefloch:ir.lfs}, \citet[][black inverted triangles]{chapman:submm.lfs}, 
    \citet[][black squares]{babbedge:swire.lfs}, 
    and \citet[][black $\times$'s]{caputi:ir.lfs}. 
    {\em Right:} Integrated IR luminosity density. Solid lines show the contributions from 
    non-merging systems (green), star formation in mergers (blue), and obscured 
    quasars in mergers (red). Blue dotted shows the total (star formation+AGN) 
    merger contribution, black dashed shows the total from all sources. Orange points 
    show observational estimates of $\rho_{\rm IR}$ from the compilation of 
    \citet[][circles; only the direct IR observations therein are plotted here]{hopkins:sfh}, 
    as well as \citet[][diamonds]{lefloch:ir.lfs}, \citet[][]{perezgonzalez:ir.lfs}, 
    and \citet[][$\times$'s]{caputi:ir.lfs}.
    Red stars show the bolometric quasar luminosity density from Figure~\ref{fig:lum.density}, 
    rescaled by a constant (mean) obscured-to-unobscured ratio of $\sim2:1$. 
    The agreement in all cases is good -- our model reproduces the star formation 
    history of the Universe and distribution of star formation rates and bolometric luminosities. 
    \label{fig:ir.lfs}}
\end{figure*}

Figure~\ref{fig:ir.lfs} compares the resulting predicted IR luminosity functions to 
observations at $z=0-2$, and to the observed IR luminosity density from $z\sim0-5$. 
At all redshifts, the agreement is good, which suggests that our model accurately 
describes the star-formation history of the Universe. This should be guaranteed, since 
at all redshifts the quiescent population dominates the $\sim L_{\ast}$ optical 
and IR luminosity functions (hence also the star formation rate and IR luminosity 
densities) -- at this level, we simply confirm that our halo occupation model is a good 
approximation. However, at high luminosities, typical of ULIRGs, the populations 
are generally dominated by mergers and (at the highest luminosities) obscured 
AGN. 

\begin{figure}
    \centering
    \figexpand
    \plotone{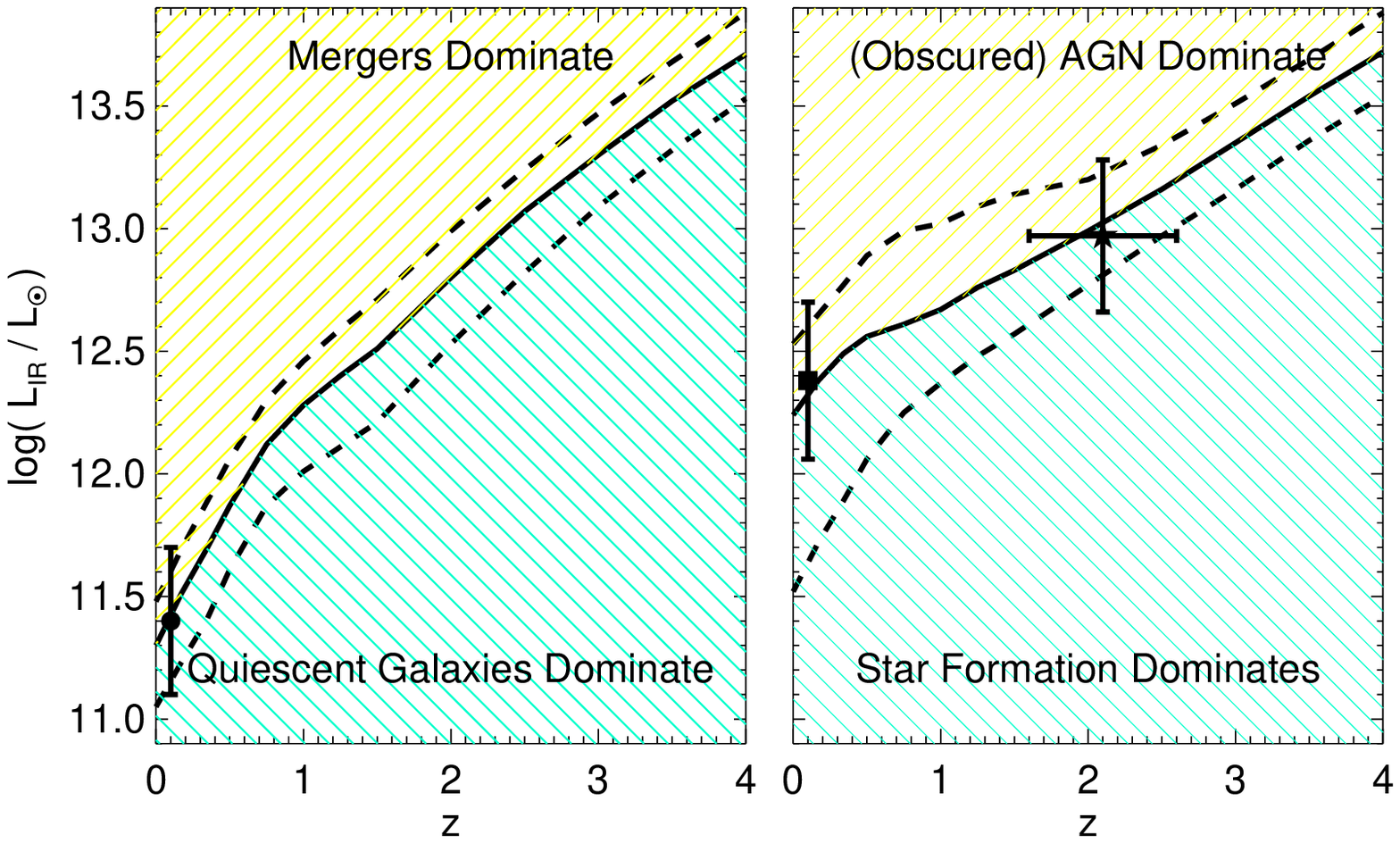}
    \caption{{\em Left:} Total IR luminosity, as a function of redshift, above which 
    mergers (star formation+AGN) dominate the total IR luminosity functions 
    (solid line, from Figure~\ref{fig:ir.lfs}; dashed lines show the range above which 
    $25/75\%$ of systems on the luminosity function are mergers). Point shows 
    the corresponding transition point (and range) observed in low-redshift systems 
    \citep{sanders:review}. {\em Right:} Same, but for the transition between star formation 
    (in non-merging+merging systems) and (obscured) AGN dominating the IR luminosity 
    functions (generally a factor $\sim{\rm a\ few}$ larger luminosity than the 
    quiescent system-merger transition). 
    Points show the observed estimates from comparison of PAH feature strengths in 
    \citet[][low redshift]{lutz:pah.qso.vs.sf.local} and \citet[][high redshift]{sajina:pah.qso.vs.sf}. 
    A similar estimate is obtained (at low redshift)
    from comparison of emission line strengths 
    \citep{sanders96:ulirgs.mergers,kewley:in.prep}, full SED template fitting 
    \citep{farrah:qso.vs.sf.sed.fitting}, or indirect comparison with Type 2 AGN luminosity 
    functions \citep{chary.elbaz:ir.lfs}.
    The model predicts the local transitions, and that by $z\gtrsim1$, the LIRG population 
    is dominated by quiescent star formation in gas-rich systems (even as the 
    total and fractional luminosity density in mergers increases rapidly). 
    \label{fig:ir.dom}}
\end{figure}

We explicitly quantify the transition point as a function of redshift in Figure~\ref{fig:ir.dom} 
(we show the comparison there just for our ``default'' model, but as is clear in Figure~\ref{fig:ir.lfs}, 
the transition between different populations dominating the LF is similar regardless of the 
exact version of our model adopted). Our comparisons generally affirm 
the conventional wisdom: at low redshift, mergers dominate the ULIRG and 
much of the LIRG populations, above a luminosity $\sim10^{11.4}\,L_{\sun}$, 
with heavily obscured (potentially Compton-thick) 
AGN (in starburst nuclei) becoming a substantial contributor to IR luminous populations 
in the most extreme $\gtrsim{\rm a\ few\ }\times10^{12}\,L_{\sun}$ systems 
(nearing hyper-LIRG $>10^{13}\,L_{\sun}$ luminosities which are common bolometric 
luminosities for $>10^{8}\,\msun$ BHs near Eddington, but would imply 
potentially unphysical $\gtrsim1000\,\msun\,{\rm yr^{-1}}$ SFRs). 
At higher redshifts, disks are more gas-rich, and thus have characteristically 
larger star formation rates, dominating the IR LFs at higher luminosities. By 
$z\sim1$, most LIRGs are quiescent systems, and by $z\sim2$, only extreme 
systems $\gtrsim{\rm a\ few\ }\times10^{12}\,L_{\sun}$ are predominantly 
mergers/AGN.

This appears to agree well with recent 
estimates of the transition between AGN and passive star formation 
dominating the bolometric luminosities of high-redshift systems. 
Interestingly, 
this shift occurs even while increasing merger rates (and higher 
gas fractions in typical mergers) lead to a larger overall contribution of 
mergers to the star formation rate and IR luminosity densities. At $z\sim0$, 
mergers contribute negligibly to the total IR luminosity density, but 
by $z\sim2$, they may contribute $\sim20-50\%$ of the IR output of the Universe, 
with that contribution owing comparably to both star formation in mergers and 
obscured BH growth \citep[which should be true, given the $\mbh-M_{\rm host}$ 
correlations and typical $\epsilon_{r}\sim0.1$ radiative efficiencies;
see, e.g.][]{lidz:proximity}. 

The integrated contribution of mergers to the star formation rate and IR luminosity 
densities agrees well with observational estimates 
\citep[available at $z\lesssim2$; see][]{bell:morphology.vs.sfr,menantau:morphology.vs.sfr}, 
and the constraint from stellar population models that only a small fraction of the 
$z=0$ stellar mass in typical early-type galaxies was formed in the 
spheroid-forming merger itself \citep[as opposed to more extended star formation in 
the pre-merger disks; e.g.][]{noeske:sfh}. For a more detailed comparison and analysis of the 
merger-induced contribution to the star formation rate density of the Universe, 
we refer to \citet{hopkins:merger.lfs}. 

We caution that the above comparisons are approximate, and intended as a broad 
check that our models are consistent with the observed abundance of 
IR luminous galaxies as a function of redshift. We have ignored a number of 
potentially important effects: for example, obscuration is a strong function of time 
in a merger, and may affect various luminosities and morphological stages 
differently. Moreover, our simple linear addition of the star formation contribution 
of mergers to the IR LF and the AGN contribution is only technically correct 
if one or the other dominates the IR luminosity at a given time in the merger; however, 
there are clearly times during the final merger stages when the contributions 
are comparable. Resolving these issues requires detailed, time-dependent 
radiative transfer solutions through high-resolution simulations that properly 
sample the merger and quiescent galaxy parameter space at each redshift, 
and is outside the scope of this work \citep[although an important subject for future, 
more detailed study; see, e.g.][]{li:radiative.transfer}.
It would be a mistake, therefore, to read too much into 
e.g.\ the detailed predictions for sub-millimeter galaxies or other extreme 
populations based on Figures~\ref{fig:ir.lfs} \&\ \ref{fig:ir.dom}. However, most of our 
predicted qualitative trends, including the evolution of the luminosity density 
(and approximate relative contribution of mergers) and the shift in where 
quiescent or merger-driven populations dominate the bright IR LF, should 
be robust. Critically, a model in which merger-driven quasar activity dominates 
the QLF predicts an abundance of IR-luminous galaxies consistent with 
the observations as a function of both luminosity and redshift.

\subsection{When Merger-Triggering Loses to Secular Processes}
\label{sec:quasars:secular}

Despite these arguments for a merger-driven origin for bright, high-redshift 
quasars, there are good reasons to believe that most local, high-Eddington ratio 
objects are {\em not} related to mergers. Most active local systems 
typically involve relatively low-mass 
BHs \citep[$\mbh\sim10^{7}\,\msun$;][]{heckman:local.mbh}, 
in Sa/b-type host galaxies, 
without significant evidence for recent major interactions 
\citep{kauffmann:qso.hosts,pierce:morphologies}, and 
have relatively low Seyfert-level luminosities 
\citep[$-21\gtrsim M_{B} \gtrsim -23$;][]{hao:local.lf}, below 
the traditional $M_{B}=-23$ Seyfert-quasar divide. Given this, it is natural to ask 
whether there are additional reasons to believe that bright quasars have 
distinct origins, and if so, when (or at what luminosities) these non-merger 
driven fueling mechanisms begin to dominate AGN populations. 

\begin{figure}
    \centering
    \figexpand
    \plotone{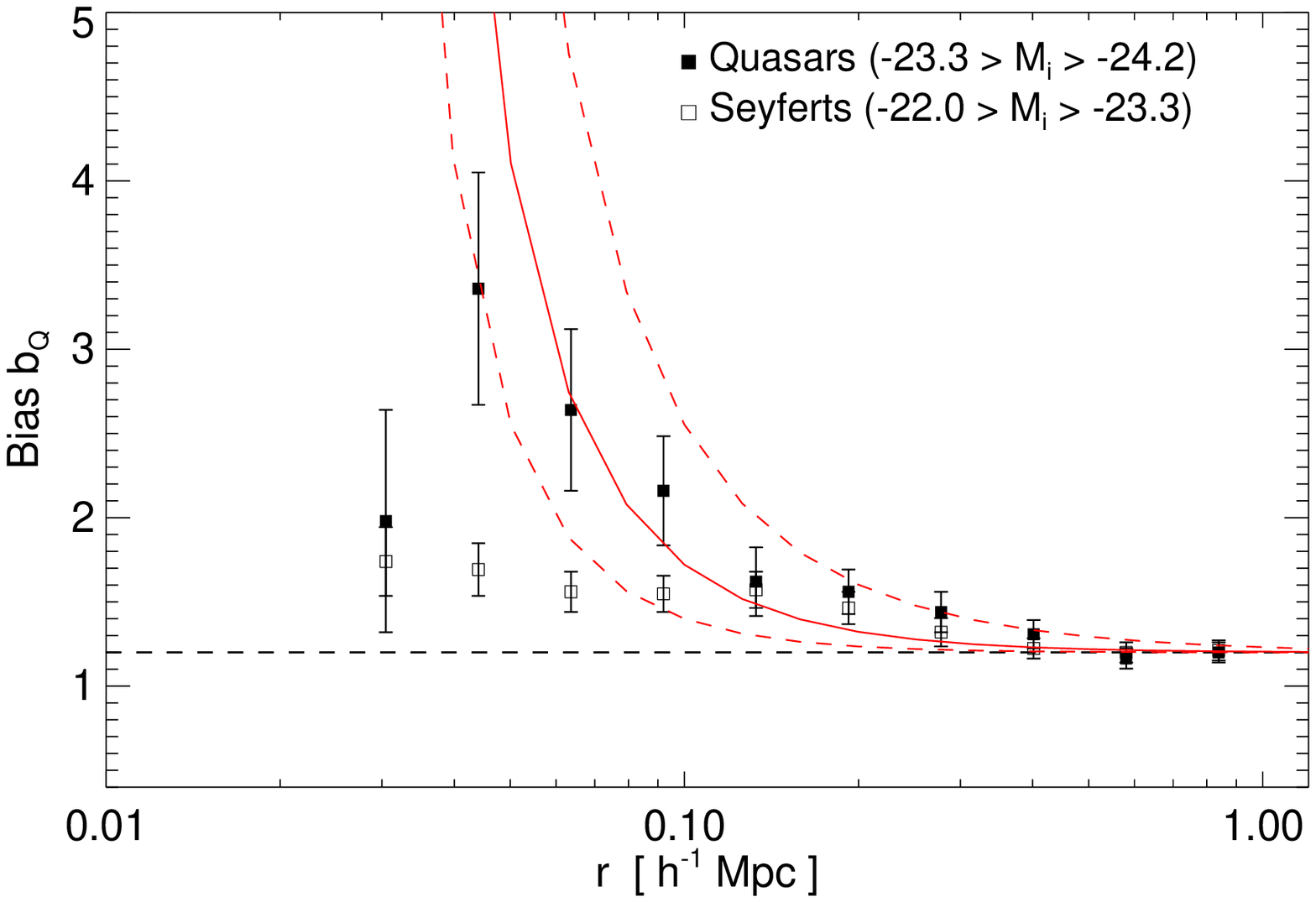}
    \caption{As Figure~\ref{fig:excess.clustering.qso} (upper center panel), but comparing 
    the clustering (quasar-galaxy cross-correlation) 
    as a function of scale measured by \citet{serber:qso.small.scale.env} 
    for bright optical quasars and dimmer Seyfert galaxies. Quasar clustering is consistent with our 
    predicted excess on small scales, indicating a merger-driven origin, but low-luminosity 
    systems show no such dependence, suggesting that processes independent of the 
    local, small-scale density (e.g.\ secular processes) may dominate at these luminosities. 
    \label{fig:excess.clustering.seyferts}}
\end{figure}
In addition to the arguments in \S~\ref{sec:quasars:mergers} \& \ref{sec:quasars:qlf}, 
there are a number of qualitative differences between bright, high-redshift quasars 
and local Seyferts. Quasars have significantly different clustering amplitudes 
\citep{hopkins:clustering} and host stellar mass distributions \citep{hopkins:transition.mass} 
from star-forming galaxies at $z\gtrsim1$, and typically have hosts 
with elliptical or merger remnant morphologies \citep{floyd:qso.hosts,falomo:qso.hosts,
zakamska:qso.hosts,letawe:qso.merger.ionization}, frequently exhibiting 
evidence of tidal disturbances \citep{bahcall:qso.hosts,
canalizostockton01:postsb.qso.mergers,
hutchings:redqso.lowz,hutchings:redqso.midz,
urrutia:qso.hosts,bennert:qso.hosts}. Figure~\ref{fig:excess.clustering.seyferts} 
compares the clustering as a function of scale measured in 
\citet{serber:qso.small.scale.env} for both bright quasars and Seyfert galaxies -- 
quasars exhibit the strong trend of excess clustering on small scales indicative of 
a triggering process which prefers small-scale overdensities, but Seyferts 
show no significant preference for local overdensities. 

\begin{figure}
    \centering
    \figexpand
    \plotter{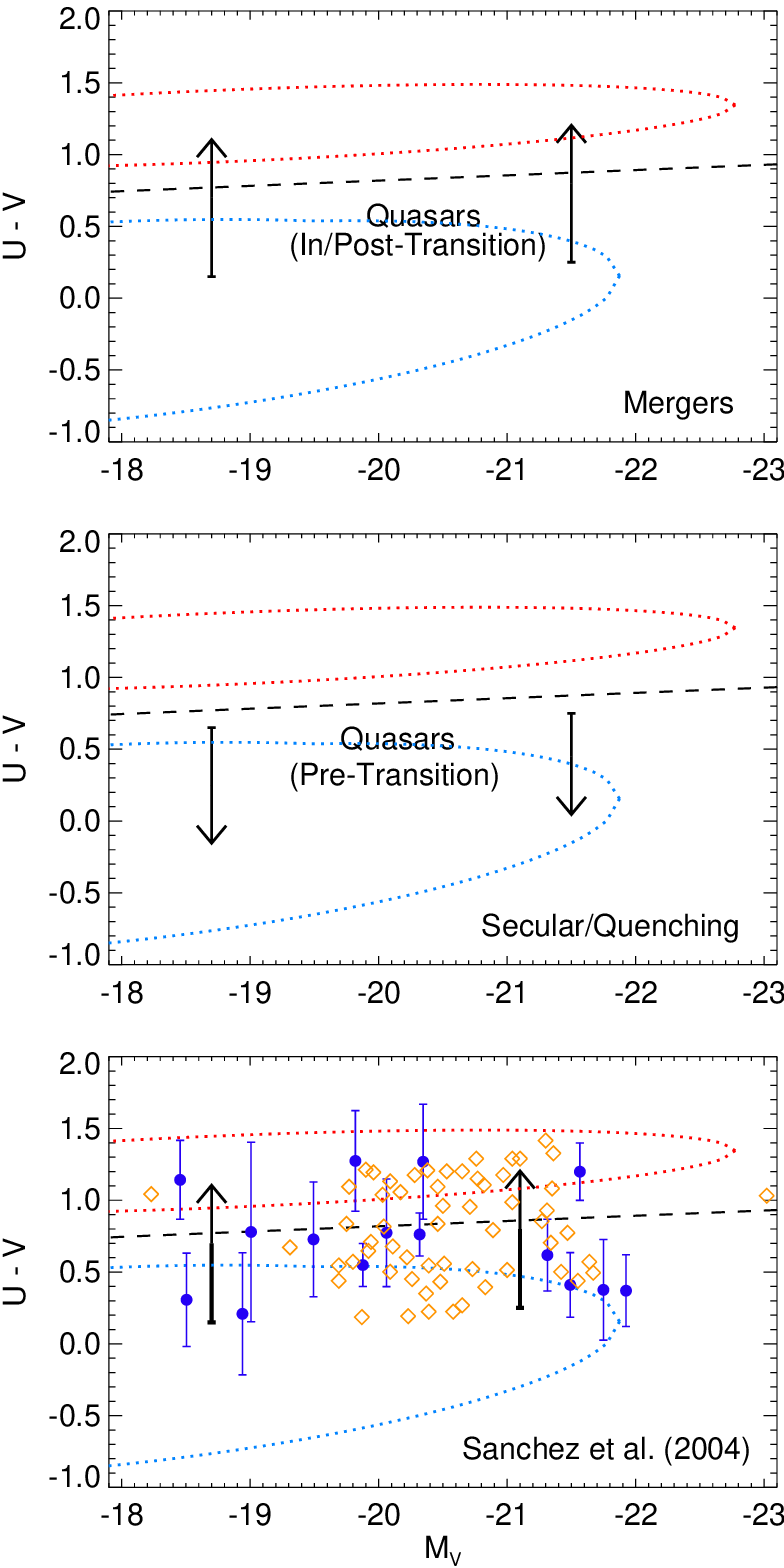}
    \caption{Location of quasars in the color-magnitude diagram, expected 
    from different models. {\em Top:} Red and blue dotted regions roughly outline the 
    red sequence and blue cloud, respectively, with the dashed line dividing the 
    bimodality \citep[from][]{bell:combo17.lfs}. Arrows show the preferred location of 
    quasar hosts in a merger driven model. At the end of a merger, a bright 
    quasar is triggered in a spheroid/merger remnant at the top of the blue cloud 
    (owing to the young stellar populations from pre-merger and merger-induced 
    star formation), and subsequently the quasar luminosity decays while the remnant 
    rapidly reddens, leaving a relatively low accretion rate remnant on the red sequence. 
    {\em Middle:} Same, but for a secular triggering scenario in which quasar 
    activity (which must still require cold gas) is uncorrelated with quenching or itself 
    exhausts the gas supply. In this case, quasars should live in the blue cloud, with 
    gas-rich systems, and their abundance rapidly drops approaching the ``green valley'' 
    as gas supplies are exhausted. {\em Bottom:} We compare to 
    observations of quasar host galaxy 
    colors at $z\sim0.7-1.1$ from \citet[][blue circles]{sanchez:qso.host.colors}. X-ray identified 
    AGN and quasar 
    hosts from \citet[][orange diamonds]{nandra:qso.host.colors} are also shown 
    (the numbers plotted should not be taken literally, as we have rescaled the authors
    $U-B$ vs.\ $M_{B}$ color-magnitude relation to that shown here for the 
    sake of direct comparison, but the result is qualitatively identical to that shown). 
    Arrows reproduce the merger expectation from the top panel. Quasars appear to 
    live in the region of color-magnitude space expected if they are triggered at the 
    {\em termination} of star formation, and subsequently decay in luminosity, as 
    expected in merger-driven scenarios. 
    \label{fig:qso.cmd}}
\end{figure}

Because galaxy mergers are also associated 
with the termination of star formation in the remnant (even if only 
temporarily), i.e.\ a rapid post-starburst phase and transition to the 
red sequence (discussed in detail in \papertwo), 
the decay of the quasar lightcurve should be associated with the 
reddening of the remnant, in a merger-driven model. 
This implies a particular preferred track for 
quasar hosts in the color-magnitude diagram, illustrated in Figure~\ref{fig:qso.cmd}. 
In this scenario, quasars should be associated with the crossing of the 
``green valley'' -- i.e.\ the triggering of a quasar occurs at the end of the merger, 
when young stellar populations imply a bluer-than-average host spheroid, and 
the quasar decays to lower luminosities as the remnant reddens onto the red sequence. 

Alternatively, if quasars were triggered in a purely secular manner, or otherwise independent 
of whatever quenching mechanism terminates the galactic supply of cold gas, 
then their natural preferred location is in the blue cloud -- i.e.\ blueward of the 
``green valley.'' Systems in this regime still have cold gas supplies and have not yet 
quenched. Because the quenching is uncorrelated with quasar triggering 
in such a model, and 
the lack of galaxies in the ``green valley'' implies that
this transition is rapid, very few quasars 
would be expected to be triggered just as the quenching occurs, and therefore 
few quasars should be present in the ``green valley.'' 

\begin{figure*}
    \centering
    \plotone{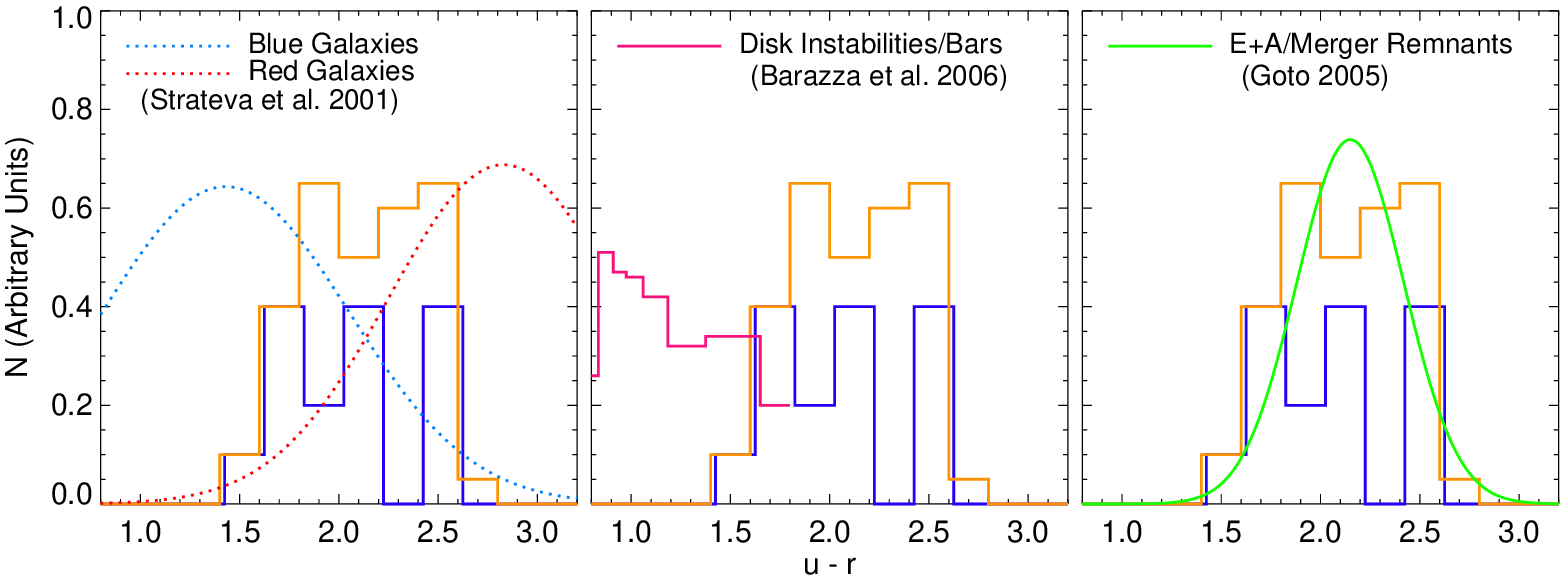}
    \caption{Distribution of quasar host galaxy colors from Figure~\ref{fig:qso.cmd}
    (histograms; from \citet{sanchez:qso.host.colors} and \citet{nandra:qso.host.colors} 
    in dark blue and orange, respectively). We compare with fitted (Gaussian) color 
    distributions of blue cloud and red sequence galaxies from \citet{strateva:color.bimodality}, 
    with the distribution of colors of barred galaxies in the SDSS from
    \citet{barazza:bar.colors} (the expected quasar hosts in a secular or instability-driven 
    quasar fueling model), and with the fitted (Gaussian) distribution of post-starburst 
    (generally merger remnant) E+A/K+A galaxies in \citet{goto:e+a.merger.connection}. 
    Quasar host colors follow the ``transition'' between blue cloud and red sequence 
    observed and expected in merger remnants, in contrast to the preferentially most 
    gas-rich, blue hosts of observed strong bars. 
    \label{fig:qso.cmd.distrib}}
\end{figure*}

Comparing these qualitative scenarios with observations appears to favor the 
former, merger-driven case. Quasars tend to live redwards of the ``top'' of the 
blue cloud, with the brightest/highest accretion rate 
quasars preferentially in bluer-than-average spheroids in 
the ``green valley'' \citep{kauffmann:qso.hosts,sanchez:qso.host.colors,nandra:qso.host.colors}. 

Figure~\ref{fig:qso.cmd.distrib} shows this quantitatively -- we plot the 
distribution of colors of quasar hosts, compared with that fitted to 
the blue cloud and red sequence, or systems with observed bars and/or 
disk instabilities (the expected quasar hosts in a secular model, regardless of 
quasar duty cycles during a bar phase), and post-starburst (E+A/K+A) 
systems, largely identified as merger remnants and ``blue spheroids'' (see 
the discussion in \S~\ref{sec:mergers:env}). The quasar hosts clearly 
lie preferentially between the blue cloud and red sequence, 
with a color distribution very similar to observed post-starburst galaxies. 

The distribution is quite distinct, however, from observed barred systems, 
which lie overwhelmingly on the blue sequence with, if anything, a bias 
towards the bluest systems (which is expected, as these are the most gas-rich 
and therefore most unstable systems). Even if one assumes that, in the most 
extreme bar instabilities, dust reddening might move the system into the 
``green valley'' as a reddened disk, this appears to contradict the observations 
above which 
find quasars to be in preferentially blue spheroids (even X-ray observations, 
which suffer less severe bias against dust-reddened systems). 
A more rigorous quantitative comparison of the tracks through 
color-magnitude space and the relative abundances in this transition region 
will be the topic of future work \citep[][in preparation]{wuyts:prep}, 
and we stress that these are all relatively low-redshift samples, but studying 
how the mean quasar luminosity and accretion rates scale/decay with the degree of 
reddening or aging of their host stellar populations can provide a powerful 
discriminant between these models.

\begin{figure}
    \centering
    \epsscale{1.07}
    \plotter{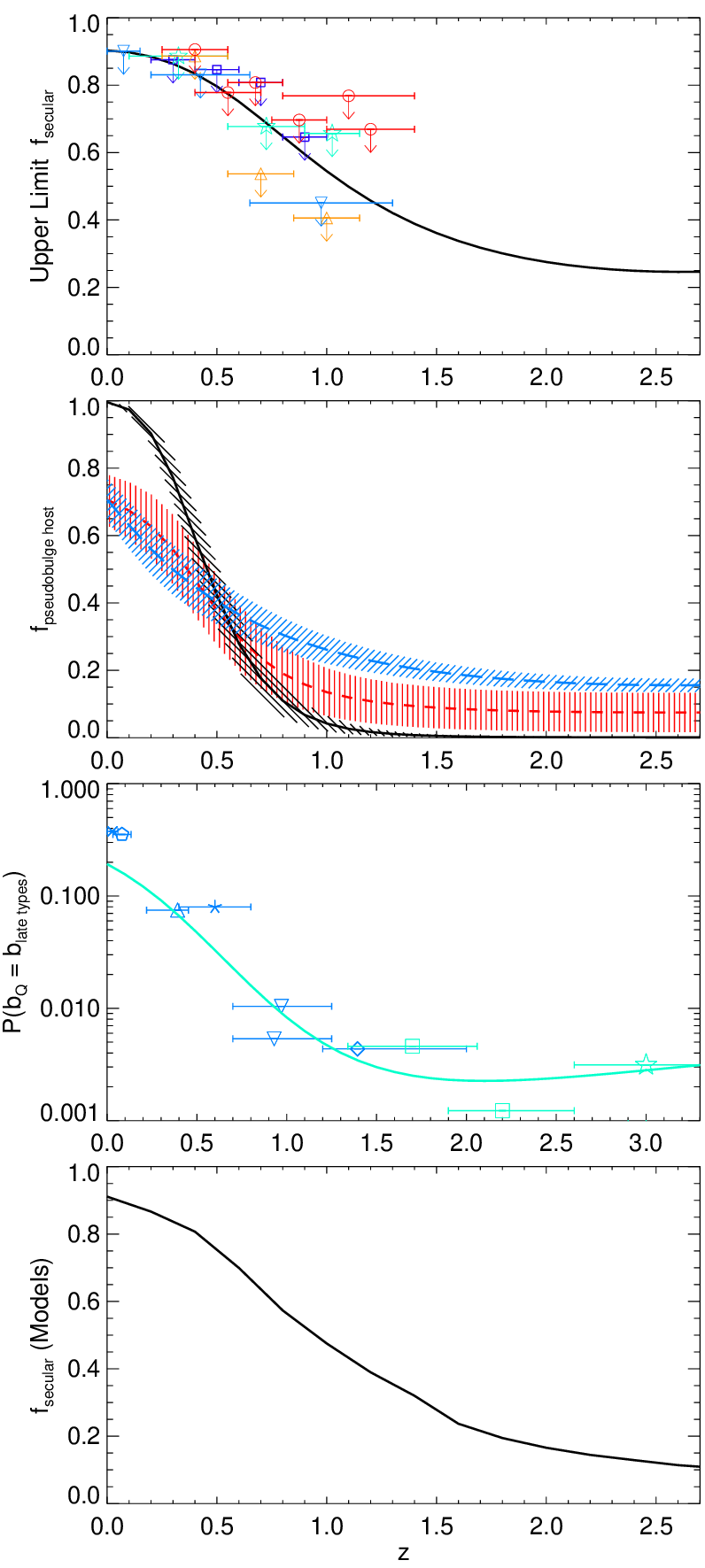}
    \caption{Fraction of the 
    integrated quasar luminosity density owing to 
    non-merger driven secular mechanisms. 
    {\em Top:} Upper limit to the contribution from 
    BHs in disk galaxy hosts at each $z$ (see text). Limits 
    are derived from the observed type-separated mass functions 
    in Figure~\ref{fig:quasar.small.groups} (same style) and \citet[][cyan stars]{franceschini:mfs}.
    Solid line assumes the disk mass function does not evolve with $z$.   
    {\em Second from Top:} Fractional 
    contribution from systems in pseudobulges at $z=0$. 
    Local distribution of pseudobulge masses is 
    estimated from the observed pseudobulge fraction versus 
    galaxy type \citep[][red dashed line, with $\sim1\sigma$
    shaded range]{noordermeer:bulge-disk}, or assuming 
    all bulges with Sersic index $n<2$ are pseudobulges 
    \citep[with the distribution of $n$ versus bulge mass from][black 
    solid line and shading]{balcells:bulge.scaling}, 
    or from directly measured pseudobulge mass functions 
    \citep[][blue long-dashed line and shading]{driver:bulge.mfs}. 
    {\em Second from Bottom:} Probability (from $\chi^{2}$) 
    that observed clustering of quasars (data in Figure~\ref{fig:quasar.bias})
    and star-forming galaxies reflect the same hosts. 
    Solid line is derived from the best-fit to the compilation of \citet{hopkins:clustering} 
    points from the individual measurements included (see Figure~\ref{fig:quasar.bias}).
    {\em Bottom:} Predicted fraction of the luminosity density from the 
    the model for secular fueling from \citet{hopkins:seyferts}, when combined with 
    the merger-driven model herein.
    \label{fig:seyferts.win}}
\end{figure}

There are a number of additional constraints we can place on the contribution to the 
QLF from secular fueling in non-merging disks. Figure~\ref{fig:seyferts.win} 
considers several of these. 
First, we place a limit on secular activity by asking: at a given $z$, what are the 
brightest QSOs possible in disk/star-forming galaxies?  For that redshift, we take the 
observed mass function of star forming galaxies, and convolve with 
$P(\mbh\,|\,\mgal)$ to obtain the hosted BH mass function (assuming 
the most massive disks are Sa/b-type galaxies). Then, assume that every such 
BH is at its Eddington luminosity. At some point (corresponding to 
$\gtrsim2-4\,\mstar$ in the disk mass function) the number density of these mock 
quasars falls below the QLF (which declines much less 
rapidly) at that luminosity and redshift. In other words, at high luminosities, the required BH masses 
from the Eddington limit are too large to live in late-type galaxies. 
To be optimistic, we assume {\em all} the quasar luminosity density below this limit 
is contributed by secular activity in disks. This then gives an upper limit to the 
fraction of the luminosity density from disks. We repeat this procedure for a 
number of different mass functions at different redshifts. In all cases, even this 
limit falls to a fraction $\ll1$ by $z\gtrsim1$, as the QLF $\lstar$ reaches large 
luminosities corresponding to $\mbh\gtrsim10^{8}\,\msun$ BHs at the Eddington limit. Given 
the BH-host spheroid mass relations, this requires a very massive spheroid, easily formed 
in a merger, but not present in even the most early-type disks. 

Second (alternatively), we assume all BHs in pseudobulges were formed via secular 
mechanisms. As discussed in \S~\ref{sec:intro}, there is good reason to believe that this is 
the case, whereas classical bulges must be formed in mergers. For a given 
$z=0$ BH population, we infer an accretion history in the standard fashion from matching the 
BH mass function and continuity equations \citep[e.g.][]{salucci:bhmf,yutremaine:bhmf}. 
We then calculate the fraction of the QLF luminosity density at a given redshift 
from systems which, at $z=0$, live in pseudobulges. We consider this for several 
different observational estimates of the pseudobulge fraction as a function of e.g.\ 
host galaxy morphological type or bulge Sersic index 
\citep{kormendy.kennicutt:pseudobulge.review,balcells:bulge.scaling,allen:bulge-disk,
noordermeer:bulge-disk}, and the directly estimated 
pseudobulge mass functions in \citet{driver:bulge.mfs}. Although the details are sensitive to 
how we define pseudobulges, we find a similar result -- massive BHs which dominate 
the luminosity density at $z\gtrsim1$ live in the most massive bulges/ellipticals, which are 
overwhelmingly classical bulges. 

Third, we calculate the probability that the observed clustering of quasars 
is consistent with that of star forming/disk galaxies (see Figure~\ref{fig:quasar.bias}). 
This is subject to some important caveats -- although quasar clustering depends 
only weakly on luminosity (see Figure~\ref{fig:bias.vs.l}), 
galaxy clustering has been shown to depend quite strongly 
on galaxy luminosity/stellar mass \citep{norberg:clustering.by.lum.type}. 
We use the compilation of 
clustering data from \citet{hopkins:clustering}, as in Figure~\ref{fig:quasar.bias}. At 
$z\lesssim1.5$, we specifically compare the clustering of $\sim\lstar$ quasars 
with that of $\sim\lstar$ blue/star-forming galaxies. For {\em any} model in which quasars 
are driven by secular activity and the statistics of quasar light curves/triggering 
are continuous as a function of host mass/luminosity (i.e.\ there is not 
a second feature in the luminosity function introduced by the 
statistics of the light curves themselves), these should roughly correspond. 
At higher redshift, galaxy clustering as a function of type and luminosity/mass 
at $\sim\lstar$ is not clearly resolved so
we can only plot combined clustering of observed 
star-forming populations (generally selected as Lyman-break galaxies); 
again caution is warranted given the known dependence of 
clustering on galaxy mass/luminosity \citep[for LBGs, see][]{allen:lum.dep.lbg.clustering}. 
Fortunately, the range of particular interest here is $z\lesssim1$, where 
we again find a similar trend -- quasar clustering is consistent with 
secular fueling at $z\sim0$, but by $z\sim1$ this is no longer true. 
As discussed in \citet{hopkins:clustering}, this 
appears to be contrary to some previous claims 
\citep[e.g.,][]{adelbergersteidel:lifetimes}; however, in most cases where 
quasars have been seen to cluster similarly to blue galaxies, either 
{\em faint} AGN populations (not $\sim L_{\ast}$ quasars) or 
bright ($\gg L_{\ast}$) blue galaxies were considered. Indeed, quasars 
do cluster in a manner similar to the {\em brightest} blue galaxies 
observed at several redshifts \citep[e.g.,][at $z\sim1$ and 
$z\gtrsim2$, respectively]{coil:agn.clustering,allen:lum.dep.lbg.clustering}. 
This should not be surprising; 
since quasars require some cold gas supply for their fueling, they cannot be significantly 
more clustered than the most highly clustered (most luminous) population of 
galaxies with that cold gas. 

Finally, we compare these with a simple model expectation. We combine our 
prediction of the merger-driven QLF with the model from \citet{hopkins:seyferts} 
for the QLF driven by secular fueling mechanisms in star-forming 
galaxies. This prediction is based on a simple model of feedback-driven 
self-regulation, calculating the rate of triggering in non-merging disks from 
the observed statistics of gas properties in the central regions of star-forming 
galaxies of different types. The result is similar to the empirical constraints. 

All of these comparisons have important caveats. For example, 
secular mechanisms could act so quickly as to completely 
transform disks to bulges, rapidly making very large BHs (although this 
conflicts with the pseudobulge constraints) from disk hosts. 
Pseudobulges could form in more systems than we 
estimated, but be subsequently transformed to 
classical bulges via major mergers. Clustering could 
be affected by a number of 
systematic uncertainties inherent in e.g.\ the mass and luminosity 
ranges considered. However, these systematics are independent, 
and there is no single loophole which can simultaneously 
reconcile the three constraints considered here with the possibility 
that secular fueling dominates bright $\sim\lstar$ quasar activity 
at $z\gtrsim1$. Although there are differences in detail, 
all the methods we have considered empirically suggest a similar 
scenario: secular (non-major merger related) fueling mechanisms 
contribute little to quasar activity at $z\gtrsim1$, which involves 
the most massive $\mbh\gtrsim10^{8}\,\msun$ BHs in the most 
massive spheroids. By $z\sim0.5$, however, the most massive 
BHs are no longer active, and a significant fraction of the quasar luminosity 
density can come from $\sim10^{7}\,\msun$ BHs in undisturbed hosts. 
By $z\sim0$, the local QLF is largely dominated by Seyfert activity in relatively 
small BHs with late-type, undisturbed host disks \citep{heckman:local.mbh}.

\begin{figure}
    \centering
    \figexpand
    \plotone{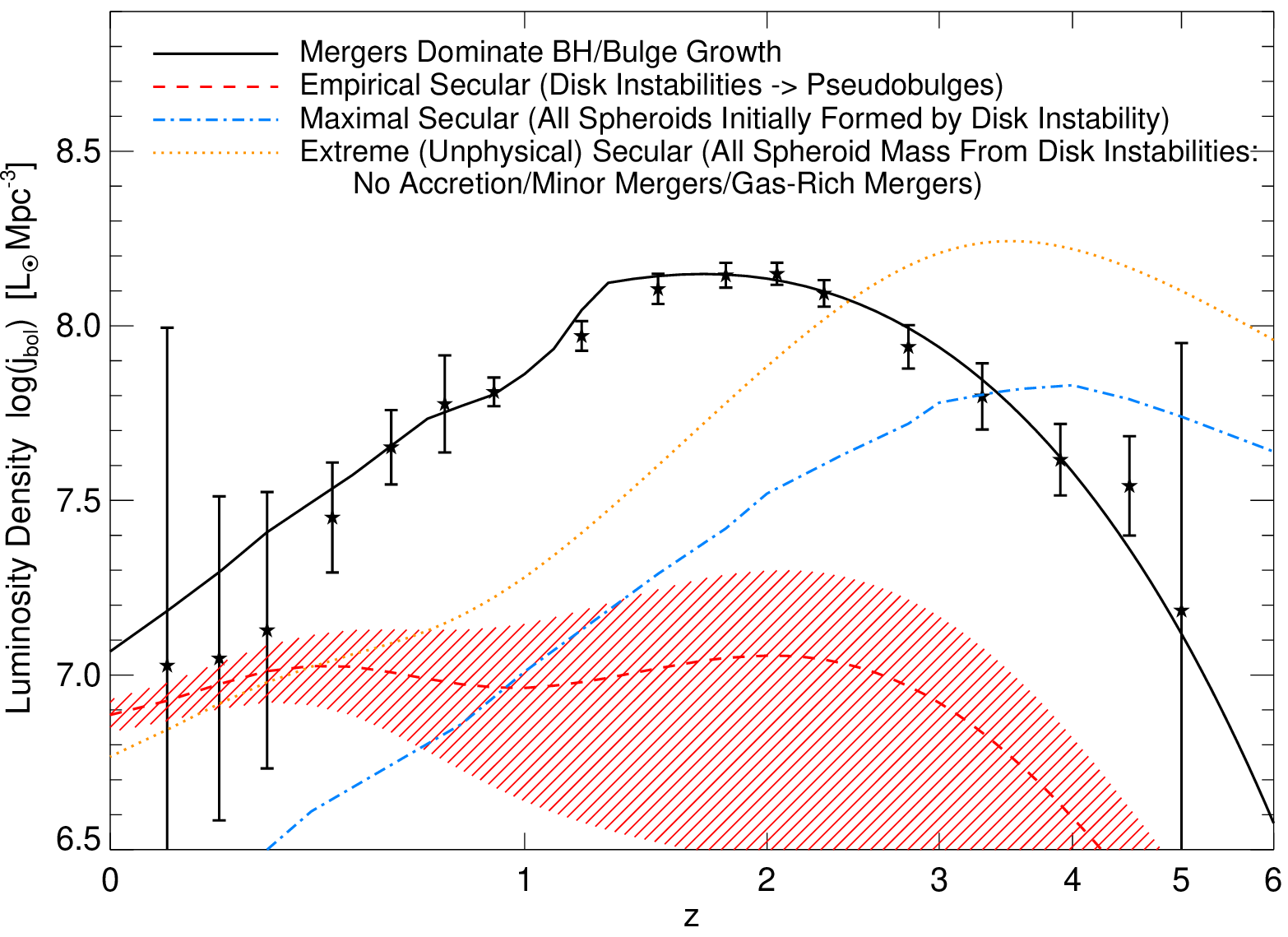}
    \caption{Bolometric quasar luminosity density as a function of redshift. Black stars 
    show the observations from \citet{hopkins:bol.qlf}. Lines show estimates from 
    different models (as labeled):  
    the prediction from a merger-driven model (as in Figure~\ref{fig:lum.density}) and 
    a moderate secular model in which BHs in pseudobulges at $z=0$ 
    were formed in disk instabilities (as in Figure~\ref{fig:seyferts.win}, line 
    in same style) are in good agreement with the luminosity density 
    evolution and empirical constraints on clustering, host galaxy colors, spheroid kinematics, 
    and disk/spheroid mass functions. We compare a maximal secular model, 
    from \citet{bower:sam}, in 
    which most BHs and (even classical) spheroids  
    are initially formed via disk instabilities, and an ``extreme'' secular model, 
    in which all $z=0$ BH mass is formed in such instabilities (same as the 
    maximal secular model, but with no BH growth from cooling, accretion, 
    or mergers; this is unphysical but serves as a strong upper limit). In order for 
    disk instabilities to dominate BH growth, they must act very rapidly, before the 
    (inevitable) major mergers can exhaust gas and form massive spheroids -- 
    this forces such models to predict a luminosity density history offset to earlier 
    times (higher redshifts) compared to the merger-driven model, in 
    disagreement with the observations. 
    \label{fig:lumden.z.models}}
\end{figure}

Even if we ignore these constraints, a model in which secular fueling dominates 
the growth of quasars and BHs has difficulty matching the observed rise and 
fall of the quasar luminosity density with cosmic time. 
Figure~\ref{fig:lumden.z.models} illustrates this. We show the observed 
bolometric quasar luminosity density as a function of redshift, compared to our 
estimate of the merger-driven luminosity density (as in Figure~\ref{fig:lum.density}). 
We also show our estimate of the luminosity density which comes from 
systems which, at $z=0$, live in pseudobulges, calculated as in 
Figure~\ref{fig:seyferts.win}. Again, this fairly moderate, empirical model of 
secular activity can account for the observed luminosity density at low 
redshifts $z\lesssim0.5$, but provides only a small contribution at high redshifts 
$z\gtrsim1$. 

We might, however, imagine a ``maximal'' secular 
model in which {\em all} spheroids 
are initially formed by disk instabilities. Equivalently (for our purposes), 
albeit highly contrived, a model might invoke secular processes to rapidly 
build up BH mass (to the final mass that will be given by the ``future'' 
$\mbh-\sigma$ relation) before a spheroid is formed in later mergers and/or instabilities. 
These have severe difficulty 
reconciling with the kinematics of observed classical bulges 
(see \S~\ref{sec:intro}) and the tightness of the BH-host spheroid correlations, respectively, 
and are not favored by simple dynamical arguments \citep[see, e.g.][]{shen:size.mass}, 
nor the constraints in Figure~\ref{fig:seyferts.win}, 
but they could in principle be invoked. In fact, the semi-analytic model of 
\citet{bower:sam} is effectively such a scenario, in which a 
very strong disk instability mode is analytically adopted, which overwhelmingly 
dominates initial bulge formation and BH growth (mergers contributing $\ll 1\%$ at all redshifts). 
We therefore compare their estimate for the total quasar luminosity density 
(accretion rate density) as a function of time. 
Finally, in the default \citet{bower:sam} model, there is still some growth of BHs 
via accretion from the diffuse ISM, cooling, and mergers (major and minor). We 
therefore also adopt an even more extreme 
secular model, in which we reproduce the \citet{bower:sam} analysis with an 
even stronger disk instability mode -- essentially renormalizing the model such 
that all $z=0$ bulge mass was formed in this ``secular'' mode (i.e.\ we allow 
{\em no} subsequent growth via other mechanisms, and demand that the observed 
integrated $z=0$ BH mass density be matched by the integrated secular mode growth). 
This latter model is of course unphysical, but yields a hard upper limit to 
secular-mode growth.

It is immediately clear that the ``maximal'' secular model predicts that the quasar luminosity 
density should peak at much higher 
redshifts $z\sim4$ than the observed $z\sim2$. In general, 
the rise and fall of the quasar luminosity density in such a model are offset to earlier 
times. The reason for this is simple: in a fully cosmological model, mergers are 
{\em inevitable}. And, whether or not most quasars are triggered by mergers, it is 
extremely difficult to contrive a major, gas-rich merger without BH accretion and 
spheroid formation, with most of the gas being consumed by star formation. The only way that 
a secular or disk instability model can dominate the integrated buildup of BH mass and 
quasar luminosity density is to ``beat mergers to the finish,'' i.e.\ to generally operate 
early and rapidly enough such that the BHs have been largely formed, and gas already 
exhausted, by the time massive galaxies undergo their first major mergers. In such models, 
then, one is forced to predict that the quasar luminosity density peaks at very early times 
and has largely declined (i.e.\ most of the gas in massive 
systems has already been exhausted) by $z\sim2$. 

Finally, this relates to a more general point. The quasar luminosity density 
\citep[and especially the number density of bright quasars corresponding to 
$\gtrsim10^{8}\,\msun$ BHs at high Eddington ratio; see][]{fan04:qlf,richards:dr3.qlf} declines 
rapidly at $z\gtrsim2-3$ (roughly as $\sim(1+z)^{4-6}$), compared to the 
global star formation rate density of the Universe, which is relatively flat 
at these redshifts \citep[declining as $\sim(1+z)^{0-1.5}$ from $z\sim2-6$;][]{hopkinsbeacom:sfh}. 
This has long been recognized, and cited as a reason why quasars and BH growth cannot 
explain reionization at high redshifts (since, similar to the global star formation history, the 
UV background declines slowly at these redshifts). It further implies that BH growth 
(at least at the masses of interest for our predictions here) cannot generically 
trace star formation. This places strong constraints on secular models, as above, as well as 
models in which essentially all high-redshift star formation is in bulges or 
some sort of dissipational collapse \citep[e.g.][]{granato:sam,lapi:qlf.sam}. Some process
must delay the formation of massive BHs, while allowing star and galaxy formation to 
proceed efficiently at high redshifts. A natural explanation is that massive BH formation 
requires major mergers. In our model, at high redshifts, low-mass galaxies can efficiently form 
(and potentially build low-mass BHs via secular instabilities), but they are 
predominantly disks, which efficiently turn gas into stars and do not form very massive 
bulges or BHs. Only later, once their hosts have grown more massive, are they likely to 
undergo major mergers, which transform the disks into spheroids and build correspondingly 
massive BHs. This automatically explains the much sharper rise and fall of the quasar 
luminosity density and number density of bright quasars, relative to the 
shallow evolution in the star formation rate density and ionizing background 
of the Universe at high redshifts. 

\section{Discussion}
\label{sec:discussion}

We have developed a theoretical model for the cosmological 
role of galaxy mergers, which allows us to make predictions for various 
merger-related populations such as starbursts, quasars, and 
spheroidal galaxies. 
By combining theoretically well-constrained 
halo and subhalo mass functions as a function of redshift and 
environment with empirical halo occupation models, we can estimate where 
galaxies of given properties live at a given epoch. This allows us to 
calculate, in an {\em a priori} cosmological manner, where major galaxy-galaxy 
mergers occur and what kinds of galaxies merge, at all redshifts. 

We compare these estimates to a number of observations, including 
observed merger mass functions; merger fractions as a function of 
galaxy mass, halo mass, and redshift; the mass flux/mass density in 
mergers; the large-scale clustering/bias of merger populations; 
and the small-scale environments of mergers, and show 
that this approach yields robust predictions in good agreement with
observations, and can be extended to predict detailed properties 
of mergers at all masses and redshifts. 
There are some uncertainties in this approach. However, we 
re-calculate all of our predictions adopting different estimates for the 
subhalo mass functions and halo occupation model (and its redshift 
evolution) and find this makes little difference (a factor $<2$) at all 
redshifts. The largest uncertainty comes from our calculation of 
merger timescales, where, at the highest redshifts ($z\gtrsim3$), merging via 
direct collisional processes might be more efficient than 
merging via dynamical friction, given the large physical densities. 
More detailed study in very high-resolution numerical simulations will 
be necessary to determine the effective breakdown between different 
merger processes.
Nevertheless, the difference in our predictions at these redshifts is still 
within the range of observational uncertainty. 
Ultimately, we find that our predictions are robust 
above masses $\mgal\gtrsim10^{10}\,\msun$, regardless of these 
possible changes to our model, as the theoretical 
subhalo mass functions and empirical halo occupation models 
are reasonably well-constrained in this regime. 

In addition to these specific observational predictions and tests, 
our model allows us to examine the physical origins of the distribution of 
major mergers of different galaxy masses and types. For example, 
there is a naturally defined major-merger scale (host halo mass $\mhalo$) for 
galaxies of mass $\mgal$ -- the ``small group scale,'' only slightly larger than 
the average halo hosting a galaxy of mass $\mgal$. This is the scale at which 
the probability to accrete a second galaxy of comparable mass $\sim\mgal$ (fuel for a 
major merger) first becomes significant. At smaller (relative) 
halo masses, the probability that the halo 
hosts a galaxy as large as $\mgal$ declines rapidly. At larger masses, the 
probability that the halo will merge with or accrete another halo hosting a comparable $\sim\mgal$ 
galaxy increases, but the efficiency of the merger of these galaxies declines rapidly. 
We stress that this small group scale is indeed small -- the 
average small group halo will still host only 1 galaxy 
of mass $\sim\mgal$, and groups will only consist of $2-3$ members of similar mass. 
We also note that this does not mean that mergers occur (in a global sense) on a specific scale, 
since the small group scale is different for different galaxy masses.
In fact, a consequence of this model is that mergers occur in halos of 
all masses and in all environments (including field and even void environments), as is observed 
\citep{alonso:groups,goto:e+a.merger.connection,hogg:e+a.env}, although 
the characteristic masses 
and star formation histories 
of galaxies merging may reflect their different environments/halo masses. 
Similarly, our model allows us to accurately predict and understand the 
(relatively weak) evolution of the merger fraction with redshift, and the 
relative evolution in merger rates as a function of mass (evolution of the 
major merger mass functions). The clustering properties and dependence of 
merger rates on both large-scale and small-scale environment are natural 
consequences of the fundamentally local nature of mergers, and we 
study in detail the effects of environment on merger rates as a function of scale. 

Having characterized mergers in this way, we examine the role 
that mergers play in triggering quasars. Even if there are other quasar ``triggers'' 
dominant at some luminosities/redshifts, it is difficult to imagine a scenario in which the 
strong nuclear gas inflows from a merger do not cause 
rapid, near Eddington-limited accretion and ultimately yield some kind of quasar 
-- and indeed such activity is ubiquitous in late-stage mergers 
\citep{komossa:ngc6240,alexander:xray.smgs,
borys:xray.ulirgs,brand:xray.ir.contrib}. We therefore make the simple 
ansatz that gas-rich, major
mergers will produce quasars (but do, in principle, allow for other 
fueling mechanisms as well). This model, with just the contribution of mergers 
to the quasar luminosity density, is able to account for 
the observed quasar luminosity density from $z=0-6$. 
The rise and fall of the luminosity density with redshift, as well as 
the shape and evolution of the quasar luminosity function, are 
accurately reproduced. This also yields predictions of the local black hole 
mass function, cosmic X-ray background \citep[see][]{hopkins:qso.all}, 
AGN fractions as a function of galaxy mass/luminosity and 
redshift, large scale quasar clustering as a function of luminosity and redshift, 
small-scale quasar clustering excesses, quasar host galaxy colors, 
and infrared luminosity functions, all in good agreement with those observed.
In particular, matching the history of the bolometric 
luminosity density of quasars requires no knowledge or assumptions about 
quasar duty cycles, light curves, or lifetimes, only our determination of the 
global mass density in gas-rich major mergers. 

In our model, the sharp rise and fall of the quasar luminosity density over 
cosmic time is the product of several factors. At high redshifts, the 
buildup of BH mass from $z\gtrsim6$ to $z\sim2$ owes in part to 
the growth of galaxy and halo mass, as most galaxies are rapidly forming, 
and the galaxy mass density involved in major mergers steadily 
increases with time. The rise is steeper than that in, for example, the 
global star formation rate density of the Universe, as it tracks 
just the major merger history (effectively, at these redshifts, the rise in the 
density of relatively massive ``small group'' sized halos), as opposed to the global buildup of 
the (relatively lower-mass) halos hosting the most rapidly star-forming galaxies. 
Below redshift $z\sim2$, merger rates begin to decline 
for all galaxies, and the exhaustion of gas in evolved systems 
slows the growth of quasars in two ways. First, major 
mergers of relatively gas-poor disks create shallower central potential 
wells for the remnant spheroid (i.e.\ lower $\sigma$ values), and 
as a consequence BH growth self-regulates at lower masses 
\citep{hopkins:bhfp}, in agreement with the observed evolution of 
the BH-host correlations with redshift \citep[e.g.,][]{peng:magorrian.evolution}. Second, an 
increasing fraction of galaxies (especially around $\sim\lstar$, where 
most of the mass density resides) have already undergone major 
mergers and exist as ``quenched'' spheroids (with very 
little remaining cold, rotationally supported gas) 
whose major mergers will not excite quasar activity. 
Recent high-resolution cosmological simulations which 
attempt to resolve the relevant merger and feedback effects 
regulating BH growth \citep{sijacki:radio,dimatteo:cosmo.bhs} further support this scenario, 
with the combination of these effects and, primarily, the merger 
history of the Universe regulating BH growth (at least at redshifts 
$z\lesssim6$). The product of these 
effects yields the observed steep rise and fall of the quasar population 
with respect to its peak at $z\sim2$, in good agreement with the 
observations and in contrast with the substantially more extended 
global star formation history of the Universe. 

We compare this model to one in which quasar fueling is primarily 
driven by secular processes -- i.e.\ disk instabilities, bars, harassment, 
or any process which operates in non-merging, gas-rich systems. 
We demonstrate that there are a number of robust, qualitatively distinct 
predictions from these models, including: 

{\em Quasar Clustering:} A merger-driven model accurately predicts 
the observed large-scale clustering of quasars (both at $\sim\lstar$ and as a detailed 
function of luminosity) as a function of redshift for the observed 
range $z\sim0.5-4$. 
The clustering is, at all these redshifts, precisely that predicted for 
``small group'' halos in which major mergers of gas-rich galaxies should proceed 
most efficiently. It is well-established empirically that quasar clustering 
traces a characteristic host halo mass
\citep{porciani2004,
wake:local.qso.clustering,croom:clustering,porciani:clustering,
myers:clustering,daangela:clustering,coil:agn.clustering,
shen:clustering,hopkins:clustering}, 
and investigations of the quasar proximity effect 
reach a similar conclusion \citep{faucher:proximity,kim:proximity,guimaraes:proximity}.
Comparing this to independent, direct measurements of the small group 
scale of $\sim\lstar$ gas-rich galaxies, and to the small group 
scale inferred from a wide variety of different halo occupation models, we show 
in all cases that these trace the same mass. 
In contrast, the clustering of typical star-forming galaxies is somewhat weaker 
(as expected relative to their small group scale), and yields an underestimate of 
quasar clustering at moderate and high redshifts. Only at low redshifts 
($z\lesssim0.5$) is there reasonable consistency between the clustering of 
$\sim\lstar$ quasars and ``secular'' populations 
\citep[for more details, see][]{hopkins:clustering}.

{\em Small-Scale Environments:} Mergers will preferentially occur in environments 
with an overdensity of galaxies on small scales, and as a consequence their 
clustering should reflect a bias (relative to a mean galaxy of the same mass) to 
excess clustering on small scales. Furthermore, triggering of binary quasars in (even 
a small fraction of) early interacting pairs can enhance this excess. 
Indeed, in a purely empirical sense, both bright quasars at all redshifts 
$z\sim0.5-3$ \citep{hennawi:excess.clustering,serber:qso.small.scale.env,
myers:clustering.smallscale} and local 
post-starburst merger remnant galaxies \citep{goto:e+a.merger.connection} are observed to 
have similar, strong excess clustering on small scales, distinct from 
quiescent (non-merger related) populations. 
This is true both in terms of the quasar-quasar autocorrelation, and for 
the quasar-galaxy cross-correlation, suggesting that it reflects a true tendency for quasars 
to reside in regions of small-scale overdensity. Our model predicts the 
magnitude of this excess clustering as a function of physical scale 
and redshift well for both populations. Interestingly, low-luminosity 
Seyfert galaxies ($M_{B}>-23$) are observed 
without such an excess on small scales \citep{serber:qso.small.scale.env}, as expected if 
AGN triggering at low luminosities (or typical $\mbh\lesssim10^{7}\,\msun$) 
is dominated by secular processes (with the true quasar populations dominated 
by mergers). However, systems of these low luminosities contribute 
significantly to the quasar luminosity density at only very low redshifts $z\lesssim0.5$, 
once more massive systems have predominantly quenched. 

{\em Host Galaxy Colors:} The stellar population colors of a 
gas-rich merger remnant will rapidly redden, at least over the $\sim$\,Gyr period 
over which subsequent infall or cooling can be ignored, and the system 
will (even if only temporarily) cross the ``green valley'' between the blue cloud and 
red sequence. If a quasar is triggered at the end of a merger, the decay of the 
quasar lightcurve should be associated with the host crossing this interval, or 
equivalently with the presence of a relatively young, blue host spheroid. 
Observed quasar hosts at $z\sim0.5-1.1$ appear to preferentially occupy this 
(otherwise relatively empty) locus in color-magnitude space 
\citep{sanchez:qso.host.colors,nandra:qso.host.colors}, 
and it is well-established that bright quasar hosts tend to be 
massive spheroids with especially young stellar or post-starburst stellar 
populations \citep[e.g.][and references therein]{canalizostockton01:postsb.qso.mergers,
jahnke:qso.host.sf,vandenberk:qso.spectral.decomposition,barthel:qso.host.sf}. 
We show that the color distribution of observed quasar hosts 
is similar to that observed for clear post-starburst merger remnant 
populations. In contrast, a secular model (regardless of the quasar duty cycle or lifetime) 
would predict that quasar hosts trace the population of systems hosting 
strong disk instabilities or bars (unless any quasar activity could somehow be suppressed 
over the entire lifetime of a relatively long-lived bar) -- these actually 
tend to be the most blue, gas-rich disk galaxies. We show that the observed colors of quasar 
hosts are distinct from those of systems observed hosting strong bars.  

{\em Host Kinematics (Pseudobulges versus Classical Bulges):} Numerical 
simulations and observations of both barred systems and merger remnants 
have established that mergers yield systems with the observed kinematic and 
photometric properties of classical bulges, whereas secular disk instabilities 
generically give rise to pseudobulges with distinct properties
(see the discussion in \S~\ref{sec:intro}). At high redshifts $z\gtrsim1$, the active 
$\sim\lstar$ quasar populations (either from direct quasar BH mass measurements or 
simply the Eddington argument) are dominated by massive BHs 
($\mbh\gtrsim10^{8}\,\msun$), which are directly observed to live in massive bulges 
at those redshifts \citep{peng:magorrian.evolution}, and whose remnants clearly live in massive bulges 
locally. These spheroids ($M_{\rm sph}\gtrsim10^{11}\,\msun$) 
are overwhelmingly classical spheroids (in particular, classical true ellipticals), 
whose kinematics argue that they were formed in mergers. To the extent that the buildup 
of BH mass traces spheroid origin (true at all redshifts observed, albeit with 
potentially redshift-dependent efficiency), this implies formation in mergers. 
Adopting a number of different estimates of e.g.\ the pseudobulge fraction as a 
function of host properties, pseudobulge mass distributions, or simply assuming 
all bulges in star-forming/disk-dominated galaxies are formed via secular instabilities, 
we compare with the distribution of active BH masses in the quasar luminosity function 
at all redshifts, and show that these populations cannot dominate the 
QLF at redshifts $z\gtrsim1$. Only at low redshifts $z\lesssim1$ are the 
global QLF and buildup of BH mass occurring mainly
in systems which typically reside in star-forming, disk-dominated hosts 
with pseudobulges potentially formed via disk instabilities or bars. 

{\em Quasar Luminosity Density versus Redshift:} As noted above, a 
merger-driven model predicts a sharp rise and fall of the quasar luminosity 
density in good agreement with observations. If, for the sake of argument, we 
adopt a model in which all BH growth is driven by disk instabilities, 
we demonstrate that, once embedded in a proper cosmological context, 
such a model is generically forced to predict a history of quasar luminosity density 
which is offset to earlier times (in each of its rise, peak, and fall), in 
conflict with the observations. This is because major mergers are dynamically 
inevitable -- one cannot simply ``remove'' the mergers a galaxy will undergo 
in a true cosmological model. In order for disk instabilities to dominate BH growth 
or spheroid formation, they must, therefore, act before massive systems undergo 
their major mergers. Since the global mass flux in gas-rich major mergers 
peaks around $z\sim2-3$, a secular-dominant model is forced to assume a sufficiently 
strong disk instability mode such that the progenitors of these systems 
rapidly exhaust their gas supplies and build up most of their final BH/spheroid mass 
at redshifts $z\gtrsim4$. By $z\sim2$, then, these models predict the quasar luminosity 
density is already in rapid decline. We demonstrate this both for current state-of-the-art 
semi-analytic models \citep{bower:sam}, constrained such that they cannot overproduce 
the $z=0$ mass density in quenched systems nor ``avoid'' major mergers, and 
simple illustrative toy models. 
The only way to avoid this is to weaken the disk 
instability criterion -- i.e.\ to assume disk instabilities are not so efficient at exhausting 
systems, and can therefore act continuously over longer times. But then, one obtains 
a prediction similar to our expectation from assuming all pseudobulges are formed 
in disk instabilities -- namely, the high rate of gas-rich mergers at high redshifts will 
dominate quasar activity at all $z>1$, and this ``gentler'' disk instability mode will 
dominate at lower luminosities (i.e.\ only dominate BH mass buildup at low 
masses $\mbh\lesssim10^{7}\,\msun$), becoming important to the 
total luminosity density only at $z<1$.

These comparisons, despite the very different possible systematic effects 
in the observations, all suggest a similar scenario. 
Secular (non-merger related) fueling mechanisms may dominate 
AGN activity in low-BH mass systems ($\mbh\lesssim10^{7}\,\msun$), 
for which mergers are relatively rare 
and hosts tend to be very gas-rich, potentially bar-unstable disks, but these 
contribute little to quasar activity at $z\gtrsim1$, which involves 
the most massive $\mbh\gtrsim10^{8}\,\msun$ BHs in the most 
massive spheroids. By $z\sim0.5$, however, the most massive 
BHs are no longer active (their hosts having primarily been gas exhausted and 
quenched, and with overall merger rates declining), 
and a significant fraction of the AGN luminosity 
density can come from $\sim10^{7}\,\msun$ BHs in undisturbed hosts, corresponding 
to relatively low-luminosity ($M_{B}>-23$) Seyfert galaxies. 
By $z\sim0$, the local QLF is largely dominated by Seyfert activity in relatively 
small BHs with late-type, undisturbed host disks \citep{heckman:local.mbh}. 
Our models allow for secular mechanisms, such as the stochastic triggering 
model of \citet{hopkins:seyferts}, to be important at low luminosities, and 
a pure comparison between this secular model and our merger-driven 
prediction here yields a transition to secular dominance at low luminosities 
in good agreement with the empirical constraints. 

Ultimately, one would like to test this by directly studying the morphology of 
true, bright quasar hosts at high redshifts. Unfortunately, 
as discussed in \S~\ref{sec:intro}, this remains extremely difficult, and 
results have been ambiguous.
As noted previously, mock observations constructed from numerical major merger simulations 
\citep{krause:mock.qso.obs}
imply that, with the best presently attainable data, the faint, rapidly 
fading tidal features associated with the quasar phase (i.e.\ final stages of the merger, 
at which the spheroid is largely formed and has begun to relax) are difficult to 
observe even locally and (for now) nearly impossible to identify at the 
redshifts of greatest interest ($z\gtrsim1$). Similarly, experiments with automated, 
non-parametric classification schemes \citep{lotz:gini-m20} suggest that the hosts will 
generically be classified as ``normal'' spheroids, even with perfect resolution and no 
surface brightness dimming. 
This appears to be borne out, as recently 
\citet{bennert:qso.hosts} have re-examined very 
low-redshift quasars previously recognized from 
deep HST imaging as having relaxed spheroid hosts, and found (after 
considerably deeper integrations) that every such object shows clear evidence for 
a recent merger. The ability to identify such features may be slightly improved if 
one considers just the population of highly dust-reddened (but still dominated by quasar 
light in the optical/near IR) or IR-luminous quasar expected to be associated with a 
(brief) ``blowout'' stage preceding the more typical optical quasar phase in a merger, and 
it does appear that observations of quasars in this stage, somewhat closer to the peak of 
merger activity, show ubiquitous evidence of recent or ongoing mergers 
\citep{hutchings:redqso.lowz,hutchings:redqso.midz,
kawakatu:type1.ulirgs,guyon:qso.hosts.ir,urrutia:qso.hosts}, albeit still requiring 
very deep integrations. 

On the other hand, it is increasingly possible to 
improve the constraints we have studied in this paper, to break the degeneracy between 
secular and merger-driven models of quasar fueling. Improving measurements of 
merger fractions, mass functions, and clustering 
at low redshifts, and extending these measurements to high redshifts, can break 
the degeneracies in our cosmological models (regarding, for example, the appropriate 
merger timescales at high redshifts) and enable more robust, tightly constrained predictions. 
We have also made a large number of predictions in this paper and previous related 
works \citep[e.g.][]{hopkins:qso.all,hopkins:clustering} which can be directly tested 
without the large ambiguities presently inherent in quasar host morphology estimates. 
Better observations of quasar 
host galaxy colors (and corresponding estimates of their recent star formation history), 
improved measurements of quasar clustering at redshifts $z\gtrsim3$ (especially 
measurements which can resolve $\sim\lstar$ quasars at these redshifts), 
detailed cross-correlation measurements of quasars and other galaxy populations 
and clustering measurements which 
can decompose the excess bias of quasars on small scales as a function of 
e.g.\ redshift and luminosity, improved constraints on the bolometric corrections of 
the brightest quasars and the history of the bolometric quasar luminosity density 
at $z\gtrsim3-4$, and estimates of the evolution with redshift of pseudobulge populations 
will all be able to test the models presented in this paper. The combination of these 
observations can greatly strengthen the constraints herein, and ultimately allow for 
more detailed modeling which attempts not just to predict the general origin of quasars in 
mergers, but to fully break down the contribution of major mergers (or mergers of different 
types) and other fueling 
mechanisms to the quasar luminosity functions as a function of luminosity and redshift. 

\acknowledgments We thank Josh Younger, Volker Springel, Gordon Richards, 
Chris Hayward, Alice Shapley, Jenny Greene, 
and Yuexing Li for helpful discussions.
This work was supported in part by NSF grant AST
03-07690, and NASA ATP grants NAG5-12140, NAG5-13292, and NAG5-13381.

\bibliography{ms}

\end{document}